\documentclass[11pt]{article}

\usepackage[textwidth=15.5cm,textheight=22.5cm]{geometry}
\usepackage{bm}
\usepackage{amsmath,amssymb,color,graphicx,amscd,amsfonts}
\usepackage{latexsym}
\usepackage{graphicx}
\usepackage{cite}
\usepackage{subfig}
\usepackage{bbm}
\usepackage{ulem}

\def\ul#1{{\underline{#1}}}
\def\div{{{\rm div}\, }}
\def\rot{{{\rm rot}\, }}

\def\a{\alpha}
\def\b{\beta}
\def\d{\delta}
\def\e{\epsilon}

\def\s{\sigma}

\def\th{\theta}

\def\hA{{\hat{A}}}
\def\ha{{\hat{a}}}
\def\hi{{\hat{i}}}
\def\hj{{\hat{j}}}
\def\hk{{\hat{k}}}
\def\hl{{\hat{l}}}
\def\hz{{\hat{z}}}
\def\hw{{\hat{w}}}
\def\hrho{{\hat{\rho}}}

\def\tA{\totalarea}
\def\totalarea{[\sigma]}
\def\Tr{\mbox{Tr}}

\newcommand{\AdStS}{AdS$_4\times S^7/\Z_k$}
\def\Jt{J_M}

\def\der{\partial}

\allowdisplaybreaks[4]

\numberwithin{equation}{section}
\newcommand{\be}{\begin{equation}}
\newcommand{\ee}{\end{equation}}
\newcommand{\ndt}{\noindent}
\def\bea{\begin{eqnarray}}
\def\eea{\end{eqnarray}}
\def\beas{\begin{eqnarray*}}
\def\eeas{\end{eqnarray*}}
\def\sla{\raise.15ex\hbox{$/$}\kern-.57em}

\newcommand\fr[1]{\frac{1}{#1}}

\newcommand {\nn} {\nonumber}

\renewcommand{\th}{\theta}

\allowdisplaybreaks[1]

\newcommand{\er}{{\rm e}}
\newcommand{\dr}{{\rm d}}
\newcommand{\tr}{{\rm tr}}
\newcommand{\R}{\mathbbm{R}}
\newcommand{\Z}{\mathbbm{Z}}

\newcommand{\CP}{\mathbbm{C}\mathbbm{P}}

\newcommand{\del}{\partial}
\newcommand{\ba}{\begin{eqnarray}}
\newcommand{\ea}{\end{eqnarray}}
\newcommand{\bdm}{\begin{displaymath}}
\newcommand{\edm}{\end{displaymath}}
\newcommand{\ra}{\rangle}

\def\b{\beta}
\def\a{\alpha}
\def\g{\gamma}
\def\s{\sigma}

\def\veps{\varepsilon}

\def\d{\delta}
\def\D{\Delta}
\def\v{\varphi}

\newcommand{\half}{\frac{1}{2}}

\newcommand{\ie}{{\it i.e.\ }}
\newcommand{\eg}{{\it e.g.\ }}

\newcommand{\calN}{{\mathcal N}}

\newcommand{\calR}{{\mathcal R}}

\DeclareMathAlphabet{\mathpzc}{OT1}{pzc}{m}{it}

\newcommand\hsp[1]{\hspace*{#1 cm}}
\newcommand\vsp[1]{\vspace*{#1 cm}}
\makeatletter
\newif\if@borderstar
\def\bordermatrix{\@ifnextchar*{%
 \@borderstartrue\@bordermatrix@i}{\@borderstarfalse\@bordermatrix@i*}%
}
\def\@bordermatrix@i*{\@ifnextchar[{\@bordermatrix@ii}{\@bordermatrix@ii[()]}}
\def\@bordermatrix@ii[#1]#2{%
\begingroup
 \m@th\@tempdima8.75\p@\setbox\z@\vbox{%
 \def\cr{\crcr\noalign{\kern 2\p@\global\let\cr\endline }}%
 \ialign {$##$\hfil\kern 2\p@\kern\@tempdima & \thinspace %
  \hfil $##$\hfil && \quad\hfil $##$\hfil\crcr\omit\strut %
  \hfil\crcr\noalign{\kern -\baselineskip}#2\crcr\omit %
  \strut\cr}}%
 \setbox\tw@\vbox{\unvcopy\z@\global\setbox\@ne\lastbox}%
 \setbox\tw@\hbox{\unhbox\@ne\unskip\global\setbox\@ne\lastbox}%
 \setbox\tw@\hbox{%
  $\kern\wd\@ne\kern -\@tempdima\left\@firstoftwo#1%
  \if@borderstar\kern 2pt\else\kern -\wd\@ne\fi%
 \global\setbox\@ne\vbox{\box\@ne\if@borderstar\else\kern 2\p@\fi}%
 \vcenter{\if@borderstar\else\kern -\ht\@ne\fi%
  \unvbox\z@\kern -\if@borderstar2\fi\baselineskip}%
 \if@borderstar\kern-2\@tempdima\kern2\p@\else\,\fi\right\@secondoftwo#1 $%
 }\null \;\vbox{\kern\ht\@ne\box\tw@}%
\endgroup
}
\makeatother
\date{}

\begin{document}

\thispagestyle{empty}

\setcounter{page}{0}

\begin{flushright} 
%\today\\
DIAS-STP-13-09, KEK-TH-1640, OIQP-13-09 \\

\end{flushright} 

\vspace{0.1cm}

\begin{center}
{\LARGE
Membranes from monopole operators in ABJM theory: \\
large angular momentum and M-theoretic AdS$_4$/CFT$_3$ 
\rule{0pt}{20pt}  }
\end{center}

\vspace*{0.2cm}

\renewcommand{\thefootnote}{\alph{footnote}}

\begin{center}

Stefano K{\sc ovacs}\,$^{\ddag\,}$\footnote
{
E-mail address: skovacs@stp.dias.ie },
Yuki S{\sc ato}\,$^{\flat,\natural\,}$\footnote
{
E-mail address: satoyuki@post.kek.jp
}
and
Hidehiko S{\sc himada}\,$^{\S\,}$\footnote
         {
E-mail address: shimada.hidehiko@googlemail.com}

\vspace{0.3cm}

${}^{\ddag}$ {\it Dublin Institute for Advanced Studies, Dublin, Ireland }\\
${}^{\flat}$ {\it High Energy Accelerator Research Organization (KEK), 
Tsukuba, Ibaraki 305-0801, Japan}\\
${}^{\natural}$ {\it Department of Particle and Nuclear Physics,
Graduate University for Advanced Studies (SOKENDAI), Tsukuba, Ibaraki 
305-0801, Japan}\\ 
${}^{\S}$ {\it Okayama Institute for Quantum Physics, Okayama, Japan}\\

\end{center}

\vspace{1cm}

\begin{abstract}
We consider states with  large angular momentum to facilitate the
study of the M-theory regime of the AdS$_4$/CFT$_3$ correspondence.
More precisely,  we study the duality between M-theory on AdS$_4
\times S^7/\Z_k$ and  $\mathcal{N}=6$ supersymmetric
Chern--Simons-matter (ABJM) theory  with gauge group
U($N$)$\times$U($N$) and level $k$,  in the regime where $k$ is of
order one and $N$ is large.  In this regime the study of both sides
of the duality is challenging: the lack of an explicit formulation of
M-theory in  AdS$_4 \times S^7/\Z_k$ makes the gravity side difficult,
while the CFT side is strongly coupled and the planar approximation is
not applicable.  In order to overcome these difficulties, we focus on
states on the gravity side  with large orbital angular momentum $J\gg 1$ 
associated with a single plane of rotation in $S^7$. We then identify the 
corresponding operators in the CFT, thereby establishing the AdS/CFT 
dictionary in this large angular momentum sector.  We show that there are 
natural approximation schemes on both sides of the correspondence as a 
consequence of the presence of the small parameter $1/J$.  On the AdS 
side, the sector we focus on is well-approximated by the matrix model of 
M-theory -- with matrices of size $J/k$ --  defined on the maximally 
supersymmetric eleven-dimensional pp-wave background.  The pp-wave 
approximation to M-theory in AdS$_4\times S^7/\Z_k$ is justified  for 
$1\ll J \ll N^{1/2}$,  while loop corrections in the matrix model are 
suppressed compared to tree-level contributions for $J\gg N^{1/3}$.  On the 
CFT side, we study the ABJM theory defined on $S^2 \times \R$ with large
magnetic flux $J/k$. Using  a carefully chosen gauge, we find an expansion 
of the Born--Oppenheimer type which arises naturally for large $J$ in spite 
of the theory being strongly coupled. The energy spectra computed on the 
two sides agree at leading order.  This provides a highly non-trivial test of 
the AdS$_4$/CFT$_3$ correspondence including near-BPS observables
associated with membrane degrees of freedom, therefore extending the 
validity of the AdS$_4$/CFT$_3$ duality beyond the previously studied 
sectors  corresponding to either BPS supergravity observables or the type 
IIA string regime.
\end{abstract}

%%%%%%%%%%%%%%%%%%%%%%%%%%%%%%%%%%%%%%%
%%%%%%%%%%%%%%%%%%%%%%%%%%%%%%%%%%%%%%%
%%%%%%%%%%%%%%%%%%%%%%%%%%%%%%%%%%%%%%%

\newpage

\setcounter{footnote}{0}

\renewcommand{\thefootnote}{\arabic{footnote}}

\section{Introduction}

Our understanding of non-perturbative aspects of string
theory is still quite limited, although important progress has been
made in recent years, thanks, in particular, to work on string
dualities and D-branes. 
It is very important to consolidate and further this progress.
M-theory~\cite{RBHullTownsend, RBWitten}, a conjectured
eleven-dimensional theory which arises as strong coupling limit of
type IIA string theory, plays a crucial role in this area.
Various general features of
M-theory are understood -- it does not contain a dimensionless
coupling constant (the only parameter in the theory is the
eleven-dimensional Planck length), it reduces to eleven-dimensional
supergravity in the low-energy limit and it contains among its
excitations M2- and M5-branes, for which a classical action is
known. These classical properties have many non-trivial
consequences and implications for non-perturbative string theory. 
However, a well established formulation of M-theory in terms
of its fundamental degrees of freedom is still lacking. In order to fully 
exploit the power of M-theory and elucidate its role in establishing a 
truly non-perturbative picture of string theory, it is crucial to develop 
a better understanding of the microscopic formulation of the theory
including a consistent framework for its quantisation.
The best candidate for such a formulation is currently the matrix model of
M-theory.

In this paper we present a proposal for the study of a sector of
M-theory combining the matrix model approach with the AdS/CFT
correspondence. We show how the AdS/CFT duality can be studied 
in a genuinely M-theoretic regime by focussing on a particular set 
of states characterised by a large orbital angular momentum. Taking 
advantage of the dual description of these states in terms of a CFT 
allows us to independently confirm the results 
of the matrix model analysis. In this way, 
we simultaneously check the validity of both the matrix model proposal 
and the AdS/CFT correspondence.  

The matrix model of M-theory can be considered as a 
regularised version of the theory describing (super)membrane degrees 
of freedom~\cite{RBGoldstoneHoppe,RBdWHN}~\footnote{More precise statements are 
the following:
(i) for a given regularisation parameter (the size of the matrices),
a sufficiently smooth configuration in the membrane theory,
which in general describes multiple membranes, 
has a corresponding configuration in the matrix model;
(ii) the classical action functionals for the configuration in
the continuum membrane theory and that for the corresponding
configuration in the discrete matrix model approximately match;
(iii) the approximation becomes better,
for a fixed configuration in the continuum theory, 
when the size of the matrices becomes larger,
provided that the parameters of the discrete theory have the
appropriate dependence on the regularisation parameter;
this dependence defines the classical continuum limit.
The above properties imply that the semi-classical approximation
to the path integral of the matrix model includes contributions which
are governed by a Boltzmann factor associated 
approximately with the action functional of the membrane theory.
In this sense the matrix model contains (multi-)membranes.
In order to have a better understanding of the relation 
between matrix model and membrane theory 
it is necessary to address questions such as
``Does the matrix model contain other degrees of freedom such
as M5-branes?'' and
``What should the quantum continuum limit be?''
}.
In this approach the 
embedding coordinates of the membrane and their fermionic 
superpartners are replaced by
$K\times K$ matrices~\footnote{In the literature the size of the
matrices in the matrix model is usually denoted by $N$. 
Here we use the letter $K$ to avoid confusion with the
parameter $N$ used in the context of the AdS/CFT correspondence.}  
and the resulting theory describes a quantum mechanical system with a
finite number of degrees of freedom. The size of the matrices plays
the role of a regulator and the quantum theory of the (super)membrane
is expected to arise in the $K\to\infty$ limit.  The same matrix model
is found in type IIA string theory as describing the low-energy
dynamics of a system of D-particles (D0-branes)~\cite{RBTownsendD0,RBBFSS}. 
In this context the size of the matrices is associated with the number of
D0-branes. 
In~\cite{RBBFSS} it was conjectured that the $K\to\infty$ limit of
this supersymmetric matrix model capture the entire dynamics 
of M-theory. 

A complete and satisfactory understanding of the large $K$ limit of
the matrix model is still lacking and this represents a major obstacle
in establishing it as a viable description of M-theory. Another
unresolved issue concerns the emergence of the eleven-dimensional
Lorentz symmetry~\cite{RBdeWitMarquardNicolai, 
RBEzawaMatsuoMurakami, RBFujikawaOkuyama, RBHoppeLorentz,
RBHoppeTrzetrzelewski}. No complete proof that a Lorentz invariant 
quantum theory arise in the large $K$ limit is known. In particular the 
construction of the matrix model is closely tied to the use of light-front 
quantisation and no manifestly Lorentz-invariant formulation is available. 

In order to substantiate the matrix model proposal it is necessary to
address the fundamental issue of identifying proper observables in
M-theory and then understanding how to realise them in the matrix 
model itself. Moreover a concrete scheme 
for the calculation of such observables should be identified and  
this is rendered challenging in particular by the absence of a 
dimensionless coupling constant. The majority of the tests of the 
matrix model approach to M-theory involve either the low-energy 
supergravity approximation or compactification to type IIA string 
theory in ten dimensions. A comprehensive review
can be found in \cite{RBTaylorReview}. 
Without considering such limits it is difficult to decide whether any
results obtained from the matrix model are correct, although strong 
constraints should come from consistency requirements associated 
with unitarity and Lorentz invariance. 

In this paper we 
propose an approach which brings the AdS/CFT
correspondence into the picture in order to overcome some of these 
limitations and make progress on these issues. 
More specifically the use of the AdS/CFT dictionary allows us to identify 
quantities which are dual to 
CFT  observables 
as ``good'' observables in the matrix model.
Moreover, being able to independently compute such
observables on the two sides of the duality, we are able to justify the 
results of the M-theory calculations. We will carry out this programme in 
a sector containing M2-brane states in M-theory, without resorting to
a limit in which eleven dimensional supergravity or type IIA string theory 
can be used.

The specific AdS/CFT duality that we focus on in this paper,
which we refer to as the AdS$_4$/CFT$_3$ correspondence hereafter, 
was proposed in~\cite{RBMaldacena, RBABJM}. It relates M-theory 
in an AdS$_4 \times S^7/\Z_k$ background to a Chern-Simons-matter 
gauge theory with $\calN=6$ supersymmetry. This theory, which we will 
refer to as the ABJM theory, has U($N$)$\times$U($N$) gauge group --
with level $k$ and $-k$ for the two factors -- and was first constructed 
in \cite{RBABJM}, following previous 
work~\cite{RBSchwarzCS, RBBaggerLambert1, RBBaggerLambert2, 
RBBaggerLambert3, RBGustavsson1,RBGustavsson2,
RBVanRaamsdonkBiFundamental,RBHosomichiLeeLeeLeePark,
RBGaiottoYin}. 
It describes the low-energy limit of the dynamics of $N$ 
coincident membranes in $\R^8/\Z_k$. 

The AdS$_4\times S^7/\Z_k$ background arises as near-horizon 
geometry of such a stack of M2-branes. The $\Z_k$ action is 
generated by $2 \pi/k$ rotations acting simultaneously in the 
$12, 34, 56$, and $78$ planes of $\R^8$ in which the $S^7$ is 
embedded. We denote the angular momentum generators associated 
with  rotations in these four planes -- which can be chosen 
as basis for the Cartan subalgebra of the SO(8) isometry group of 
$S^7$ -- by $J_1, J_2, J_3$ and $J_4$ respectively. 
The $S^7$ can be described as an $S^1$ fibration over $\CP^3$, where 
the $S^1$ has constant radius and is generated at each point by 
$\Jt=J_1+J_2+J_3+J_4$. This is the $S^1$ which is identified as the 
M-theory circle~\cite{RBNilssonPope,RBABJM} and the $\Z_k$ quotient 
has the effect of dividing the circumference of this circle by $k$. For
$k\to\infty$ with $N/k$ fixed the theory is compactified to ten dimensions 
and reduces to type IIA string theory in 
AdS$_4\times\CP^3$~\cite{RBABJM}. 
This limit has been extensively studied after the original 
proposal~\cite{RBABJM} and corresponds to the 't Hooft limit in the 
CFT, where $N$ is large with $\lambda=N/k$ fixed. 

We are instead interested 
in studying a genuinely eleven-dimensional, M-theoretic, regime where 
$k$ is of order $1$ and $N$ is large.

One reason to study the M-theory regime of the AdS$_4$/CFT$_3$
duality is that one hopes to learn about M-theory in this way,
as already discussed above. In particular, since the ABJM theory is 
conjectured to describe the low-energy dynamics of M2-branes, it is 
natural to ask whether there is a direct connection between this theory 
and the matrix model. One of the main results in this paper is to establish 
a natural and very direct connection between a certain sector of the ABJM 
theory and the pp-wave matrix model first formulated in \cite{RBBMN}.

Another motivation for our work comes from the possibility of gaining
new insights into fundamental aspects of the AdS/CFT correspondence
by studying it in a regime which is essentially different 
from what has been considered before. Although the AdS/CFT
duality has been extensively studied, especially in its canonical version
relating the $\calN=4$ supersymmetric Yang--Mills (SYM) theory in four 
dimensions to type IIB string theory in an AdS$_5\times S^5$ 
background, important open questions remain concerning its
foundations. In particular the fundamental mechanism underlying the 
correspondence is not fully understood. Analysing a non-stringy 
AdS/CFT, of which the M-theoretic regime of the AdS$_4$/CFT$_3$ 
duality is a prime example, should help to shed light on this aspect, 
as certainly in this case the correspondence cannot be explained in terms 
of open/closed string duality.  Another important feature of the regime 
we focus on is that it is not compatible with the use of the planar 
approximation, since it requires large $N$ but $k\sim 1$, so that 
$\lambda=N/k$ cannot be fixed. This is natural as the 't Hooft 
expansion suggests that the gauge theory should have a description 
in terms of string-like degrees of freedom, which is not the case in the 
M-theory regime. Therefore the sector we consider allows us to 
analyse the gauge/gravity duality independently
of the special role played by the planar approximation.

The duality in the M-theoretic regime is considered to be rather 
non-tractable. On the CFT side, the theory is strongly coupled as 
$k\sim 1$. Furthermore, one cannot focus on the planar diagrams 
and all non-planar contributions are in principle relevant.
On the AdS side, one has to face the problem of formulating 
M-theory in AdS$_4 \times S^7$, in particular
when trying to calculate observables including quantum corrections.

In this paper, we present evidence that when one introduces a large 
orbital angular momentum, $J$, the presence of the small parameter 
$1/J$ makes it possible to identify good approximation 
schemes on both the CFT and the AdS sides. We discuss the relevant 
observables on both sides and establish a dictionary between them. The 
spectra computed on the two sides match, verifying the AdS/CFT 
conjecture in an M-theoretic regime.

The idea of using a large angular momentum to obtain 
a workable approximation is natural as the WKB approach is
usually applicable in cases where one has large quantum numbers
(in our case $J$). In the AdS$_5$/CFT$_4$ context this idea 
has been put forward 
in~\cite{RBBMN,RBGKP2,RBFrolovTseytlin}.
As first shown by Berenstein, 
Maldacena and Nastase (BMN) in~\cite{RBBMN},  
focussing on a large angular momentum sector 
leads to a situation in which both sides of the duality are weakly coupled 
and the AdS/CFT correspondence is directly testable. 
Our work is in many ways analogous to the BMN analysis, although with 
some important differences. 
We construct operators in the ABJM theory, which play a role 
analogous to the BMN operators. The construction of such operators is,
however, totally different and this reflects the fact that they correspond 
to excited states of membranes rather than strings.

On the gravity side of the correspondence we describe the physics 
of states in AdS$_4\times S^7/\Z_k$ which belong to a sector 
characterised by large angular momentum. M-theory states are 
classified by the eigenvalues of the Cartan generators $J_1, J_2, J_3$ 
and $J_4$. We focus on states which have large $J_4$ and the other 
components of the angular momentum of order one. 
The dynamics of such states can be described
using the maximally supersymmetric eleven dimensional pp-wave geometry
to approximate the AdS$_4\times S^7/\Z_k$ background. Following the
proposal to use the matrix model as a microscopic formulation of
M-theory, it is then natural to adopt as framework for our
calculations the pp-wave matrix model \cite{RBBMN}.  An important 
aspect of our proposal is that the size of the matrices in this matrix 
model should be  identified with $\Jt/k$. 

The possible vacuum states in the large angular momentum sector are
the BPS states of the pp-wave matrix model, which were studied in
\cite{RBBMN,RBDSVR1,RBDSVR2}. 
The simplest such state is a fuzzy sphere configuration corresponding to a
spherical membrane which extends in the AdS$_4$ directions and is
point-like in $S^7$, where it moves along a great 
circle with large angular momentum $J$. 
In general the BPS
states correspond to a collection of concentric fuzzy spheres, labelled
by a set of integers corresponding to the portion of the total angular
momentum carried by the individual membranes. 
The radii of the fuzzy spheres are proportional to their angular momentum. 
The use of the pp-wave approximation is justified if these radii are much 
smaller than the radius of curvature of the AdS$_4$ and $S^7$ factors in 
the original background. This leads to the condition
\begin{equation}
1\ll J \ll N^{1/2}
\label{RFUpperboundJ}
\end{equation}
for the applicability of the pp-wave approximation.  

After describing the ground state in the large $J$ sector, we discuss
the spectrum of fluctuations around the classical vacuum
configurations following \cite{RBDSVR1, RBDSVR2, RBKimPlefka}.
We present the tree level spectrum, which is
determined by the pp-wave matrix model Hamiltonian at quadratic order
in the fluctuations. We then discuss the behaviour of quantum
corrections associated with cubic and quartic terms in the
fluctuations~\footnote{
The spherical configurations discussed above can also be obtained
as solutions to the equations of motion derived from the classical membrane action.
However, we emphasise that
the matrix model is a formulation at the quantum level and
this is a definite advantage because 
it provides a framework  
to compute quantum corrections to the spectrum and to compare them 
with the dual CFT.}.
The condition that one-loop effects produce small
corrections to the tree level result turns out to be 
\begin{equation}
J \gg N^{1/3} \, .
\label{RFLowerboundJ}
\end{equation}
It is crucial for our proposal
that both conditions, (\ref{RFUpperboundJ}) 
and (\ref{RFLowerboundJ}), can be satisfied for large $N$ choosing 
the parameter $J$ so that
\begin{equation}
N^{1/3} \ll J \ll N^{1/2} \quad\mbox{\ie}\quad J^2 \ll N \ll J^3 \, .
\end{equation}

Having discussed
the large angular momentum sector 
on the gravity side using the pp-wave matrix model, we then 
describe the dual large $J$ observables in the CFT.
These are gauge-invariant operators in the ABJM theory with 
quantum numbers matching those of the membrane states we discussed. 
The requirement of gauge invariance leads to identify 
monopole operators as dual to membrane states in the large $J$ sector.
Monopole operators~\cite{RBtHooftMonopoleOperators}, 
which play a crucial role in the ABJM theory
and also in three dimensional gauge theory
in general~\cite{RBBorokhovKapustinWu1,
RBBorokhovKapustinWu2, RBABJM}, 
are classified by a set of integers, the 
so-called GNO charges~\cite{RBGNO}, which satisfy a Dirac 
quantisation condition~\cite{RBDiracMonopole, 
RBWuYangQuantisationCondition}. 
The BPS operators we consider in this paper are special 
cases, characterised by a large R-charge,
of those already considered in \cite{RBABJM} and further studied 
in \cite{RBSKim, RBSheikhJabbariSimon,
RBBennaKlebanovKlose, RBBerensteinPark, RBBashkirovKapustin}.
We show,
by focussing on BPS or ground states,
that it is possible to
identify the GNO charges of the relevant 
CFT operators with the angular momenta of the dual membrane states
associated with motion along the great circle in $S^7$. 
This correspondence was also observed in \cite{RBSheikhJabbariSimon}.

Monopole operators are associated 
with a Dirac monopole singularity at the insertion point. As such 
they do not have a simple manifestly local description in 
terms of the elementary fields in the theory. 
In order to deal with this 
complication it is convenient to use radial quantisation and study the 
ABJM theory on 
$S^2\times\R$ in Hamiltonian formulation in the presence of magnetic flux 
through the 
$S^2$~\cite{RBBorokhovKapustinWu1,RBBorokhovKapustinWu2}.
Using the state-operator map 
we identify the states in the radially quantised ABJM theory -- in a sector
characterised by large magnetic flux, $J$ -- which are
dual to membrane excitations in the bulk. 
An important ingredient in this construction is the
identification of a suitable gauge.

In this framework the dictionary relating the gravity and gauge sides 
arises in a natural way, leading to a very direct correspondence. 
Bulk states corresponding to spherical membranes and their excitations
have a dual description in terms of states of the ABJM theory on $S^2$.
Therefore states on the two sides of the duality are described
in terms of the same spherical harmonics. 
The energy spectrum of the membrane excitations, which are in 
general non-BPS, corresponds to the energy
spectrum of the ABJM theory in 
radial quantisation. 

In the case where the ground state on the gravity side is a single 
membrane, we verify that the tree-level spectrum obtained from the 
matrix model calculation agrees with the leading order result on the 
CFT side for all types of bosonic and fermionic excitations.

Despite the fact that the ABJM theory is strongly coupled for $k\sim 1$, 
we argue that a perturbative expansion is possible using a 
Born-Oppenheimer type approximation. The presence of a large 
magnetic flux, $J$, induces a separation of energy scales which leads to a 
natural identification of slow (or low-energy) modes and fast (or 
high-energy) modes. Integrating out the fast modes leads to an effective 
low-energy Hamiltonian for the slow modes which is weakly coupled 
for large $J$. We propose that this approach provides a framework 
for the systematic study of quantum corrections in the ABJM theory 
in the large $J$ sector that we defined. 

In our construction leading to the 
formulation of the Born-Oppenheimer approximation for the large $J$ 
sector of the ABJM theory we will assume that it is possible to use the 
classical action as a starting point to identify the BPS states even for small
$k$. This assumption is partially justified by supersymmetry and by the 
consistency of the results of related work which uses localisation 
techniques~\cite{RBSKim} in combination with a similar premise. A full
justification of this assumption will be provided a posteriori by the emergence of 
an expansion in which the effective coupling constant controlling quantum 
corrections is not the bare $1/k$, but a combination involving inverse powers
of $J$. A more detailed discussion of these issues is presented
in the sections devoted to the analysis of the CFT side.

We also discuss the generalisation to the case in which the ground 
state contains multiple membranes. The dual CFT sector involves 
monopole operators characterised by multiple non-zero GNO
charges, corresponding to the angular momenta of the individual 
membranes. In this case the pp-wave matrix model vacuum consists of 
block-diagonal matrices~\cite{RBBMN, RBDSVR1}. 
The excited states built on such vacua involve 
fluctuations in off-diagonal blocks, which do not correspond to 
degrees of freedom associated with individual membranes in the continuum. 
We will identify the dual states in the ABJM 
theory and show that in some cases -- specifically when there are two 
membranes of approximately the same size and hence close to each other 
-- these extra degrees of freedom on the two sides of the correspondence 
can be compared reliably and quantitatively within the limits of validity of 
our approximations. The agreement between the corresponding spectra is a 
strong indication that these states describe true degrees of freedom of 
M-theory, which are captured by the matrix model, but are not present in 
the conventional continuum membrane theory.

The AdS$_4$/CFT$_3$ duality proposed in~\cite{RBABJM} has
been extensively studied in the type IIA regime. Many of the techniques
originally developed for the AdS$_5$/CFT$_4$ correspondence
have been adapted to this case. In particular integrability has been exploited 
in the ABJM theory following the early results in~\cite{RBMinahanZarembo,
RBGaiottoGiombiYin, RBGrignaniHarmarkOrselli}. For a review see chapter 
IV.3~\cite{RBKlose} of \cite{RBIntegrabilityReview}.
Also, large angular momentum operators 
(with vanishing total monopole charge) in the type IIA limit were first studied 
in~\cite{RBNishiokaTakayanagiPP, RBGaiottoGiombiYin}. In the small $k$
regime, on the other hand, localisation techniques were successfully 
applied to the calculation of the superconformal index in~\cite{RBSKim}. 
Similar methods have been used to obtain exact results for other BPS 
observables such as the free energy starting with the work 
of~\cite{RBKapustinWilletYaakov,RBDrukkerMarinoPutrov, 
RBDrukkerMarinoPutrov2,RBFujiHiranoMoriyama}. 
Our analysis is also devoted to the small $k$ (M-theoretic) regime, 
however,  we focus on non-BPS quantities. In the large $J$ sector 
described above, we develop an approach which makes it possible to 
systematically study quantum corrections to certain non-BPS observables on 
both sides of the correspondence.

This paper is organised as follows. In section \ref{RSAdS} we describe the 
AdS side of the correspondence. We discuss the pp-wave approximation for 
membranes in AdS$_4\times S^7/\Z_k$ and present the associated matrix 
model and its energy spectrum. In section \ref{RSCFT} we describe the 
CFT side. We first discuss the Hamiltonian formulation of the ABJM theory 
in $S^2\times\R$. We then explain the separation between fast and slow 
modes in the framework of the Born-Oppenheimer approximation and
present the energy spectrum in the large $J$ sector. Particular attention 
is devoted to the discussion of gauge-fixing which plays an essential role
in our analysis.  In the discussion of both sides of the duality we first 
consider BPS states (ground states) and then near-BPS states (fluctuations
around the ground state), which are not protected 
and receive quantum corrections. In section \ref{RSMultiMembrane} we 
discuss the case of multi-membrane vacua. We conclude in section 
\ref{RSConclusion} with a discussion of our results and an outline of 
possible extensions and generalisations.

\section{AdS side} \label{RSAdS}

In this section, we describe the AdS side of the correspondence.
We begin by recalling some basic formulae in  
M-theory and the AdS$_4$/CFT$_3$ duality.

M-theory has only one length scale and the membrane tension, $T$,
is directly related to the eleven dimensional Planck length.
We use the conventions of \cite{RBMaldacena, RBABJM} in which 
the Planck length is defined 
so that the Einstein-Hilbert part of the $D=11$ supergravity action reads
\begin{equation}
S= - \frac{1}{2^8 \pi^8 l_P^9} \int \dr^{11}x\, \sqrt{-g}\, \calR + \cdots.
\end{equation}
The relation between the membrane tension and 
the Planck length
is then~\cite{RBKlebanovTseytlin}
\begin{equation}
T=\frac{1}{4\pi^2 l_P^3} \, .
\label{RFM2tension}
\end{equation}
The AdS$_4$/CFT$_3$ correspondence proposed 
in~\cite{RBMaldacena,RBABJM} was constructed considering the near 
horizon limit of a stack of $N$ M2-branes in $\R^8/\Z_k$ (which may be 
understood as a certain projection of $Nk$ M2-branes in flat space). 
The resulting geometry is AdS$_4\times S^7/\Z_k$, where the radius 
of the $S^7$, $R$, in terms of the eleven-dimensional Planck length satisfies
\begin{equation}
2^5 \pi^2 Nk = \frac{R^6}{l_P^6} \, , 
\label{RFRInTermsOfNk}
\end{equation}
while the radius of curvature of the AdS$_4$ factor is
\begin{equation}
R'=\frac{1}{2} R \, .
\end{equation}

We shall now specify the kinematical regime we study in this paper. 
Corresponding to rotations in the $12, 34, 56$ and $78$ planes
of $\R^8$ in which the $S^7$ is embedded, there are four angular 
momentum quantum numbers, $J_1, J_2, J_3$ and $J_4$.
The states we focus on are those for which one of them, which 
conventionally we take to be $J_4$, is large and the other angular 
momentum quantum numbers are of order $1$. 

Another important 
quantum number is $\Jt=J_1+J_2+J_3+J_4$.
This is related to the momentum along the M-theory circle,
which in the 
AdS$_4\times S^7/\Z_k$ background is identified with the great circle 
(or rather the family of great circles) corresponding to the orbit of the 
$\Jt$ generator~\footnote{For points in the $78$ 
plane, the M-theory circle coincides with the equator generated 
by $J_4$ rotations.}~\cite{RBABJM, RBNilssonPope}.
The states we are interested in 
have $J_4 \gg 1$, $\Jt  \gg 1$ and $\Jt-J_4\sim 1$.
In most instances we will simply write
$J$ to refer to either $J_4$ or $\Jt$. 
Since we focus on the leading order terms in a
$1/J$ expansion,
the difference will often be irrelevant.
When the distinction between the two
is relevant, we will explicitly specify whether we are referring to $J_4$ or
$\Jt$.

In the following we first consider $k=1$ and then generalise to the case 
of $k \neq 1$, which is obtained via a certain projection. Since the 
$\Z_k$ quotient acts on the M-theory circle, the projection requires the 
$\Jt$ quantum number of any individual object to be a multiple of $k$ (while 
of course $J_4$ can take any integer value). 

The dynamics of objects (both point-like and extended, such as strings or 
membranes) propagating in a curved geometry with large spatial 
momentum can be described using an approximation scheme referred 
to as the pp-wave approximation~\cite{RBBMN, RBGKP2, 
RBFrolovTseytlin}. 
This can be understood as an extension of the familiar 
infinite momentum frame argument (or the ultra-relativistic limit) 
in flat space to the case of a curved background.
As is well known, the dynamics of an object 
having very large spatial momentum in flat space-time is approximately  
governed by a free non-relativistic Hamiltonian. 
If the background space-time is curved, 
the dynamics of objects
with very large spatial momentum, proportional to a parameter $J$, is 
instead approximately controlled by a non-relativistic Hamiltonian 
containing an external harmonic oscillator potential, whose strength is
determined by the curvature radius and by $J$. 

The same Hamiltonian with  a suitable identification of parameters
also describes the dynamics of objects in a so-called pp-wave geometry.
There is a limiting process, referred to as Penrose 
limit~\cite{RBPenrose,RBGueven,RBppBlauHulletal2}, 
which produces the pp-wave geometry starting from the original 
background. However, we stress that the point of view that we take in this 
paper is to treat the procedure as an approximation scheme
 to describe the dynamics of special states in the AdS/CFT correspondence.
Rather than viewing the pp-wave background as arising from a formal limit
between two geometries, we consider it as an approximation which allows to 
capture the dynamics of states with large spatial momentum propagating in 
the original space-time~\cite{RBGKP2, RBFrolovTseytlin}. 

Let us recall the essential points of the pp-wave approximation by
using a simple example, a massless particle in the space-time 
$\R \times S^n$ with metric 
\begin{equation}
\dr s^2= -(\dr x^0)^2 + R^2 \dr\Omega^2_{n}, \label{RFMetricStR}
\end{equation}
where $\dr\Omega^2_{n}$ is the line element on the $n$-dimensional unit 
sphere $S^n$ with $n\geq 2$.
The dynamics of the particle is governed by the mass shell condition 
\begin{equation}
g^{ij} P_i P_j = 0. \label{RFDispertionForSphere}
\end{equation}
We temporarily use the indices $i, j= 0, \ldots, n$
to label the coordinates of the space-time (\ref{RFMetricStR}). 
We focus on a great circle in $S^n$.
We then assume that the 
particle has large momentum along this fixed circle
and does not deviate far from it.
Let the spatial coordinate $x^1$ be defined as 
the angle around the fixed large circle multiplied by the radius $R$.
The longitudinal momentum $P_1 >0$ conjugate to $x^1$ is by 
assumption large.
We choose the transverse coordinates $x^\alpha$, $\alpha=2, \ldots, n$,
in the directions orthogonal to the great circle. 
In terms of these coordinates the metric is approximately 
\begin{equation}
\dr s^2\approx -\left(\dr x^0\right)^2 + \left(1-\frac{(x^\alpha)^2}{R^2}\right)
\left(\dr x^1\right)^2 +\left(\dr x^\alpha\right)^2,
\label{RFMetricStRApp}
\end{equation}
neglecting higher order terms 
in $x^\alpha/R$.
Using (\ref{RFMetricStRApp}), the dispersion relation 
(\ref{RFDispertionForSphere}) becomes
\begin{equation}
(-P_0)^2\approx\left(1+\frac{(x^\alpha)^2}{R^2}\right)(P_1)^2 
+(P_\alpha)^2.
\end{equation}
Large longitudinal momentum $P_1$ therefore implies large energy 
$-P_0>0$. The finite difference, which plays a role analogous to the 
light-cone gauge Hamiltonian, is given by
\begin{equation}
(-P_0)-P_1 \approx \frac{(P_\alpha)^2+(P_1)^2
\frac{(x^\alpha)^2}{R^2}}{2 P_1} \, .
\label{RFPpExampleFinal}
\end{equation}
where we used $(-P_0)+P_1 \approx 2P_1$. 
Equation (\ref{RFPpExampleFinal}) shows that, for fixed (large)
longitudinal momentum, $P_1$, the dynamics of 
the particle in curved space is approximately 
that of a non-relativistic harmonic oscillator.
Notice that the longitudinal momentum is actually quantised -- $P_1=J/R$, 
where $J$ is an integer -- because its conjugate coordinate $x^1$ is 
periodic with  period  $2 \pi R$. Therefore in this approximation
\begin{equation}
(-P_0)+P_1 \approx 2 P_1 = 2 \,\frac{J}{R}.
\end{equation}
Equation (\ref{RFMetricStRApp}) is valid if  $|x^\a|/R\ll 1$. 
Classically one can assume a particle to remain arbitrarily close to the 
fixed great circle. However, in the quantum theory the wave function of 
the particle has finite extension. For the $n$-th excited state, the extension 
can be estimated using~(\ref{RFPpExampleFinal}) and $P_1=J/R$,
\begin{equation}
\langle x\rangle \sim R\sqrt{\frac{2n+1}{J}}.
\label{RFEstimateFluctuationPPApprox}
\end{equation}
Hence we see that the condition $\langle x\rangle \ll R$, which  
validates the use of the pp-wave approximation, 
gives an upper bound on the excitation number,  
\begin{equation}
n \ll J,  \label{RFValidityPPApproxExcitationNumber} 
\end{equation}
and also implies 
\begin{equation}
J \gg 1.  \label{RFValidityPPApproxZeroPoint}
\end{equation}

Another way of understanding the above formulae is in terms of a
centrifugal potential. Because of the large angular momentum, the
particle experiences a strong centrifugal force confining it around the
equator (where the radius of the trajectory is the largest -- the centrifugal 
force pushes objects in the direction where the radius becomes larger). 
The pp-wave approximation keeps the leading order term in this 
centrifugal potential, which as expected has the harmonic oscillator form. 
The strength of the potential is determined by the 
curvature radius of the background and the (angular) momentum.

The use of the pp-wave approximation in the context of the AdS/CFT
correspondence involves an additional subtlety. In order to have a 
consistent dictionary between the gravity and CFT sides, it is necessary 
to change the space-time picture on the AdS side to the one given in 
\cite{RBDobashiShimadaYoneya, RBShimada3} which is
particularly suited for studying holographic aspects 
(\ie the computation of correlation functions
following the prescription in~\cite{RBGKP, RBWittenHolography}). 
More specifically, one should 
not consider objects (particles, strings or membranes) 
propagating in the AdS space (with oscillating wave functions), 
but rather one should consider objects undergoing a tunnelling process
(with exponentially decreasing or increasing wave functions).
In practice this is achieved by a certain double Wick rotation. 
This prescription was proposed in~\cite{RBDobashiShimadaYoneya}
for the pp-wave approximation to string theory in 
AdS$_5\times S^5$. It solves various puzzles regarding the signature 
of the bulk/boundary, including the identification of energy and conformal 
dimension and the signature of vector type fluctuations. Although 
the new interpretation is different leading to a better, consistent 
correspondence, the end result of the pp-wave approximation is 
mathematically equivalent~\cite{RBDobashiShimadaYoneya}. Both 
of the interpretations, with or without the double Wick rotation, lead to 
the same effective Hamiltonian in the pp-wave approximation. This is 
the case even for more general backgrounds corresponding to near 
horizon limits of D$p$-brane configurations~\cite{RBAsanoSekinoYoneya}.
The same interpretation has also been applied to the computation
of correlation functions 
using methods derived from the study of integrable systems
in~\cite{RBTsuji, RBJanikSurowkaWereszczynski}.
We shall not elaborate on this issue any further and we refer the reader 
to~\cite{RBDobashiShimadaYoneya,RBShimada3} for additional details.
In the following we assume that the identification of observables between 
the gravity and CFT sides of the correspondence is made adopting the
prescription discussed in these papers. 

Applying the above considerations to the study of M-theory in 
AdS$_4 \times S^7/\Z_k$, we conclude that the dynamics of states 
with large angular momentum in this background can be described 
using a suitable pp-wave approximation. Combining this idea with 
the matrix model proposal leads us naturally to use a matrix model 
which has the same form as the one arising in the maximally 
supersymmetric eleven-dimensional pp-wave 
geometry~\cite{RBKowalskiGlikman, RBppBlauHulletal2}. 
This matrix model was first proposed in~\cite{RBBMN} and it was 
later derived in~\cite{RBDSVR1, RBSugiyamaYoshida1} from the 
regularisation of the supermembrane theory in the pp-wave background. 
This matrix model is the main ingredient in our analysis of the gravity 
side of the AdS$_4$/CFT$_3$ duality.

Our discussion in this section is based on a reinterpretation of 
previous results on the pp-wave matrix model~\cite{RBDSVR1, RBDSVR2, 
RBKimPlefka}.
In the spirit of using the pp-wave background as an approximation 
scheme to study a large angular momentum sector of M-theory in 
AdS$_4 \times S^7$, we will write 
the matrix model in terms of
parameters characterising the original geometry, \ie the radii $R$ and 
$R^\prime=R/2$, and the angular momentum parameter 
$J$~\footnote{In particular 
we do not introduce a mass parameter, $\mu$, as commonly done in the 
literature. This introduction of $\mu$ is not necessary for
the comparison between observables on the AdS side and the CFT side
and moreover it makes the analysis of the limits of validity of the pp-wave 
approximation less transparent.}.  
We first consider the membrane theory in \AdStS\ in the pp-wave 
approximation and then regularise it to obtain the matrix model. 
Rather than providing a detailed derivation of the membrane Hamiltonian
starting from the supermembrane theory in \AdStS\ (analogous to that in 
\cite{RBFrolovTseytlin, RBDobashiShimadaYoneya} for the type IIB string in 
the AdS$_5 \times S^5$ background), 
we will justify its form based on the same arguments that led to 
(\ref{RFPpExampleFinal}).
The physics of membranes in AdS$_4 \times S^7/\Z_k$ 
can be captured by simply restricting the attention to those special states 
of the supermembrane theory in 
AdS$_4 \times S^7$~\cite{RBdeWitPeetersPlefkaSevrin} 
for which all individual membranes have a $\Jt$ quantum number which is 
a multiple of $k$.

The bosonic part of the membrane 
Hamiltonian~\footnote{We will often refer to the combination 
$-P_0-P_1$ as the Hamiltonian on the AdS side
of the correspondence.} in the pp-wave 
approximation is 
\begin{eqnarray}
-P_0-P_1&\!\!=\!\!&
\int \dr^2 \s \left(
\frac{[\s] }{2  P_1} (p_\a)^2
+
\frac{[\s] }{2P_1} \,\frac{1}{2} 
T^2
\left(
\{x^m ,x^n \}^2 + \{y^i,y^j\}^2 + 2\{x^m ,y^i \}^2
\right) \right. \nn \\
&& \left. \hsp{1.1} +
\frac{1}{2}\frac{P_1}{[\s]} \frac{(x^m)^2}{R^2} 
+ \frac{1}{2}\frac{P_1}{[\s]} \frac{(y^i)^2}{(R')^2} 
- \frac{T}{2R'} \e_{ijk} y^i \{y^j, y^k\} 
\right) \, ,
\label{RFM2ppwaveHamiltonian1}
\end{eqnarray}
%FORMULA WITHOUT FIXING A SIGN CONVENTION
%\begin{eqnarray}
%(-P_0)-P_1&\!\!=\!\!&
%\int \dr^2 \s \left(
%%\frac{[\s] }{2 ? P_1} p^2
%\frac{[\s] }{2  P_1} (p_\a)^2
%+
%\frac{[\s] }{2P_1} \,\frac{1}{2} 
%T^2
%\left(
%\{x^m ,x^n \}^2 + \{y^i,y^j\}^2 + 2\{x^m ,y^i \}^2
%\right) \right. \nn \\
%&& \left. \hsp{1.1} +
%\frac{1}{2}\frac{P_1}{[\s]} \frac{(x^m)^2}{R^2} 
%+ \frac{1}{2}\frac{P_1}{[\s]} \frac{(y^i)^2}{(R')^2} 
%\mp \frac{T}{2R'} \e_{ijk} y^i \{y^j, y^k\} 
%\right) \, ,
%\label{RFM2ppwaveHamiltonian1}
%\end{eqnarray}
where $T$ is the membrane tension (\ref{RFM2tension}) and the nine 
transverse coordinates have been denoted by $x$ and 
$y$, with $y^i$, $i=1,2,3$, indicating three scalars originating from 
AdS$_4$ directions and $x^m$, $m=4,\ldots,9$, referring to six scalars 
originating from $S^7$ directions. We also use $\a=1,\ldots,9$ to refer to 
the set of all nine transverse directions. In the following we will use the 
notation $x^\a$ to collectively denote all the membrane coordinates when
we do not need to distinguish between AdS$_4$ and $S^7$ directions. 
The Lie bracket, $\{\,.\,,\,.\,\}$, in 
(\ref{RFM2ppwaveHamiltonian1}) is defined as 
\begin{equation}
\{f,g\} = \frac{\del f}{\del\s_1}\frac{\del g}{\del\s_2}
- \frac{\del f}{\del\s_2}\frac{\del g}{\del\s_1} \, ,
\label{RFLieBracket}
\end{equation}
for any functions, $f(\s_1,\s_2)$ and $g(\s_1,\s_2)$, on the membrane 
world-volume. 
The constant $[\s]$ is the total area of the base space,
\begin{equation}
[\s]=\int \dr^2 \s.
\end{equation}
It should not of course appear in observable quantities
and we will see later that $[\s]$ does not appear after the regularisation.
$P_1$ is the momentum along the equator of the $S^7$.
It is related to the (integer-valued) quantum number $J$ by
\begin{equation}
P_1=\frac{J}{R} \,,
\label{RFRelationP1J}
\end{equation}
where to be precise $J$ in the numerator should be understood
as the value of $J_4$.
$-P_0 >0$ is a similar quantity associated with a ``time-like'' direction 
in AdS$_4$, which, by the conventional dictionary of the AdS/CFT duality, 
is related to the conformal dimension $\Delta$ of the dual CFT operators 
by~\footnote{The identification becomes quite direct and transparent in the 
interpretation discussed in \cite{RBDobashiShimadaYoneya, RBShimada3}.}
\begin{equation}
-P_0=\frac{\Delta}{R'} = 2 \frac{\Delta}{R} \, .
\label{RFRelationP0Delta}
\end{equation}

The various terms in (\ref{RFM2ppwaveHamiltonian1})
can be understood as follows. The quadratic terms in the $x$ and $y$ 
coordinates come from the harmonic oscillator potential arising in the 
pp-wave approximation, analogous to the quadratic term appearing in 
(\ref{RFPpExampleFinal}). The cubic term for the $y$'s is induced by 
the coupling of the membrane to the three-form potential, which has 
non-zero background value in the AdS$_4$ space. The remaining 
terms are those appearing in the membrane Hamiltonian in flat space 
in the ultra-relativistic limit~\footnote{Those familiar with the light-cone 
gauge formulation of the membrane theory might wonder whether we 
are working in the light-cone gauge or using the ultra-relativistic limit 
(also called the infinite momentum frame 
in the case of flat space-time). 
Arguably, it makes sense to distinguish the two points of view
in flat space since the light-cone gauge gives exact results and it is
applicable to generic states, whereas the ultra-relativistic limit is an 
approximation valid only for special states.
However, this distinction is meaningless in the present case of a curved 
space-time in which we have to make an approximation -- the pp-wave 
approximation -- and consider special states with large angular momenta.}.
We have partially fixed 
the reparametrisation invariance of the membrane
in a way analogous to that used in the light-cone gauge for membranes in 
flat space-time~\cite{RBdWHN, RBGoldstoneHoppe}.
The Hamiltonian (\ref{RFM2ppwaveHamiltonian1}) can be rewritten in the 
form of a sum of squares, which simplifies the study of the minima of the 
potential,
\begin{eqnarray}
-P_0-P_1&\!\!=\!\!&
\int \dr^2 \s \left(
\frac{[\s] }{2  P_1} (p_\a)^2
+
\frac{[\s] }{2 P_1} \,\frac{1}{2} 
T^2
\left(
\{x^m ,x^n \}^2+2\{x^m ,y^i \}^2
\right)
+
\frac{1}{2}\frac{P_1}{[\s]} \frac{(x^m)^2}{R^2} \right. \nn \\
&& \left. \hsp{1.1} +
\frac{[\s]}{2P_1}
\left(
\frac{1}{2} T \e_{ijk} \{y^j, y^k\} - \frac{P_1}{[\s]} \frac{y^i}{R'}
\right)^{\!2}
\right) \, .
\label{RFM2ppwaveHamiltonian2}
\end{eqnarray}
%FORMULA WITHOUT FIXING A SIGN CONVENTION
%\begin{eqnarray}
%(-P_0)-P_1&\!\!=\!\!&
%\int \dr^2 \s \left(
%\frac{[\s] }{2  P_1} (p_\a)^2
%+
%\frac{[\s] }{2 P_1} \,\frac{1}{2} 
%T^2
%\left(
%\{x^m ,x^n \}^2+2\{x^m ,y^i \}^2
%\right)
%+
%\frac{1}{2}\frac{P_1}{[\s]} \frac{(x^m)^2}{R^2} \right. \nn \\
%&& \left. \hsp{1.1} +
%\frac{[\s]}{2P_1}
%\left(
%\frac{1}{2} T \e_{ijk} \{y^j, y^k\} \mp \frac{P_1}{[\s]} \frac{y^i}{R'}
%\right)^{\!2}
%\right) \, .
%\label{RFM2ppwaveHamiltonian2}
%\end{eqnarray}
There is also the phase space constraint 
\begin{equation}
\{x^\a, p_\a\}=0, \label{RFM2ppwaveConstraint}
\end{equation}
associated with the residual gauge symmetry 
corresponding to the
area preserving diffeomorphisms.
The membrane Hamiltonian (\ref{RFM2ppwaveHamiltonian1}), 
(\ref{RFM2ppwaveHamiltonian2}) and the constraint
(\ref{RFM2ppwaveConstraint}) are mathematically equivalent 
to those of the membrane theory on the pp-wave 
background~\cite{RBDSVR1, RBSugiyamaYoshida1}
by appropriate rewriting of the parameters.

The matrix model which we will use in the following was 
obtained by regularising the Hamiltonian described in the previous 
paragraphs. An essential element of our proposal is that the proper 
matrix model regularisation, suitable to describe the large $J$ sector 
of the AdS$_4$/CFT$_3$ duality, should use matrices of size $K=\Jt/k$. 

One way to understand this identification is to notice that the D0-brane 
charge, which should be the matrix size~\cite{RBBFSS}, 
is equal to $\Jt/k$. 
This follows from the identification of the M-theory circle in
the AdS$_4$/CFT$_3$ correspondence with the orbits of the $\Jt$ 
generator acting on the \AdStS\ space-time. 

Another way to understand the identification of the matrix size with
the angular momentum, which is based on the interpretation of the matrix 
model as regularised membrane theory, is the following. 
In our gauge fixing of the membrane theory, we choose the space-like 
coordinates on the world-volume so that the longitudinal momentum 
density is constant on a time-slice of the world-volume.
This implies that the (base-space) area of a certain portion
of the time-slice of the world-volume
is proportional to the longitudinal momentum contained in that portion.
The longitudinal momentum is approximately equal to the momentum 
along the M-theory circle to leading order in our approximation.
Because of the periodicity of the angle along the M-theory circle,
the associated momentum has a minimum, $k/R$.
This minimum of the momentum implies a minimum for the area in 
the time-slice of the world-volume of the membrane. 
The total area is proportional to the total momentum, $J/R$, and the
minimum of the area is proportional to $k/R$ with the same coefficient of  
proportionality. Hence the time-slice of the world-volume 
is divided into $J/k$ pieces. This is achieved by regularising the 
membrane world-space by matrices, as the matrix regularisation 
corresponds to dividing the membrane world-space into equal 
area pieces. The number of these pieces is equal to the matrix size 
$K=J/k$. This can be understood using an analogy with the quantisation 
of a system with a single degree of freedom: the Bohr-Sommerfeld 
quantisation says that the minimum area of the phase space (the 
membrane world-space) is quantised in units of $2 \pi \hbar$ which 
is equal to the total area divided by the matrix size in the membrane 
context. Therefore the size of the matrices should be $J/k$~\footnote{
A similar interpretation can be applied to the case of the BMN analysis
of the AdS$_5$/CFT$_4$ duality. In that case fixed-time slices of the string 
world-sheet are discretised to a lattice with $J$ sites, conforming with the 
construction of BMN operators on the CFT side.}.
The use of finite dimensional matrices in the
presence of a compact longitudinal direction
is reminiscent of the DLCQ argument presented 
in \cite{RBSusskindDLCQ}.

Let us recall some basic relations used in the matrix regularisation. 
A comprehensive review can be found in~\cite{RBTaylorReview}. In this
paper we follow the conventions of~\cite{RBShimadaBetaMM}.
Functions on the membrane world-volume at fixed time, $f(\s^1, \s^2)$, 
$g(\s^1, \s^2), \ldots$,  are replaced by $K\times K$ matrices,  
$\hat{f}=\rho(f)$, $\hat{g}=\rho(g), \ldots$, where the map $\rho$ is linear. 
These matrices provide a discrete approximation to the corresponding 
functions. The basic operations on functions have counterparts on the 
associated matrices. This correspondence can be summarised as 
follows,
\begin{align}
\rho(f g) &\approx \frac{1}{2} \Big(\rho(f) \rho(g) + \rho(g) \rho(f)\Big),
\label{RFMRMultiplication} \\
\rho\left(\{f, g\}\right) &\approx \frac{2\pi K}{i \tA} 
\left[ \rho(f), \rho(g) \right] \, ,
\label{RFMRCommutator} \\
\frac{1}{\tA}\int f \, \dr^2 \s &\approx \frac{1}{K} \tr \big(\rho(f)\big).
\label{RFMRTr}
\end{align}
The symbol $\approx$ indicates that the two sides of these relations
are equal up to higher order corrections in $1/K$.
The first equation simply states that the product of two functions 
becomes the multiplication (or more precisely one half the anti-commutator) 
of the corresponding matrices. The second equation relates the Lie bracket
of two functions to the commutator of the associated matrices
multiplied by a factor proportional to $K$. 

Following this procedure, one introduces
the matrix version of the membrane coordinates, 
\begin{equation}
X^m=\rho(x^m)\, , \qquad Y^i=\rho(y^i).
\end{equation}
The canonical conjugates of these matrices, $P_\a$, are related to
the matrix version of the continuum momentum, $p_\a$, by 
\begin{equation}
P_\a = \frac{\tA}{K} \rho(p_\a) \, .
\end{equation}
Using 
$K= J/k$, the complete matrix model Hamiltonian takes the form
\begin{eqnarray}
-P_0 - P_1 &\!\!\!=\!\!\!&\tr \Bigg\{ 
\frac{R}{2 k } \left(P_\a\right)^2
-(2\pi T)^2 \frac{R}{2 k }
\frac{1}{2} \left(
[X^m, X^n]^2 + 2 [X^m,Y^i]^2
+ [Y^i, Y^j] ^2
\right) \nonumber \\
&& \hsp{0.4} 
+ \frac{k}{2R^3} (X^m)^2
+ \frac{k}{2R R'^2} (Y^i)^2
+ i 2\pi T \frac{1}{R} \e_{ijk}Y^i[Y^j, Y^k]
\nn 
\\
&& \hsp{0.4}
+ 2\pi T \frac{R}{k} \frac{1}{2} \left(\Psi^T \gamma^m[ X^m, \Psi]
+\Psi^T \gamma^i[ Y^i, \Psi] \right)
- \frac{3i}{4} \frac{1}{R} \, \Psi^T \gamma^{123} \Psi
\Bigg\} ,
\label{RFMMHamiltonian1}
\end{eqnarray}
%FORMULA WITHOUT FIXING A SIGN CONVENTION
%\begin{eqnarray}
%-P_0 - P_1 &\!\!\!=\!\!\!&\tr \Bigg\{ 
%%\Tr \Bigg\{ 
%\frac{R}{2 k } \left(P_\a\right)^2
%-(2\pi T)^2 \frac{R}{2 k }
%\frac{1}{2} \left(
%[X^m, X^n]^2 + 2 [X^m,Y^i]^2
%+ [Y^i, Y^j] ^2
%\right) \nonumber \\
%&& \hsp{0.4} 
%+ \frac{k}{2R^3} (X^m)^2
%+ \frac{k}{2R R'^2} (Y^i)^2
%\pm i 2\pi T \frac{1}{R} \e_{ijk}Y^i[Y^j, Y^k]
%\nn 
%\\
%&& \hsp{0.4}
%+ 2\pi T \frac{R}{k} \frac{1}{2} \left(\Psi^T \c^m[ X^m, \Psi]
%+\Psi^T \c^i[ Y^i, \Psi] \right)
%\mp \frac{3i}{4} \frac{1}{R} \, \Psi^T \c^{123} \Psi
%\Bigg\} ,
%\label{RFMMHamiltonian1}
%\end{eqnarray}
where we have also included the fermionic terms which were omitted in 
the membrane theory. Here $\g^\a$ are SO(9) gamma-matrices and 
$\g^{123}=\g^1\g^2\g^3$.
As in the case of the membrane Hamiltonian, the bosonic part of
(\ref{RFMMHamiltonian1}) 
can be rewritten as a sum of squares,
\begin{eqnarray}
-P_0 - P_1 &\!\!\!=\!\!\!&\tr \Bigg\{ 
\frac{R}{2 k } \left(P_\a\right)^2
-(2\pi T)^2 \frac{R}{2 k }
\frac{1}{2} \left(
[X^m, X^n]^2 + 2 [X^m,Y^i]^2
\right) \nonumber \\
&& \hsp{0.4} 
+ \frac{k}{2R^3} (X^m)^2
- (2\pi T)^2 \frac{R}{2k}
\left( \frac{1}{2} \epsilon_{ijk} [Y^j, Y^k]
- i \frac{1}{2\pi T} \frac{2 k }{R^2} Y^i
\right)^{\!\!2} \nn \\
&& \hsp{0.4}
+ 2\pi T \frac{R}{k} \frac{1}{2} \left(\Psi^T \gamma^m[ X^m, \Psi]
+\Psi^T \gamma^i[ Y^i, \Psi] \right)
- \frac{3i}{4} \frac{1}{R} \, \Psi^T \gamma^{123} \Psi
\Bigg\} ,
\label{RFMMHamiltonian}
\end{eqnarray}
%FORMULA WITHOUT FIXING A SIGN CONVENTION
%\begin{eqnarray}
%-P_0 - P_1 &\!\!\!=\!\!\!&\tr \Bigg\{ 
%\frac{R}{2 k } \left(P_\a\right)^2
%-(2\pi T)^2 \frac{R}{2 k }
%\frac{1}{2} \left(
%[X^m, X^n]^2 + 2 [X^m,Y^i]^2
%\right) \nonumber \\
%&& \hsp{0.4} 
%+ \frac{k}{2R^3} (X^m)^2
%- (2\pi T)^2 \frac{R}{2k}
%\left( \frac{1}{2} \epsilon_{ijk} [Y^j, Y^k]
%- i \frac{1}{2\pi T} \frac{2 k }{R^2} Y^i
%\right)^{\!\!2} \nn \\
%&& \hsp{0.4}
%+ 2\pi T \frac{R}{k} \frac{1}{2} \left(\Psi^T \c^m[ X^m, \Psi]
%+\Psi^T \c^i[ Y^i, \Psi] \right)
%\mp \frac{3i}{4} \frac{1}{R} \, \Psi^T \c^{123} \Psi
%\Bigg\} ,
%\label{RFMMHamiltonian}
%\end{eqnarray}
where we used $R'=R/2$.

The use of $K=\Jt/k$ implies
that the M-theory charge $\Jt$ should be a multiple of $k$
for any state in the matrix model. 
Based on this observation, we propose that the matrix 
model describes physics in \AdStS, rather than AdS$_4\times S^7$, in 
the pp-wave approximation.

The canonical (anti-)commutation relations are
\begin{equation}
[X^\a{}^r{}_s, P_\b{}^u {}_v ] = i \delta^{\a}{}_{\b} \d^r{}_v \d^u{}_s \, ,
\end{equation}
\begin{equation}
[\Psi^a{}^r{}_s, \Psi^b{}^u {}_v ]_+ = \delta^{a b} \d^r{}_v \d^u{}_s \, ,
\end{equation}
where $X^\a$, $\a=1, \ldots, 9$, collectively denotes the matrices 
associated with all nine membrane coordinates, 
$a,b=1, \ldots, 16$ are SO(9) Majorana spinor indices and 
$r,s,u,v=1,\ldots ,K$ are matrix indices.

The phase space constraints are
\begin{equation}
[X^\a, P_\a] - i \Psi^T \Psi = 0 \, ,
\label{RFMMConstraint}
\end{equation}
where again the sum over $\a$ runs from 1 to 9.

\subsection{BPS states}
\label{RSAdSBPS}
The classical stable solutions of the pp-wave matrix model with zero energy 
are known~\cite{RBBMN, RBDSVR1}. 
They are the BPS states (or the ground states) 
in the sector we are studying in this paper. They are given by a collection of 
so-called fuzzy spheres extending in the 3 transverse directions originating 
from AdS$_4$ and are point-like in the $S^7$ directions.

From the form of the matrix model Hamiltonian 
(\ref{RFMMHamiltonian}), which is written as a sum of squares, it is clear
that minimum energy configurations have $X^m=0$ for $m=4,\ldots,9$, 
so that the only non-vanishing fields in the classical solution are 
$Y^i$, $i=1,2,3$. They should satisfy
\begin{equation}
\frac{1}{2} \epsilon_{ijk} [Y^j, Y^k]
- i \frac{1}{2\pi T} \frac{2 k }{R^2} Y^i = 0 \, .
\label{RFFuzzySphereEq}
\end{equation}
%FORMULA WITHOUT FIXING A SIGN CONVENTION
%\begin{equation}
%\frac{1}{2} \epsilon_{ijk} [Y_0^j, Y_0^k]
%\mp i \frac{1}{2\pi T} \frac{2 k }{R^2} Y_0^i = 0 \, .
%\label{RFFuzzySphereEq}
%\end{equation}
This equation is solved by taking the $Y^i$'s to be proportional 
to $K\times K$ generators, $L^i$, of a representation of SU(2). 
The explicit form of the solution is
%SIGN CONVENTION IS FIXED USING BELOW
\begin{equation}
Y_0^i = \frac{2k}{(2 \pi T) R^2} \, L^i \, ,
\label{RFVacuumMM}
\end{equation}
where the $L^i$'s obey
\begin{equation}
[ L^i, L^j ] = i \epsilon^{ijk} L^k \, .
\end{equation}
The simplest solution corresponds to choosing the $L^i$'s to be the 
generators of the irreducible $K$ dimensional SU(2) representation.  
Taking the proportionality constant so that $Y^i_0$ is written as
\begin{equation}
Y^i_0 = r \sqrt{\frac{4}{K^2-1}} \, L^i \, ,
\label{RFFuzzyS2Sol}
\end{equation}
one finds for the parameter $r$
\begin{equation}
r = \frac{k\sqrt{K^2-1}}{2\pi T R^2} \approx \frac{J}{2\pi T R^2} \, ,
\label{RFFuzzyS2Radius}
\end{equation}
where we used $K=J/k$ and $J\gg 1$.
Equations (\ref{RFFuzzyS2Sol}) and (\ref{RFFuzzyS2Radius}) have then 
a simple geometric interpretation. A spherical membrane of unit radius, 
described by coordinates $y^i$, $i=1,2,3$, with $\sum (y^i)^2=1$, is
approximated in the matrix model (with matrices of size $K$) by the 
configuration $Y^i = \sqrt{4/(K^2-1)}\, L^i$, referred to as a fuzzy 
sphere of unit radius~\cite{RBGoldstoneHoppe,RBdWHN}. Therefore 
the solution (\ref{RFFuzzyS2Sol}) corresponds to a fuzzy sphere of radius 
$r$ given by (\ref{RFFuzzyS2Radius}).

A more general solution to (\ref{RFFuzzySphereEq}) can be obtained 
considering a reducible $K$ dimensional representation of SU(2). Equation 
(\ref{RFFuzzySphereEq}) can be satisfied taking the $Y$'s to be 
block-diagonal matrices,
\begin{equation}
Y^j_0 = \left[ \begin{array}{cccc} \!Y^j_{0(1)} & & & \\
& \!\!\ddots & & \\ & & \!\!Y^j_{0(i)} & \\
& & & \!\!\ddots \!\end{array} \right] ,
\label{RFBlockDiagMMSol}
\end{equation}
where the $i$-th block on the diagonal, $Y^j_{0(i)}$, $i=1,\ldots,n$, is of 
size $K_{(i)} = J_{(i)}/k$. It is given by 
\begin{equation}
Y^j_{0(i)} = r_{(i)} \sqrt{\frac{4}{K_{(i)}^2-1}}\,L_{(i)}^j \, ,
\label{RFiBlockMMSol}
\end{equation}
where
\begin{equation}
r_{(i)} = \frac{k\sqrt{K_{(i)}^2-1}}{2\pi T R^2} 
\approx \frac{J_{(i)}}{2\pi T R^2}
\label{RFMultiM2Radii}
\end{equation}
and $L^j_{(i)}$ are the generators of the irreducible SU(2) 
representation of dimension $K_{(i)}$. 

Block-diagonal configurations in the matrix model (for which 
the equations of motion factorize into those for the individual blocks)
are interpreted as describing collections of classically 
independent objects. In the present case,  
(\ref{RFBlockDiagMMSol})-(\ref{RFiBlockMMSol}) 
represent distinct fuzzy spheres. More precisely the block diagonal 
matrices (\ref{RFBlockDiagMMSol}) describe a collection of concentric 
fuzzy spheres of radii $r_{(i)}$ given in (\ref{RFMultiM2Radii}). They 
extend in the AdS$_4$ directions and carry momentum $J_{(i)}/R$
along a great circle in $S^7$. 

The general solution minimising the matrix model Hamiltonian is therefore
characterised by a set of integers, $J_{(i)}$, $i=1,\ldots,n$, satisfying
\begin{equation}
\sum_{i=1}^n J_{(i)} = J \, .
\label{RFVacuumClassificationMM}
\end{equation}
The $J_{(i)}$'s must be multiples of $k$, because the size of the $i$-th 
block in (\ref{RFBlockDiagMMSol}), $K_{(i)}=J_{(i)}/k$, 
$i=1,\ldots,n$, is necessarily an integer. This gives further support 
to our proposal that the matrix size should be $J/k$, since the projection 
associated with the $\Z_k$ quotient implies that the angular momentum 
of each membrane in a multi-membrane configuration should be a 
multiple of $k$.

In~\cite{RBDSVR2} it was shown that the states in the pp-wave matrix 
model can be organised into multiplets of the SU(2$|$4) supergroup and 
special states belonging to BPS multiplets were identified. The vacua 
we described in this section were shown to belong to multiplets termed 
doubly-atypical in~\cite{RBDSVR2}. These multiplets have energies which
are non-perturbatively protected. Therefore the degeneracy of the vacua
corresponding to different numbers of spherical membranes is not lifted
in the full quantum theory.

The theory contains distinct sectors associated with the vacuum 
configurations~(\ref{RFBlockDiagMMSol})-(\ref{RFMultiM2Radii}) and 
the fluctuations around them. It would be interesting to study the possibility 
of tunnelling connecting these sectors corresponding to different 
perturbative vacua~\footnote{
The fact that the energies of the ground states are non-perturbatively 
protected~\cite{RBDSVR2} suggests that the tunnelling processes may be
allowed only between excited states and not between pairs of ground states.
Some properties of instanton solutions associated with tunnelling 
processes were studied in \cite{RBDSVR1, RBYeeYi, 
RBBachasHoppePioline}.
}.
Such an effect should be understood as corresponding to
the interaction of membranes. For example, in a two-membrane 
vacuum, interactions can lead to a transfer of longitudinal momentum 
between the two membranes. This corresponds to a transition between an 
initial state characterised by two angular momenta, $J_{(1)}$ and $J_{(2)}$, 
and a final state in which the angular momenta are $J_{(1)}'$ and 
$J_{(2)}'$, with $J_{(1)}+J_{(2)} = J_{(1)}'+J_{(2)}'=J$. Similarly it is possible 
to have tunnelling processes corresponding to the splitting or joining of 
membranes. 
For example a single membrane with angular momentum $J$ could split 
into two membranes with angular momenta $J_{(1)}$ and $J_{(2)}$, with 
$J_{(1)}+J_{(2)} = J$. Since the angular momenta are quantised (being 
integers and multiples of $k$), these transitions are not allowed in 
perturbation theory. We expect the effect of these tunnelling processes
to be negligible compared to the leading order perturbative corrections 
to the spectrum which will be discussed in section \ref{RSAdSPerturbation}.

Let us more closely examine the formula (\ref{RFMultiM2Radii}) for the 
radii of the minimal energy fuzzy spheres,  
\begin{equation}
r_{(i)}=\frac{J_{(i)}}{2\pi T R^2} \, , 
\label{RFSphericalM2Radius}
\end{equation}
where $i=1,\ldots,n$ in a vacuum with $n$ membranes.
This shows that the size of the spherical membranes grows with their 
angular momentum, $J_{(i)}$. However, for our analysis to be valid we 
should require that the membranes do not extend beyond the region in 
which the AdS$_4\times S^7/\Z_k$ background is well approximated by 
the pp-wave geometry. More precisely for the pp-wave approximation to 
be applicable we should require that the radii $r_{(i)}$ satisfy $r_{(i)}\ll  R$. 
Using (\ref{RFRInTermsOfNk}) this amounts to $J_{(i)}\ll (Nk)^{1/2}$. 
Combining this result with the requirement 
(\ref{RFValidityPPApproxZeroPoint}) that the $J_{(i)}$'s be large we 
obtain the condition 
\begin{equation}
1\ll J_{(i)} \ll (Nk)^{1/2} \, 
\label{RFValidityPPApproxRadiusM2}
\end{equation}
for the pp-wave approximation to be valid. 
In section 
\ref{RSAdSPerturbation} we will discuss how a stricter lower bound on 
$J$ arises if one requires that quantum corrections in the matrix 
model be small.

The pp-wave approximation we have discussed so far can be considered 
as keeping the leading order terms in an expansion in 
powers of 
\begin{equation}
\frac{r}{R} \sim \left( \frac{J^2}{Nk} \right)^{\frac{1}{2}}.
\end{equation}
It should be possible to compute corrections to the pp-wave 
approximation and incorporate higher orders in this 
expansion into the matrix model.

As observed above, the various perturbative vacua are expected to be 
non-perturbatively connected through tunnelling processes. Therefore it 
may be more natural to require that the pp-wave approximation be 
applicable to all possible vacua and not just to a particular one 
corresponding to a given set of $J_{(i)}$'s. If we take this point of view, 
considering the perturbative vacuum consisting of a single membrane, it 
follows that the total $J$ should satisfy
\begin{equation}
J\ll (Nk)^{1/2} \, . 
\label{RFValidityPPApproxRadiusSingleM2}
\end{equation}
This condition in turn implies a bound on the number, 
$n$, of membranes. Since the individual $J_{(i)}$'s are integers and 
multiples of $k$, the vacuum with the largest number of membranes with a 
given total $J$ corresponds to the case in which $J_{(i)}=k$ for all 
$i=1,\ldots,n$. Combining (\ref{RFValidityPPApproxRadiusSingleM2}) and 
(\ref{RFVacuumClassificationMM}) for this vacuum we get 
\begin{equation}
J = \sum_{i=1}^n J_{(i)} = nk \ll (Nk)^{1/2} \, ,
\end{equation}
and thus
\begin{equation}
n \ll (N/k)^{1/2} \, .
\label{RFBoundOnNumberOfM2}
\end{equation}
This condition is consistent with the fact that we are describing 
configurations of membranes in a fixed background, obtained as 
near-horizon limit of a black brane solution corresponding to $N$ 
coincident membranes, without including any back-reaction.  

At first sight, requiring the validity of the pp-wave approximation for all 
possible perturbative vacua may appear to be incompatible with the lower 
bound in (\ref{RFValidityPPApproxRadiusM2}). Considering for simplicity 
$k=1$, in the extreme case in which $J_{(i)}=1$ for all $i$, the condition 
$J_{(i)}\gg 1$ is not satisfied, implying that the vacuum fluctuations of the 
centre of mass of the membranes will invalidate the use of the 
pp-wave approximation, as explained 
in the general discussion around (\ref{RFValidityPPApproxZeroPoint}). 
However, this problem may be resolved if we use 
a dual description of this membrane configuration in terms of M5-branes, 
using the proposal in~\cite{RBBMN, 
RBMaldacenaSheikhJabbariVanRaamsdonk}. 
According to these papers 
the vacuum corresponding to the partition
$J=1+\cdots+1$, should be identified with a configuration 
of a single M5-brane. Since the angular momentum of the M5-brane is $J$,
(\ref{RFValidityPPApproxZeroPoint}) is satisfied from the M5-brane point 
of view. The size of this M5-brane 
is given by~\cite{RBBMN, RBMaldacenaSheikhJabbariVanRaamsdonk} 
\begin{equation}
r_{M5}^4 \sim \frac{J}{R^2} l_P^6 \, .
\label{RFM5Radius}
\end{equation}
Similarly to the condition (\ref{RFValidityPPApproxRadiusM2}), the validity 
of the pp-wave approximation for the M5-brane requires $r_{M5}/R \ll 1$. 
Using (\ref{RFRInTermsOfNk}), this amounts to
\begin{equation}
J \ll Nk,
\end{equation}
which is satisfied automatically in our regime.

Similar considerations can be applied to other states containing multiple 
membranes with small angular momentum, which can be identified 
with configurations of M5-branes carrying large angular 
momentum, 
satisfying the conditions of applicability of the pp-wave approximation. 
For intermediate values of the angular momenta, more complicated 
configurations, such as the five-branes discussed 
in~\cite{RBLozanoPrinsloo},
may be relevant.

\subsection{Near BPS fluctuations}
\label{RSAdSFluctuation}

We next consider the fluctuations around the ground states discussed 
in the previous section. The spectrum of such fluctuations for the pp-wave 
matrix model has been studied in detail in~\cite{RBDSVR1,RBKimPlefka}. 
We will present the results in terms of parameters, $R$, $J$ and $k$, which 
are suitable for the comparison with the ABJM theory to be discussed in 
section \ref{RSCFT}.

We focus on the single membrane vacuum, \ie the case in which the 
minimal energy configuration corresponds to matrices $Y_0^i$ of the 
form (\ref{RFVacuumMM}), where the $L^i$'s are the generators of the 
$K=J/k$ dimensional irreducible representation of SU(2). 
The case of multi-membrane vacua will be discussed in section 
\ref{RSMultiMembrane}. 

In order to study the spectrum of excited states in the single membrane 
sector all the fields are expanded 
in terms of fluctuations around the classical 
solution $(X_0^m=0, Y_0^i,\Psi_0=0)$. For the $Y^i$ scalars, which are the 
only variables with a non-trivial background value, we denote the fluctuation 
by ${Y'}^i$ and write
\begin{equation}
Y^i = Y_0^i + {Y'}^i \, .
\end{equation}
Substituting into the matrix model Hamiltonian 
one obtains quadratic, cubic 
and quartic terms in the fluctuations,
\begin{equation}
H = H^{(2)} + H^{(3)} + H^{(4)} \, .
\end{equation}
The tree-level spectrum is determined by computing the eigensystem of 
the quadratic Hamiltonian, $H^{(2)}$, which takes the form
\begin{eqnarray}
H^{(2)} &\!\!\!=\!\!\!&\tr \left\{ 
\frac{R}{2 k } \left(P_\a\right)^2
+\frac{2k}{R} \left[ \left( \frac{{Y'}^i}{R} + i(2\pi T) \frac{R}{2k}
\veps^{ijk}[Y_0^j,{Y'}^k]\right)^{\!2} + \fr{4R^2} \left(X^n\right)^2
\right. \right. \nn \\
&& \hsp{0.3} \left. \left. -(2\pi T)^2 \frac{R^2}{4k^2}
[X^n,Y^i_0]^2\right] + (2\pi T)\frac{R}{2k} \Psi^T \gamma^i [Y^i_0,\Psi]
- \frac{3i}{4R} \Psi^T \gamma^{123} \Psi \right\} .
\label{RFMMHamiltonianQuad}
\end{eqnarray}
%FORMULA WITHOUT FIXING A SIGN CONVENTION
%\begin{eqnarray}
%H^{(2)} &\!\!\!=\!\!\!&\tr \left\{ 
%\frac{R}{2 k } \left(P_\a\right)^2
%+\frac{2k}{R} \left[ \left( \frac{{Y'}^i}{R} \pm i(2\pi T) \frac{R}{2k}
%\veps^{ijk}[Y_0^j,{Y'}^k]\right)^{\!2} + \fr{4R^2} \left(X^n\right)^2
%\right. \right. \nn \\
%&& \hsp{0.3} \left. \left. -(2\pi T)^2 \frac{R^2}{4k^2}
%[X^n,Y^i_0]^2\right] + (2\pi T)\frac{R}{2k} \Psi^T \gamma^i [Y^i_0,\Psi]
%\mp \frac{3i}{4R} \Psi^T \gamma^{123} \Psi \right\} .
%\label{RFMMHamiltonianQuad}
%\end{eqnarray}
This Hamiltonian is diagonalised by expanding the fluctuations, ${Y'}^i$, 
$X^m$ and $\Psi$, in a basis of $K\times K$ matrices, which consists of 
discretised versions of the spherical 
harmonics~\cite{RBGoldstoneHoppe, RBdWHN}.
This should be expected, since the matrix model is the regularised 
version of the continuum membrane theory. In the continuum the 
vacuum solution is a spherical membrane and the spherical harmonics 
are the natural basis to use to expand its fluctuations. The discretised 
versions of the spherical harmonics are referred to as matrix spherical 
harmonics. They are classified by a pair of quantum numbers, $(l,m)$, 
where $l=0,1, \ldots, K-1$, and $m=-l,-l+1,\ldots,0, 1, \ldots, l$.
The excited states in the matrix model spectrum are correspondingly  
labelled by integers $l, m$.

For the scalars associated with $S^7$ directions, $X^n$, $n=4,\ldots,9$, 
there are $6$ polarisations and the spectrum is
\begin{equation}
\omega=\frac{1}{R}\sqrt{1+ 4 l(l+1)} = \frac{2}{R} \left(\half+l\right) \, ,
\qquad l=0,1,\ldots, K-1 \, .
\label{RFS7ScalarSpectrum}
\end{equation}
The upper bound on the quantum number $l$ reflects the effect of 
discretisation introduced by the matrix regularisation:  
matrix spherical harmonics constructed from the generators of the $K$ 
dimensional irreducible representation of SU(2) exist only with $l<K$. 
Each level in (\ref{RFS7ScalarSpectrum}) has a degeneracy 
$(2l+1)$, corresponding to the allowed values of the quantum number 
$m$.

We note that the mass term and the contribution from the Laplacian 
($1$ and $4l(l+1)$ respectively under the square root 
in~(\ref{RFS7ScalarSpectrum})) combine in such a way as to result in a 
rational energy spectrum. The same is true for the spectrum of the 
${Y'}^i$ fluctuations and the fermions that we present below. This fact 
does not seem to have a simple explanation in the matrix model. 
However, we will see in section~\ref{RSCFTNearBPS} that it has a 
simple interpretation on the CFT side.

The three scalars coming from AdS$_4$ directions, ${Y'}^i$, contain only 
two physical transverse polarisations. This is because of the presence of 
the constraint~(\ref{RFMMConstraint}) associated with the residual gauge 
symmetry corresponding to area preserving 
diffeomorphisms~\footnote{
Since the membranes are point-like in the
$S^7$ directions, fluctuations of all $X^m$'s are
transverse and there is no similar reduction of degrees of freedom. 
}. 
Diagonalising the quadratic Hamiltonian in this sector yields energies
\begin{equation}
\omega = \frac{2}{R} (1+l) \, , \qquad l=0,1,\ldots, K-2 
\label{RFAdS4ScalarSpectrum1}
\end{equation}
and
\begin{equation}
\omega = \frac{2}{R} \, l \, , \qquad l=1,2,\ldots, K
\label{RFAdS4ScalarSpectrum2}
\end{equation}
respectively for the two sets of states. For each of the energies 
(\ref{RFAdS4ScalarSpectrum1}) and (\ref{RFAdS4ScalarSpectrum2}) 
the degeneracy of the corresponding states is $(2l+1)$. 

In order to study the fermionic fluctuations one first decomposes the SO(9) 
Majorana spinors according to the SO(3)$\times$SO(6) isometries of the 
pp-wave matrix model. Diagonalising the quadratic Hamiltonian yields two 
sets of states with energies respectively 
\begin{equation}
\omega = \frac{2}{R} \left(\frac{3}{4} + j\right) \, , \qquad 
j = \half,\frac{3}{2},\ldots, K-\frac{3}{2} 
\end{equation}
and
\begin{equation}
\omega = \frac{2}{R} \left(\frac{1}{4} + j\right) \, , \qquad 
j = \half,\frac{3}{2},\ldots, K-\frac{1}{2} \, .
\end{equation}
The multiplicity of the corresponding states is $4\times (2j+1)$ for both 
sets, with the factor of $4$ coming from the fact that the fermions are
spinors of the SO(6) isometry group associated with rotations in 
the transverse directions in $S^7$. 

The spectrum of the pp-wave matrix model is summarised in 
table~\ref{RTAdSNBPSTreeSpectrum}.

\begin{table}[htb] 
\begin{center}
\begin{tabular}[htb]{| l || c | c | c |}
\hline
Type \raisebox{-6pt}{\rule{0pt}{19pt}}
& Label & Energy ($\omega$) & Multiplicity \\
\hline \hline
$S^7$ scalars, $X^{n}$ & $l=0,1,\ldots, K-1$ &
$\displaystyle \frac{2}{R} \left(\half+l\right)$ & $6 \!\times\! (2l+1)$ 
\raisebox{-13pt}{\rule{0pt}{31pt}} \\
\hline
\raisebox{-14pt}{AdS$_4$ scalars, ${Y'}^i$} & $l=0,1,\ldots, K-2$ & 
$\displaystyle \frac{2}{R} (1+l)$ & $(2l+1)$ 
\raisebox{0pt}{\rule{0pt}{18pt}} \\ 
 & $l=1,2,\ldots, K$ & 
$\displaystyle \frac{2}{R}\,l$ & $(2l+1)$ 
\raisebox{-12pt}{\rule{0pt}{12pt}} \\ 
\hline 
\raisebox{-14pt}{Fermions, $\Psi$} & 
$\displaystyle j = \half,\frac{3}{2},\ldots, K-\frac{3}{2}$ & 
$\displaystyle \frac{2}{R}\!\left(\frac{3}{4}+j\right)$ &
$4 \!\times\! (2j+1)$
\raisebox{0pt}{\rule{0pt}{18pt}} \\ 
 & $\displaystyle j = \half,\frac{3}{2},\ldots, K-\frac{1}{2}$ &
$\displaystyle \frac{2}{R}\!\left(\frac{1}{4}+j\right)$ &
$4 \!\times\! (2j+1)$
 \raisebox{-14pt}{\rule{0pt}{14pt}} \\ 
\hline 
\end{tabular}
\caption{Spectrum of pp-wave matrix model near-BPS excitations}
\label{RTAdSNBPSTreeSpectrum}
\end{center}
\end{table}

In section~\ref{RSCFT} we will compare these results with the energies of 
the dual states in the radially quantised ABJM theory. The comparison is 
done using~(\ref{RFRelationP1J}) and~(\ref{RFRelationP0Delta}) which
imply the relation
\begin{equation}
\omega= \frac{\Delta}{R'} - \frac{J_4}{R} = \fr{R}(2\Delta -J_4)
\end{equation}
between the matrix model energies, $\omega$, and the parameters 
$\Delta$ and $J_4$ characterising the CFT operators.

\subsection{Perturbation theory}
\label{RSAdSPerturbation}

Quantum corrections to the energy spectrum reviewed in the previous 
subsection are computed using standard quantum mechanics perturbation 
theory~\cite{RBDSVR1, RBDSVR2, RBKimPlefka}.  
The majority of the fluctuations are non-BPS and therefore their 
spectrum will be corrected, but there are some BPS fluctuations whose
spectrum is protected~\cite{RBDSVR2}. 
The situation is reminiscent of the 
open string spectrum around giant gravitons 
in the pp-wave approximation~\cite{RBBalasburamanianEtal}.
Leading order corrections for some of the states in the spectrum 
were computed in~\cite{RBDSVR1,RBKimPlefka}. 

The perturbation part of the Hamiltonian consists of cubic and quartic terms 
in the fluctuations around the classical solution. Expanding the Hamiltonian 
(\ref{RFMMHamiltonian}) one gets
\begin{eqnarray}
H^{(3)} &\!\!\!=\!\!\!&  \tr \left\{-(2\pi T)^2 \frac{R}{k}\left(
[X^m,Y^i_0][X^m,{Y'}^i]+[Y^i_0,{Y'}^j][{Y'}^i,{Y'}^j]\right)
+ i(2\pi T)\frac{1}{R} \veps_{ijk} {Y'}^i[{Y'}^j,{Y'}^k] \right. \nn \\
&&\hsp{0.5} \left. \rule{0pt}{16pt} + (2\pi T)\frac{R}{2k}
\left(\Psi^T\gamma^m [X^m,\Psi] 
+\Psi^T\gamma^i[{Y'}^i,\Psi] \right)\right\}
\label{RFMMHamiltonianCub}
\end{eqnarray}
%FORMULA WITHOUT FIXING A SIGN CONVENTION
%\begin{eqnarray}
%H^{(3)} &\!\!\!=\!\!\!&(2\pi T) \frac{R}{2k} \, \tr \left\{
%2[X^m,Y^i_0][X^m,{Y'}^i]+2[Y^i_0,{Y'}^j][{Y'}^i,{Y'}^j]
%\pm i\frac{2k}{R^2} \veps_{ijk} {Y'}^i[{Y'}^j,{Y'}^k] \right. \nn \\
%&& \hsp{1.7} \left. \rule{0pt}{16pt} + \Psi^T\gamma^m [X^m,\Psi] 
%+\Psi^T\gamma^i[{Y'}^i,\Psi] \right\}
%\label{RFMMHamiltonianCub}
%\end{eqnarray}
and
\begin{equation}
H^{(4)} = -(2\pi T)^2 \frac{R}{4k} \, \tr \left\{ [X^m,X^n]^2 + 
2[X^m,{Y'}^i]^2 + [{Y'}^i,{Y'}^j]^2 \right\} .
\label{RFMMHamiltonianQuart}
\end{equation}
The leading order correction to the 
energy of a generic state, $|n\ra$, is computed using the familiar formula
\begin{equation}
\Delta E_n = \sum_{n'} \frac{\langle n |H^{(3)}| n' \rangle
\langle n' |H^{(3)}| n \rangle}{E_n - E_n'} + \langle n |H^{(4)}| n \rangle \, . 
\label{RFMMPertCorrection}
\end{equation}
Note in particular that the degeneracy of the un-perturbed states due to the
SO(3) symmetry does not require the use of degenerate perturbation theory 
as the perturbed Hamiltonian still possesses the SO(3) symmetry.
We will not present explicit perturbative calculations in the pp-wave matrix 
model. We will limit ourselves to recalling the relative weight of the 
perturbative corrections compared to the tree level result. This was studied 
in~\cite{RBDSVR1,RBKimPlefka} and we present here the result in terms of 
parameters which are more suitable in the AdS/CFT context for the 
comparison with the ABJM theory. The tree level energies summarised in 
table~\ref{RTAdSNBPSTreeSpectrum} are of order $1/R$. The ratio of the 
one loop corrections (\ref{RFMMPertCorrection}) to the tree level result 
is of order~\cite{RBDSVR1, RBDSVR2, RBKimPlefka}
\begin{equation}
\frac{T^2R^6}{J^3} \sim \frac{Nk}{J^3} \, ,
\label{RFTreeTo1LoopRatio}
\end{equation}
where we used (\ref{RFM2tension}) and (\ref{RFRInTermsOfNk}) and 
omitted numerical factors. Hence, quantum corrections in the pp-wave
matrix model are small when
\begin{equation}
J \gg (Nk)^{1/3} \, . 
\label{RFValidityLoopExpansionMM}
\end{equation}
The first term in 
(\ref{RFMMPertCorrection}) involves a sum over intermediate states and 
both terms contain sums over the $l$ and $m$ quantum numbers arising 
from the expansion in matrix spherical harmonics. 
Each summand is the matrix element between individual states.
The matrix elements, estimated using the Hamiltonian 
(\ref{RFMMHamiltonian1}),
are of order $Nk/J^3$, which is the same as (\ref{RFTreeTo1LoopRatio}).
The fact that the matrix elements are small for large $J$
is expected since
the strong centrifugal force for large $J$ suppresses 
the fluctuations 
(see (\ref{RFEstimateFluctuationPPApprox})) 
making the interaction terms smaller than the quadratic terms. However,
the sums in (\ref{RFMMPertCorrection}) can
potentially produce factors of $K=J/k$ and alter
(\ref{RFTreeTo1LoopRatio}). 
Hence the dependence of the loop corrections on $J$ 
is the result of the competition of two effects:
as $J$ grows, each matrix element is suppressed,  
but at the same time the number of degrees of freedom increases.
The explicit calculations in~\cite{RBDSVR1,RBKimPlefka} show that at 
leading order no extra factors of $K$ arise from the summations, thanks to 
non-trivial cancellations due to supersymmetry.  
This was proven in~\cite{RBDSVR2} for all states in the 
single membrane vacuum and it is natural to 
expect~(\ref{RFTreeTo1LoopRatio}) 
to hold for the leading order corrections in all vacua. 
The absence of extra factors of $K$ in the perturbative 
expansion at leading order is related to the one loop finiteness of the 
membrane world-volume theory in the matrix regularisation, where
the size of the matrices, $K$, plays the role of UV cut-off.
Further work is needed to establish whether similar cancellations persist 
at higher orders, so that (\ref{RFTreeTo1LoopRatio}) can be considered 
a genuine coupling constant for the pp-wave matrix model.

The ratio (\ref{RFTreeTo1LoopRatio}) can also be related to the ratio of 
the eleven dimensional Planck length to the size of the spherical 
membranes,
\begin{equation}
\frac{Nk}{J^3}\sim \left(\frac{l_P}{r}\right)^{\!3} \, ,
\end{equation}
where $r$ is given in (\ref{RFSphericalM2Radius}). 
This is natural, since in a theory of quantum gravity, such as M-theory, 
loop corrections should be suppressed when the relevant length scale is 
much larger than the Planck scale.
Only when $J$ is sufficiently large such that $Nk/J^3 \ll 1$,
it is possible to distinguish the extended spherical membranes we are 
discussing from point-like gravitons. 

We can also estimate the size of the fluctuations of the membrane 
coordinates around the stable fuzzy sphere. These fluctuations should be 
small compared to the radius of the sphere. The magnitude of the  
fluctuations of the membrane coordinates is of order $R/\sqrt{J}$, as can 
be deduced from the simple particle picture of the pp-wave approximation 
presented at the beginning of section \ref{RSAdS}. The ratio of this to the 
size of the spherical membrane, $J/ T R^2$, is therefore $T R^3/J^{3/2}= 
(Nk/J^3)^{1/2}$. 
Hence one gets the same condition, $J \gg (Nk)^{1/3}$, 
that we deduced from the suppression of loop corrections.

Combining the condition of applicability of the pp-wave approximation 
(\ref{RFValidityPPApproxRadiusM2}) with the requirement that quantum 
corrections be small leads to the condition 
\begin{equation}
(Nk)^{1/3} \ll J \ll (Nk)^{1/2} \, .
\label{REJWindow}
\end{equation}
Therefore the sector of M-theory states in AdS$_4\times S^7/\Z_k$ 
characterised by angular momentum $J$ satisfying (\ref{REJWindow}) 
can be reliably studied perturbatively using the pp-wave matrix model. In 
the next section we will argue that in this regime a suitable perturbative 
expansion scheme can be developed for the ABJM theory as well.

In section~\ref{RSAdSBPS} we noticed that the use of the pp-wave 
approximation for all possible perturbative vacua can be justified if a 
dual description of certain vacua in terms of M5-branes 
following~\cite{RBMaldacenaSheikhJabbariVanRaamsdonk} is employed. 
The quantum fluctuations of the M5-brane should be suppressed when 
\begin{equation}
r_{M5} \gg l_P \,.
\label{RFM5WeakCoupling}
\end{equation}
Recalling the formula~(\ref{RFM5Radius}) for the radius of the M5-brane, 
we find that~(\ref{RFM5WeakCoupling}) is satisfied when the condition 
$J \gg (Nk)^{1/3}$ found above holds.
This does not contradict the fact that 
the $J=1+\ldots+1$ vacuum cannot be treated perturbatively,
since we lack a direct classical description
of the degrees of freedom of M5-branes in the matrix model.

We note the formula
\begin{equation}
\frac{Nk}{J^3} \frac{J^2}{Nk} = \frac{1}{J}
\label{RF1overJ}
\end{equation}
which suggests that $1/J$ corrections to the computation we have 
described may be understood in terms of a double expansion in powers of 
$Nk/J^3$, which controls the loop corrections, and of $J^2/Nk$, which 
controls the corrections to the pp-wave approximation. 
A simple class of $1/J$ corrections arises from the distinction 
between the $J_M$ and $J_4$ generators.
The parameter that we denoted by $J$ in the continuum membrane 
Hamiltonian is the eigenvalue of $J_4$, 
whereas in the matrix regularisation we used matrices of size $J/k$, 
related to the eigenvalue of $J_M$. 
On the states discussed in section~\ref{RSAdSFluctuation} $J_M$ and 
$J_4$ differ by an amount of order $1$, which 
can be neglected in our analysis. 
Keeping track of this difference would result in 
$1/J$ corrections to the spectrum.

Our prescription of using $K\times K$ matrices in the regularisation of 
the large angular momentum sector we focus on implies that the number 
of degrees of freedom in the resulting quantum mechanical 
system is of order $K^2\sim (J/k)^2$. 
In the next section we will show that in the dual 
sector of the ABJM theory a number of states of order $K^2$ arises 
naturally within the Born-Oppenheimer scheme. This matching of the 
number of degrees of freedom between the two sides of the 
correspondence lends additional support to our proposal. However, 
this observation should be taken cautiously because the matrix model can 
be expected to provide a good approximation to the continuum theory only 
for low-lying states with small quantum number $l$. For $l$ approaching
$K$ one expects the discretised description of membranes in terms of 
matrices to provide a poor approximation. To be more precise, in order 
for the approximation (\ref{RFMRCommutator}) of the Lie bracket 
$\{f_1, f_2\}$ by the matrix commutator to be valid, we need the condition 
$k_1 k_2 \ll K$, where $k_1$ and $k_2$ refer to the wave numbers (in 
the sphere case, the label of the spherical harmonics $l$)
characterising the scale of variation of the functions $f_1$ and 
$f_2$~\footnote{
The simplest way to understand this condition is
to recall the situation for toroidal membranes~\cite{
RBFairlieFletcherZachos1, RBFairlieFletcherZachos2, 
RBFloratos}. In this case the approximate equality between the Lie brackets 
and the matrix commutator follows from the condition
$\sin (\frac{\pi}{K} \vec{k}_1 \times \vec{k}_2) 
\approx \frac{\pi}{K} \vec{k}_1 \times \vec{k}_2$, where $\vec{k}_1$
and $\vec{k}_2$ denote the two dimensional wave number
vectors on the torus. 
}.
This at first sight seems to put a rather stringent condition $l \ll \sqrt{K}$.
However, the leading order terms contributing to the dynamics of the
fluctuations around the stable solution are given by the Lie brackets
between the background solution itself (which has wave number of 
order $1$) and the fluctuation, with wave number $l$. This 
leads to the condition $l \ll K$.
On the CFT side a similar restriction arises from the fact that the effective 
description we use is good only  for states with energies much smaller than 
that of the high energy states that we will argue should be
integrated out.

\section{CFT side} 
\label{RSCFT}

In this section we will study the CFT side of the correspondence 
in the large $J$ regime.

Let us first recall some general features of the theory which was proposed 
in~\cite{RBABJM} as CFT dual to M-theory in \AdStS. The ABJM theory is a 
U($N$)$\times$U($N$) Chern-Simons gauge theory with matter in the 
bi-fundamental representation. The field content consists of four 
bi-fundamental complex scalar fields,  $(\phi^{A})^{i}{}_{\hat{j}}$, 
four bi-fundamental complex spinors, $(\psi^{A})^{\hat{i}}{}_{j}$, 
$A=1,\ldots,4$,
in addition to the
gauge fields, $(A_{\mu})^{i}{}_{j}$ and $( \hat{A}_{\mu}  )^{\hi}{}_{\hj}$, 
associated with the two U($N$) factors. Here $i,j=1,\ldots,N$ and 
$\hat{i},\hat{j}=1,\ldots,N$ are colour indices referring to the two U($N$)'s. 
The conformal dimension of fermions and gauge fields is $1$, while the 
scalars have dimension $1/2$. The two Chern-Simons terms in the theory 
have (integer) level $k$ and $-k$ respectively and $1/k$ plays the role of 
expansion parameter in a standard perturbative treatment.

For generic $k$ the theory has an SU(4)$\times$U(1) R-symmetry 
corresponding to the isometry group of $S^7/\Z_k$ in the dual M-theory 
background. The U(1) factor is generated by $\Jt$ corresponding to 
translations along the M-theory circle and the SU(4) factor 
is the remaining part of the SO(8) isometries of $S^7$,
which commute with $\Jt$.
Because of the $\Z_k$ quotient the full SO(8) symmetry is present only 
for $k=1,2$. We will use a component formulation of the ABJM theory 
similar to that given in~\cite{RBBandresLipsteinSchwarz}, in which the 
SU(4) symmetry and the corresponding $\calN=6$ supersymmetry are 
manifest~\footnote{This SU(4)$\sim$SO(6) symmetry
should not be confused with the SO(6) symmetry of the
pp-wave matrix model in section \ref{RSAdS}. 
They are embedded into the full SO(8) symmetry
in inequivalent ways.}. The full $\calN=8$ 
supersymmetry is believed to be recovered in the special case of $k=1,2$. 
 
The matter fields $\phi^A{}^i{}_\hj$, $\psi^A{}^\hi{}_j$
transform in the fundamental of SU(4).
They transform under the symmetry 
generated by $\Jt$ as 
\begin{equation}
\left(\phi^A{}^i{}_\hj\right)'=e^{i\a}
\phi^A{}^i{}_\hj
\end{equation}
\begin{equation}
\left(\psi^A{}^\hi{}_j\right)'=e^{-i\a} 
\psi^A{}^\hi{}_j
\label{RFJtABJMTransformation}
\end{equation}
where $\alpha$ is the parameter of the 
transformation~\cite{RBABJM}~\footnote{This may be understood as the 
matter part of a constant gauge transformation
in which the U(1) parts of the two U($N$) gauge groups
are assigned opposite charges.
When the transformation parameter is equal to $2 \pi /k$,
the two states related by the transformation are 
indistinguishable~\cite{RBABJM}. 
}.
The symmetry generators $J_1, J_2, J_3$ and $J_4$ which we introduced 
in section~\ref{RSAdS} (corresponding to rotations in the 12, 34, 56 and 78 
planes of $\R^8$ in which the $S^7$ is embedded)
are realised in the ABJM theory as certain linear combinations 
of $\Jt=J_1+J_2+J_3+J_4$ and the Cartan generators of 
SU(4). Even for $k \neq 1, 2$
the currents associated with the generators $J_1, J_2, J_3$ and $J_4$ 
are conserved, although the full SO(8) symmetry is broken to 
U(1)$\times$SU(4).
It will be important for our construction that 
the scalar field $\phi^4$ has unit charge under 
$J_4$ and is not charged with respect to $J_1, J_2$ and $J_3$.

In section \ref{RSAdS}, we studied a sector of M-theory in 
AdS$_4\times S^7/\Z_k$ consisting of states for which 
$J_4$ is large and $J_1, J_2, J_3$ are of order $1$.
We now wish to construct the corresponding 
gauge-invariant operators on the CFT side. 
They are 
characterised by having large R-charge $J_4$. This can be achieved by 
considering operators involving a large number of 
$\phi^4$ insertions, 
since this field has unit charge under $J_4$. By construction 
this results in a large $\Jt=J_1+J_2+J_3+J_4$ as well. 
It is known that 
the definition of gauge invariant 
composite operators with non-zero $\Jt$ charge in the ABJM theory 
involves the use of so-called monopole 
operators~\cite{RBABJM, RBBorokhovKapustinWu1, 
RBBorokhovKapustinWu2}, 
closely related to the disorder operators introduced 
in~\cite{RBtHooftMonopoleOperators}. Monopole operators play an 
important role in the ABJM theory and more generally in three dimensional 
gauge theories~\cite{RBABJM, RBBorokhovKapustinWu1, 
RBBorokhovKapustinWu2}.
They are crucial for example for the enhancement of supersymmetry to 
$\mathcal{N}=8$ for $k=1, 2$ in the ABJM 
theory~\cite{RBBennaKlebanovKlose, RBBashkirovKapustin, 
RBGustavssonRey}. 

Monopole operators have no simple realisation as local polynomials in the 
elementary fields and the most convenient way of describing them in a 
conformal field theory is using radial quantisation and the state-operator 
map~\cite{RBBorokhovKapustinWu1, RBBorokhovKapustinWu2}. 
In the case of a three-dimensional CFT this involves 
mapping local operators inserted at the origin to 
states in a Hamiltonian formulation with the radial direction interpreted as 
Euclidean time. The Hamiltonian in radial quantisation,
for which each time slice is an $S^2$, is equivalent to the 
dilation
operator of the theory in $\R^3$ and thus operators with definite 
scaling dimension are in one to one correspondence with eigenstates of the 
Hamiltonian. The requirement of gauge-invariance for the composite 
operators is translated into Gauss law constraints 
which the physical states 
in radial quantisation should satisfy. 

In the present case the use of radial quantisation has the added benefit 
of leading to a more direct and natural correspondence between states 
in the matrix model discussed in section~\ref{RSAdS} -- describing 
fluctuations of spherical membranes -- and states in the ABJM theory 
on $S^2$. 

The relevance of monopole operators in the ABJM theory has been 
observed by various authors 
and the use of radial quantisation has been advocated 
before~\cite{RBSKim, RBBashkirovKapustin, RBBennaKlebanovKlose,
RBBerensteinTrancanelli, RBBerensteinPark, RBSheikhJabbariSimon, 
RBKimMadhu}. 
However, the approaches proposed so far are not suitable to study the
aspects of the AdS$_4$/CFT$_3$ duality we are interested in. We consider 
small $k$ (in order to be in a genuinely M-theoretic regime)
and therefore perturbation theory in $1/k$ is not applicable. Moreover we 
are especially interested in studying non-BPS states for which we cannot 
rely on exact non-renormalisation properties induced by supersymmetry. 
We will argue, however, that focussing on a large $J$ sector makes a 
quantitative comparison with the matrix model possible. 

In the remaining part of this section we will 
construct and study operators corresponding to the 
states on the AdS side discussed in section \ref{RSAdS},
using monopole operators and the state-operator map.
Since parts of the following discussion will be rather technical, we first 
present a brief summary of our analysis 
in order to highlight the essential points and 
emphasise the main line of ideas. 

\subsection*{Summary of analysis on the CFT side}
\label{RSCFTSummary}

We begin section \ref{RSCFTBPS} with the description of the Hamiltonian 
formulation of the radially quantised ABJM theory. In order to construct 
states corresponding to operators with large $J_4$ charge, one has to 
excite the $\phi^4$ field $J$ times. The colour charge density associated 
with the U($N$)$\times$U($N$) gauge group will have expectation value of 
order $J$ on the resulting state. In Chern-Simons theories coupled to matter 
fields, the Gauss law constraints equate the charge density to the magnetic 
field. Hence, in our states we should have a magnetic flux with strength of 
order $J$ through the $S^2$ space corresponding to a fixed (Euclidean) 
time slice in radial quantisation. The presence of this magnetic
flux, which satisfies a Dirac quantisation condition, 
defines the monopole operators in 
the framework of the state-operator map.

As we explain in section \ref{RSCFTBPS}, the magnetic flux we consider
is in the Abelian, diagonal, part of the two U($N$) gauge groups.
The total magnetic flux, which equals $J$ (up to $O(1)$ factors),
can be distributed among the $N$ entries of the diagonal part of the field 
strength (which is an $N\times N$ matrix). The integers characterising 
this partition of $J$ are referred to as the GNO charges.
The set of possible GNO charges gives the classification of the BPS 
operators/states on the CFT side (\ref{RFVacuumClassificationABJM}). 
This characterisation of the BPS states in the ABJM theory is in direct 
correspondence with the classification of the ground states on the AdS 
side (\ref{RFVacuumClassificationMM}).

In section \ref{RSCFTNearBPS} we consider the fluctuation spectrum 
around a particular ground state. In order to do this, it is necessary to fix 
a gauge. Our gauge fixing conditions (which involve a combination of 
background, unitary and Coulomb gauges) are 
specified in (\ref{RFGaugeFixBKG1})-(\ref{RFGaugeFixBKG2}),
(\ref{RFGaugeFixUnitary1})-(\ref{RFGaugeFixUnitary3}) and 
(\ref{RFGaugeFixRad1})-(\ref{RFGaugeFixRad3}).
We then compute the Hamiltonian of the gauge-fixed theory by solving
the Gauss law constraints. The part of the gauge-fixed Hamiltonian
necessary for the computation of the spectrum is given in 
(\ref{RFGaugeFixedHamiltonianCFT}). We find that the spectrum 
contains (i) low-energy modes with eigenfrequencies of order $1$
residing in the diagonal entries of the $N\times N$ fields, summarised 
in table \ref{RTCFTSlowModeSpectrum}, and (ii) high energy
modes with eigenfrequencies of order $J$ 
associated with off-diagonal entries, summarised in table 
\ref{RTCFTFastModeSpectrum}.
The spectrum of low-energy modes agrees 
with the spectrum we found on the AdS side, which is summarised in 
table~\ref{RTAdSNBPSTreeSpectrum}.

We then explain in section \ref{RSCFTPerturbation} that this large
separation 
of energy scales between diagonal and off-diagonal modes 
suggests that a Born-Oppenheimer type approximation is applicable.
In this scheme the off-diagonal, 
high-energy, modes should be integrated out.
The calculation of the spectrum in section 
\ref{RSCFTNearBPS} is then justified as arising from the leading order 
in this approximation.

\subsection{BPS states} 
\label{RSCFTBPS}

We begin this section by presenting the radial quantisation of the 
ABJM theory and introducing our notation and conventions. 
Properties of BPS observables in the (deformed) ABJM theory have been 
studied by various authors using radial 
quantisation~\cite{RBSKim, RBBashkirovKapustin,RBBennaKlebanovKlose,
RBSheikhJabbariSimon, RBBerensteinTrancanelli, RBBerensteinPark}. 
The derivation of the radially quantised ABJM theory in the Hamiltonian 
formalism requires special care with regard to the complex conjugation 
of fermionic fields. We have worked it out starting from the action of the 
ABJM theory in the component form given 
in~\cite{RBBandresLipsteinSchwarz} using slightly different conventions. 
In our conventions the Hamiltonian is
\begin{align}
H &= \int \dr\theta \dr\varphi \  \Tr \biggl[ \frac{1}{\sin \theta } \pi^{\ast}{}^{A} 
\pi_{A} + \sin \theta  g^{ \alpha \beta  } D_{\alpha} \phi^{\ast}{}_{A} D_{\beta}  
\phi^{A} 
    + \frac{1}{4} \sin \theta   \phi^{\ast}{}_{A} \phi^{A}   +  \sin \theta V_{6} 
    +   \sin \theta   V_{Y}  \notag \\
    &\ \ \ \ \ \ \ \ \ \ \ \ \ \ \ \ \ \  +   \sin \theta   \psi^{\ast T}{}_{A}  \left( 
    \sigma^{r} \sigma^{\alpha} D_{\alpha}  - 1 \right)  \psi^{A}    \biggl],
\label{RFABJMHamiltonianRadialQuantised}
\end{align}
where the sextic scalar potential, $V_6$, and the Yukawa couplings, $V_Y$, 
are
\begin{align}
V_{6} 
&=
\left( \frac{2 \pi }{k}  \right)^{\!2} 
\biggl(
\frac{1}{3} \phi^{B} \phi^{\ast}{}_{D} \phi^{D}   \phi^{\ast}{}_{C} \phi^{C} 
\phi^{\ast}{}_{B} 
+\frac{1}{3} \phi^{B}\phi^{\ast}{}_{B} \phi^{C}  \phi^{\ast}{}_{C} \phi^{D} 
\phi^{\ast}{}_{D}   
+  \frac{4}{3} \phi^{B} \phi^{\ast}{}_{D} \phi^{C} \phi^{\ast}{}_{B} \phi^{D} 
\phi^{\ast}{}_{C} \nonumber 
\\
&\ \ \ \ \ \ \ \ \ \ \ \ \ \ \ \  
-2  \phi^{B} \phi^{\ast}{}_{B} \phi^{D} \phi^{\ast}{}_{C} \phi^{C} 
\phi^{\ast}{}_{D}
\biggl), \label{RFV6} 
\\
V_{Y} &=   \frac{2 \pi}{k} \biggl(
    -2  \psi^{\ast T}{}_{A}   \sigma^{r}  \psi^{B} \phi^{A} \phi^{\ast}{}_{B}     
    +2  \psi^{\ast T}{}_{A}   \sigma^{r}  \phi^{\ast}{}_{B} \phi^{A} \psi^{B}   
    -  \psi^{\ast T}{}_{A}  \sigma^{r} \phi^{\ast}{}_{B} \phi^{B} \psi^{A} 
    +  \psi^{\ast T}{}_{A}   \sigma^{r}  \psi^{A} \phi^{B} \phi^{\ast}{}_{B} \notag 
\\
& \ \ \ \ \ \ \ \ \ \ \  
- i  \epsilon^{ABCD} 
\left(\psi^{ \ast T}{}_{D} B^{\ast T}\right) \phi^{\ast}{}_{C}   \psi^{\ast}{}_{B}  
\phi^{\ast}{}_{A} 
+ i  \epsilon_{ABCD}   \phi^{D} \psi^{CT}    \phi^{B} \left(B \psi^{A} 
 \right)  \! \biggl). 
\label{RFYukawa}
\end{align}
The conventions used in the above expression are as follows.
The $\pi$'s are the canonical conjugate variables of the $\phi$'s.
The indices $\a, \b=1,2$ are used for the $\th, \varphi$ coordinates of
the $S^2$ time-slice in radial quantisation.
In Chern-Simons theory the gauge fields $A_1$ ($A_\th$) 
and $A_2$ ($A_\v$) are canonically 
conjugate to each other and the same is true for the components of 
$\hA_\a$. We use Pauli matrices in polar coordinates, 
\begin{align}
&\hspace*{-0.25cm}\sigma^{r} \!=\! 
\begin{bmatrix}
\cos \theta & \sin \theta \,\er^{-i \varphi} \\
\sin \theta\, \er^{i \varphi} & - \cos \theta
\end{bmatrix} 
, \ \ \ \sigma^{\theta} \!=\! 
\begin{bmatrix}
- \sin \theta & \cos \theta\, \er^{-i \varphi} \\
\cos \theta\, \er^{i \varphi} &  \sin \theta
\end{bmatrix}, \ \ \  \sigma^{\varphi} \!=\!  \frac{1}{\sin \theta}
\begin{bmatrix}
0 & -i  \er^{-i \varphi} \\
i \er^{i \varphi} & 0
\end{bmatrix}. 
\end{align}
In the last line of~(\ref{RFYukawa}) 
$B$ is the charge conjugation matrix and in our conventions
$B=\s^2$,
where $\s^2$ is the usual Pauli matrix.
The superscript $T$ indicates transposition of the spinors.
We use the $*$ symbol to signify the operation 
in which one takes the complex conjugation 
(the adjoint for quantum mechanical operators) 
and the transpose 
in the colour indices.
For example, we have
\begin{equation}
\phi^*{}_A{}^{\hat{i}}{}_{j}=( \phi^{A}{}^{j}{}_{\hat{i}} )^{\ast}.
\end{equation}
The $*$ symbol is redundant as one
can see immediately from the position of the flavour index 
whether the complex conjugate is implied. We will hereafter omit the $*$
symbol when it is appropriate.
Our conventions for the covariant derivative
and the field strength are 
\begin{equation}
D_{\a} \phi^{A} = \partial_{\a} \phi^{A} - i A_{\a} \phi^{A} + i \phi^{A} 
\hat{A}_{\a} \, ,
\label{RFCovariantDer} 
\end{equation}
\begin{equation}
D_{\a} \psi^{A} = \partial_{\a} \psi^{A} - i \hA_{\a} \psi^{A} + i \psi^{A} A_{\a} 
\, ,
\end{equation}
\begin{equation}
F_{\a \b} = \partial_{\a}A_{\b} - \partial_{\b} A_{\a} - i [ A_{\a} , A_{\b}  ], \ \ \ 
\hat{F}_{\a \b} = \partial_{\a}\hat{A}_{\b} - \partial_{\b} \hat{A}_{\a} - i [ 
\hat{A}_{\a} , \hat{A}_{\b}  ].
\end{equation}
The Hamiltonian contains -- in addition to the kinetic terms, the scalar
potential~(\ref{RFV6}) and the Yukawa couplings~(\ref{RFYukawa}) --
mass terms for scalars and fermions. These mass terms, arising from 
radial quantisation, reflect the conformal dimensions of the fields. For
example scalar fields have mass $1/2$ (when the radius of the $S^2$ is
normalised to $1$).

The canonical (anti-)commutation relations are given by
\begin{equation}
[\phi^{A i}{}_\hj (x^{\prime}) ,  
\pi_B{}^\hk{}_l (x^{\prime \prime})  ]
= i \delta^{A}{}_{B} 
\delta^i{}_l \delta^\hk{}_\hj
\delta^2(x'-x'')
,
\end{equation}
\begin{equation}
[\psi^{A \hi}{}_j{}^a (x^{\prime}) , 
\psi^{T}{}_B{}^k{}_{\hl b} (x^{\prime \prime})  ]_{+}
=\frac{1}{\sin \theta}  
\delta^{A}{}_{B} \delta^\hi{}_\hl \delta^k{}_j \delta^a{}_b
\delta^2(x'-x'')
,
\end{equation}
\begin{equation}
[A_1{}^i{}_j(x'), A_2{}^k{}_l(x'')]
= i \frac{2\pi}{k} \delta^2(x'-x'')
\delta^i{}_l \delta^k{}_j,
\end{equation}
\begin{equation}
[\hA_1{}^\hi{}_\hj(x'), \hA_2{}^\hk{}_\hl(x'')]
= -i\frac{2\pi}{k} 
\delta^2(x'-x'') \delta^\hi{}_\hl \delta^\hk{}_\hj,
\end{equation}
where $\delta^2(x'-x'')= \delta ( \theta^{\prime} - \theta^{\prime \prime} ) 
\delta (\varphi^{\prime} - \varphi^{\prime \prime})$
and $a,b=1,2$ are spinor indices.

The Gauss law constraints are 
\begin{equation}
\frac{k}{2\pi} F_{12} = \rho, \label{RFGaussLaw}
\end{equation}
\begin{equation}
\frac{k}{2\pi} \hat{F}_{12} = -\hat{\rho}, \label{RFGaussLawHat}
\end{equation}
where $\rho$ and $\hrho$ are the colour charge densities for the 
two U($N$) gauge groups,
\begin{equation}
\rho^i{}_j = \left( i \phi^A \pi_A - i \pi^{A} \phi_A
+\sin\th \,\psi^{T}{}_A \psi^A \right)^i{}_j,
\label{RFDefRho}
\end{equation}
\begin{equation}
\hat{\rho}^\hi {}_\hj = \left( i \phi^{}_A \pi^{ A}-i \pi_A \phi^A 
\right)^\hi{}_\hj - \sin\th \,\psi^{T}{}_A{}^k{}_\hj \psi^{A\hi}{}_k.
\label{RFDefRhoHat}
\end{equation}
The relative sign difference in the Gauss law constraints 
reflects the fact that the Chern-Simons levels for the two U($N$) gauge 
fields are $k$ and $-k$.

The idea of the state-operator map is that the eigenstates of the 
Hamiltonian (\ref{RFABJMHamiltonianRadialQuantised})
correspond to operators with definite scaling dimensions,
which are given by the eigenvalues of the Hamiltonian.
The Gauss law constraints, (\ref{RFGaussLaw}) and 
(\ref{RFGaussLawHat}), give the condition that
the states, and hence the operators, be 
gauge invariant.

We now discuss the construction of states corresponding to
BPS operators with large charge $J_4$, 
which will involve the introduction of monopole 
operators.
The role of monopole operators in the definition and classification of certain 
BPS operators in the ABJM theory was first discussed in~\cite{RBABJM}.
For further discussion, see \cite{RBSKim, RBBennaKlebanovKlose, 
RBBashkirovKapustin,RBBerensteinTrancanelli, RBSheikhJabbariSimon, 
RBKimMadhu, RBBerensteinPark}.
The structure of BPS operators studied in these papers is consistent with 
the dual AdS picture.
The BPS operators we discuss below 
are a special case, characterised by a large $J_4$ charge, 
of the operators considered in these papers.
The main novelty in our work will be in extending the construction to 
non-BPS operators. This will be presented in the next subsection after 
a suitable gauge fixing. 

The arguments in the remainder of this subsection leading to the 
identification of BPS states should be considered as heuristic.
They involve assumptions, which may be difficult to directly justify.
In section \ref{RSCFTNearBPS}, 
however, we will give an alternative description of these states.
This is achieved by a suitable choice of gauge
in which each BPS state 
becomes the simple
perturbative vacuum.
The gauge choice is motivated by the description of BPS states in this 
subsection. The analysis in section \ref{RSCFTNearBPS} of these BPS ground states
-- and also of the non-BPS excited states around them --
is reliable in the large $J$ sector, provided that the approximation scheme
we propose, which will be explained in section \ref{RSCFTPerturbation}, 
is valid.

The BPS operators we are interested in have minimum conformal 
dimensions for given R-charges. Therefore, using the state-operator map, 
we will look for states which have minimum energies, 
\ie the ground states in the sector with given $J_4$. 
Our first assumption is that these ground states can be identified using
a free field description in which we treat the theory as if it consisted of
a collection of harmonic oscillators, neglecting the interactions among them.
One basis for this assumption is that we are considering BPS 
operators, for which observable quantities such as conformal dimensions
are protected. Under this assumption, 
states for which the R-charge $J_4$ takes a given value $J$ can be 
obtained by exciting the field $\phi^4$ $J$ times.

We should excite only zero modes of $\phi^4$ on the $S^2$
time-slice in radial quantisation, since non-zero modes have larger energy 
and hence their excitation should be avoided.
The state 
thus obtained has constant colour charge densities, $\rho$
and $\hat{\rho}$, of order $J$, because the $\phi^4$ field 
carries non-zero colour charge~\footnote{This state may be 
interpreted as the Bose-Einstein condensate, resulting from the requirement 
that the charge $J_4$ be large.}.
Hence because of the Gauss law constraints, (\ref{RFGaussLaw}) and 
(\ref{RFGaussLawHat}), 
one has to introduce a constant magnetic flux 
through the $S^2$ 
time-slice in radial quantisation.
This magnetic flux defines the monopole operator.

Since the $\phi^4$ field is a matrix,
$\phi^4{}^i{}_\hj$, we should specify which elements
of the matrix contribute to the ground state.
Our second assumption is that only diagonal elements of $\phi^4{}^i{}_\hj$
should be excited. The need of a similar assumption was pointed out 
in~\cite{RBSKim}. 
The states for which off-diagonal elements 
are excited should either have larger energy or be gauge-equivalent to 
the states containing diagonal excitations only.
The rationale for this assumption is the fact that 
the non-negative scalar potential 
(\ref{RFV6}) is classically zero 
when the fields consist of diagonal matrices (or 
configurations gauge-equivalent to diagonal matrices). 
Moreover, classically any configuration of the
matrix field $\phi^{4 i}{}_{\hat{j}}$ can be diagonalised by 
a U($N$)$\times$U($N$) gauge transformation, as can be proven 
using the so-called polar decomposition of 
matrices~\cite{RBGolubVanLoan}.

By the above assumptions, it is sufficient to excite  
only zero modes in the diagonal entries of $\phi^4$ and we denote by 
$J_{(i)}$ the number of excitations of the $i$-th diagonal component.
Since the total number of excitations of $\phi^4$ should be $J$, 
the possible BPS states are labelled by a set of integers 
satisfying
\begin{equation}
J=\sum_{i=1}^N J_{(i)},
\label{RFVacuumClassificationABJM}
\end{equation}
where each $J_{(i)}$ turns out to be a multiple of $k$ as we will see below.
As pointed out above, this excitation of zero modes induces a constant flux, 
through the Gauss law 
constraints~(\ref{RFGaussLaw})-(\ref{RFGaussLawHat}). In order to satisfy
these equations we need gauge fields $A_\a$ and $\hat{A}_\a$ with
diagonal components given by the vector potential of a Dirac 
monopole~\cite{RBWuYangHarmonics} with magnetic charge 
$J_{(i)}/2k$,
\begin{equation}
A_\a= \mbox{diag} \left[\frac{J_{(i)}}{2 k} \right] \times \label{RFAGNO1} 
(A_\a \mbox{ for Dirac monopole with unit magnetic charge}) \,~
\end{equation}
\begin{equation}
\hA_\a=\mbox{diag} \left[ \frac{J_{(i)}}{2 k} \right] \times \label{RFAGNO2}
(A_\a \mbox{ for Dirac monopole with unit magnetic charge}) \,.
\end{equation}
Because of the Dirac quantisation 
condition~\cite{RBDiracMonopole, RBWuYangQuantisationCondition}, 
the gauge fields (\ref{RFAGNO1}) and (\ref{RFAGNO2}) are 
consistent only if all $J_{(i)}/2 k$ are integers divided by two,
\ie only if all the $J_{(i)}$'s are multiples of $k$.
We define
\begin{equation}
2 k q_{(i)} = J_{(i)} \, ,
\end{equation}
where $2q_{(i)}$ is an integer~\footnote{
We follow the convention in \cite{RBWuYangHarmonics} for the definition
of the magnetic flux $q$, where $2q$ is an integer. The spectrum of the
Laplace operator on $S^2$ in the presence of the magnetic flux is
given by $l(l+1)-q^2$ in this convention. In the recent literature, \eg in 
\cite{RBBashkirovKapustin, RBSKim}, $q'=2q$ is usually denoted by $q$.}.
These $q_{(i)}$'s, which characterise the configuration 
of flux in radial quantisation (or equivalently the  monopole operator), 
are referred to as GNO charges
\cite{RBGNO, RBBorokhovKapustinWu1, RBBorokhovKapustinWu2}.

This classification of BPS states in terms of a set of integers, $J_{(i)}$, 
satisfying~(\ref{RFVacuumClassificationABJM}) is in direct correspondence 
with the classification of vacua in the pp-wave matrix model discussed in 
section~\ref{RSAdSBPS}. In the matrix model the integers $J_{(i)}$ 
characterise the angular momenta of concentric spherical membranes and 
satisfy the condition~(\ref{RFVacuumClassificationMM}), which is the same 
as~(\ref{RFVacuumClassificationABJM}). 

The energy $\Delta$ of the state
resulting from the excitations of $\phi^4$ described above is 
\begin{equation}
\Delta= \frac{J}{2} \, , 
\label{RFCFTEnergyGroundState}
\end{equation}
since the only contribution to the energy comes from the mass of the 
field $\phi^4$, which is $1/2$. 
We note that in (\ref{RFABJMHamiltonianRadialQuantised}) 
there is no direct contribution to the energy from the magnetic field.
This is as expected in view of the BPS nature of the state and
it also agrees with the property of the dual ground state in the matrix model.
For a more generic class of BPS operators 
relations analogous to (\ref{RFCFTEnergyGroundState}) were verified
in \cite{RBBashkirovKapustin, RBBennaKlebanovKlose} 
using a method based on an appropriate deformation of the ABJM theory.

A difference between the two sides of the duality, at least at first sight, is 
that in the CFT the rank of the gauge groups, $N$, sets an upper bound on 
the number of non-zero integers in the partition of $J$: the number of GNO 
charges cannot exceed $N$, which in particular implies $J\le N$. Such a 
bound need not be satisfied on the gravity side. For example, in a vacuum 
with a very large number $n$ of membranes, for which the individual 
angular momenta $J_{(i)}$ obey~(\ref{RFValidityPPApproxRadiusM2}), the 
sum $\sum_{i=1}^n J_{(i)}=J$ can be larger than $N$. 
This is not an inconsistency, as can be understood using the following 
observation. The states on the AdS side violating the upper bound 
would contain more than $N$ 
spherical membranes with angular momentum in $S^7$. However, the 
original AdS$_4\times S^7$ background is produced by a stack of $N$ 
membranes and it is natural to expect that neglecting the back-reaction of 
a number of rotating membranes larger than $N$ on this background 
would be inconsistent. Therefore, the study of configurations of this type 
would be outside of the validity of the usual AdS/CFT correspondence.
In fact, this is a general feature common to all 
examples of AdS/CFT duality. In a gauge theory for 
finite $N$ there is an upper bound on the number of independent 
gauge-invariant combinations of fields, 
which in general should be understood on the AdS side as being related to 
the effect of the back-reaction on the background.

A more concrete resolution of the apparent contradiction can be given 
recalling the considerations of section~\ref{RSAdSBPS} on the applicability 
of the pp-wave approximation. As discussed in~\ref{RSAdSBPS}, the 
possibility of tunnelling between different perturbative sectors leads to a 
compelling argument for requiring the applicability of the pp-wave 
approximation to all possible matrix model vacua. This in turn leads to the 
condition $J\ll (Nk)^{1/2} <N$, so that the upper bound on the CFT side 
ceases to be meaningful. 

The correspondence between 1/2 BPS operators of the ABJM theory and 
ground states of the pp-wave matrix model was also pointed out 
in~\cite{RBSheikhJabbariSimon}.
However, the importance of focussing on a large $J$ sector,
which is the essential ingredient that allows us to 
extend the analysis beyond the BPS sector,
was not noticed before.

On the AdS side, as discussed in section \ref{RSAdSBPS}, 
the vacuum states classified by the partition of $J$
should be connected non-perturbatively by tunnelling effects,
which physically should be interpreted as interaction of membranes.
Therefore we expect the ground states
considered in this section
to be connected non-perturbatively as well, 
provided that the sum of the GNO charges, $\sum_{(i)} q_{(i)}$, is 
conserved. This presumably means that there should be classical solutions 
in the Euclidean theory connecting two given Dirac monopoles (with the 
same total GNO charge), corresponding to a tunnelling between 
the two configurations.
It would be interesting to identify and explicitly construct such 
instanton-like solutions interpolating between vacua 
corresponding to different sets of GNO charges.

The construction in this paper
is analogous to the construction of BPS operators
in the BMN sector of $\calN=4$ super Yang-Mills~\footnote{
A matrix model approach to the type IIB superstring theory 
in the pp-wave approximation was proposed in \cite{RBSheikhJabbari}.
}. The scalar field
$\phi^4$ plays a role similar to the complex scalar $Z$ in~\cite{RBBMN}.
However, one cannot define simple gauge invariant 
BMN-like operators by taking a trace because 
$\phi^4$ is a bi-fundamental field. 
This leads to the necessity of using monopole operators
in the way discussed here.
This is related to a crucial difference between the Chern-Simons and 
Yang-Mills theories, \ie the fact that in the former the Gauss law constraints 
(\ref{RFGaussLaw})-(\ref{RFGaussLawHat})
equate the charge density to the magnetic part of field strength, 
rather than the divergence of the electric field. As a consequence 
one can have a non-zero (although quantised by Dirac's 
condition) total colour charge even on a compact space (in the present 
case the $S^2$ in radial quantisation).

\subsection{Near BPS excitations}
\label{RSCFTNearBPS}

In this section we study the near BPS fluctuations around the BPS states
described in the previous section after presenting an alternative description
of the latter.

Before going into the details of the gauge fixing procedure we use,
we discuss some basic ideas behind it.
In the state-operator map, 
the gauge invariance of an operator 
translates into the 
Gauss law constraints imposed on the corresponding state.
A clean approach to study these physical states
is to first quantise the Hamiltonian formalism described
in section \ref{RSCFTBPS} and
then later impose the first class constraints (the Gauss law 
constraints) on state vectors, 
following Dirac's approach to the quantisation of constrained 
systems~\cite{RBDiracConstraint}.
We found that, for the ABJM theory in the sector we are considering, 
it is not easy to carry out this program, 
due to technical difficulties 
associated with ordering ambiguities in the quantum Hamiltonian  
and the constraints.
In the following, 
we employ another standard approach
(used for example 
in the quantisation of string theory in light-cone gauge~\cite{RBGGRT})
to study physical states in a theory with gauge symmetries.
We first eliminate some of the degrees of freedom 
by introducing gauge fixing conditions. We then 
express the conjugate momenta of the eliminated variables
in terms of the remaining physical variables using the Gauss law constraints.
These two steps are carried out in the classical theory, and 
the resulting gauge-fixed theory is then quantised.
The information of the Gauss law constraints
is already taken into account at the classical level 
and the states in the quantum theory are by construction
physical. 
(If there is a
residual gauge symmetry, the constraints cannot be solved completely.
The remaining part of the constraints 
should be imposed on the states of the 
(partially) gauge-fixed theory.)
The ordering problem does not arise in this approach.
The price one has to pay
is that there is no guarantee that the gauge-fixed theory 
will have all the (global) symmetries of the original theory.
Although we believe that 
all global symmetries of the original ABJM theory
are properly realised in the gauge-fixed theory we describe below,
it is important to explicitly verify this.

The gauge we choose is a combination 
of the background, unitary and Coulomb gauges. 
This choice is particularly well-suited to
clarify the physical content of the theory in the sector we consider.
We focus on the case where only one of the GNO charges is non-zero,
$J=J_{(1)}$, corresponding to the case where 
there is only one spherical membrane. 
We discuss some aspects of the more general case in
section \ref{RSMultiMembrane}.
In the presence of a single GNO charge, it is convenient to introduce 
indices $i', j'=2, \ldots, N$ and $\hi', \hj'=2, \ldots, N$. 
Elementary 
$N\times N$  fields in the ABJM theory can be decomposed into blocks
of size $1\times 1$, $(N-1)\times (N-1)$, $1\times (N-1)$ and
$(N-1)\times 1$ respectively. 
For example, the field $\phi^A{}^i{}_\hj$
is decomposed as
\begin{equation}
\phi^A{}^i{}_\hj=
\left[ \begin{array}{c|c}
\phi^A{}^1{}_{\hat{1}}
& 
\rule{30pt}{0pt} 
\phi^A{}^1{}_{\hat{j}'}  
\rule{30pt}{0pt}
\raisebox{-10pt}{\rule{0pt}{0pt}} \\
\hline
\phi^A{}^{i'}{}_{\hat{1}}  
&
\rule{12.5pt}{0pt} 
\phi^A{}^{i'}{}_{\hj'}
\raisebox{-40pt}{\rule{0pt}{80pt}}\rule{12pt}{0pt}
\rule{0pt}{47pt}
\end{array} \right].
\end{equation}
Since we wish to treat fluctuations around the ground state
as perturbations,
it is necessary to
separate the large, order $J$, contribution of the ground state 
from the small, order one, contribution of the fluctuations.
We accordingly split the gauge field into a background part, $B_\a$, 
corresponding to the 
constant monopole
flux as explained around (\ref{RFAGNO1}) and 
(\ref{RFAGNO2}),
and 
the fluctuation about it, $a_\a$, 
\begin{equation}
(A_{\a})^{i}{}_{j} = 
(B_{\a})^{i}{}_{j} +  (a_{\a})^{i}{}_{j} \, ,
\label{RFGaugeFixBKG1} 
\end{equation}
as is done in the usual  background gauge. Similarly we decompose 
$\hA_\a$ as
\begin{equation}
(\hat{A}_{\a})^{\hat{i}}{}_{\hat{j}} = 
(B_{\a})^{ \hat{i}}{}_{ \hat{j}} +  (\hat{a}_{\a})^{ \hat{i} }{}_{ \hat{ j  }} \, .
\label{RFGaugeFixBKG2}
\end{equation}
Here the background field $B_\a$ 
about which both $A_\a$ and $\hat A_\a$ are expanded, is
\begin{equation}
\arraycolsep5pt
( B_{2} )^{i}{}_{j} = ( B_{2} )^{\hat{i}}{}_{\hat{j}} =
\left[
\begin{array}{@{\,}c|cccc@{\,}}
q(1-\cos \theta ) & 0  & 0 & \cdots  &  0 \\
\hline
0 &&&&\\
0 &  \multicolumn{4}{c}{\raisebox{-10pt}[0pt][0pt]{ \large{ $0$}} }\\
\vdots &&&&\\
0 &&&&\\
\end{array}
\right], \ \ \ (B_{1})^{i}{}_{j} = (B_{1})^{\hat{i}}{}_{\hat{j}} =0. 
\label{RFBKGGaugeField}
\end{equation}
In \cite{RBWuYangHarmonics} the background gauge field 
is given for two patches excluding either the north-pole or the south-pole 
of $S^2$. We will use the patch excluding the south-pole. 
The observable quantities we compute in the following should not 
depend on the choice of patch.

We require that the fluctuation fields $a_\a$ and $\ha_\a$ 
do not have 
singularities present in the gauge field for the Dirac monopole, 
as small fluctuations cannot satisfy the Dirac quantisation condition.
Therefore formulae (\ref{RFGaugeFixBKG1}), (\ref{RFGaugeFixBKG2}) 
and (\ref{RFBKGGaugeField}) 
imply that we focus on a particular sector of the
ABJM theory. More precisely, 
we have focussed on a perturbative vacuum associated 
with a particular choice of GNO charges, although
as remarked in section \ref{RSCFTBPS}, two given vacua
with the same total GNO charge should be non-perturbatively connected 
by tunnelling effects.

The Gauss law  
constraints, (\ref{RFGaussLaw}) and  (\ref{RFGaussLawHat}),
equate the magnetic field to the charge density.
Therefore we should 
separate the contribution of the ground state
and the fluctuation to the charge density as well.
The charge density for our ground state
(the state with lowest energy for
given charge)
is produced 
by the excitation of the zero-mode oscillator 
associated with the field $\phi^{41}{}_{\hat{1}}$,
as explained in section~\ref{RSCFTBPS}.
However, the separation cannot be achieved  
in a straightforward manner
by 
introducing a background value for $\phi^4{}^1{}_{\hat{1}}$; 
the expectation value of $\phi^4{}^1{}_{\hat{1}}$ is zero
for the ground state  we are considering, although 
the expectation values of the composite operators $\rho$ and $\hrho$ are 
non-zero. This is a consequence of the fact that the phase of 
$\phi^{4\,1}{}_{\hat{1}}$ is undetermined (whereas 
$ |\phi^4{}^1{}_{\hat{1}}|^2$, as well as $|\pi_4{}^{\hat{1}}{}_1|^2$, 
have definite non-zero expectation value of order $J$) 
and thus averaging over in the path integral produces a vanishing
expectation value for $\phi^{4\,1}{}_{\hat{1}}$.
Our idea is to gauge away the 
phase degrees of freedom of $\phi^4{}^1{}_{\hat{1}}$
by choosing the unitary gauge in order to deal with this issue.

We define the real fields $u$, $v$ such that
\begin{equation}
\phi^{4 1}{}_{\hat{1}} = \frac{1}{\sqrt{2}} (f + u+ iv). \label{RFDeffuv}
\end{equation}
We will see 
that an effective potential is generated such that
the minimum of the potential is at a non-zero value of the real
part of $\phi^{4 1}{}_{\hat{1}}$, which will be determined later.
In (\ref{RFDeffuv}) 
we have separated the vacuum expectation value, $f$, 
and the fluctuation around it, $u$, for convenience. 
The canonical conjugate momenta $p_u$ and $p_v$ satisfy
\begin{equation}
\pi_4{}^{\hat{1}} {}_1=\frac{1}{\sqrt{2}} (p_u - ip_v).
\end{equation}
By requiring 
\begin{eqnarray}
&&
v=0,
\label{RFGaugeFixUnitary1}
\end{eqnarray}
together with the condition $f+u\geq 0$,
we gauge away the phase degrees of freedom of $\phi^{4 1}{}_{\hat{1}}$.
Other unitary gauge conditions we choose are
\begin{eqnarray}
&&
{\phi^{41}}_{\hi'}=0,
\label{RFGaugeFixUnitary2}
\\
&&
{\phi^{4 i'}}_{\hat{1}}=0, 
\label{RFGaugeFixUnitary3}
\end{eqnarray}
which similarly eliminate the gauge freedom associated with
$(1, i')$ gauge transformations.
One can prove, using the polar decomposition of
matrices~\cite{RBGolubVanLoan},
that a configuration satisfying these gauge conditions
can be obtained by performing a U($N$)$\times$U($N$) gauge 
transformation
from any configuration of $\phi^4$.

We fix the remaining gauge freedom by imposing Coulomb 
gauge conditions. We decompose the fluctuation part of the gauge fields 
$a_\a$ and $\ha_\a$ as 
\begin{equation}
( a_{\a} )^{i}{}_{ j }=
\left[ \begin{array}{c|c}
z_\a
& 
\rule{30pt}{0pt} 
w_{\a j'}  
\rule{30pt}{0pt}
\raisebox{-10pt}{\rule{0pt}{0pt}} \\
\hline
(w_{\a i'})^*
&
\rule{12.5pt}{0pt} 
\left( a_{\a } \right)^{ i' }{}_{ j' } 
\raisebox{-40pt}{\rule{0pt}{80pt}}\rule{12pt}{0pt}
\rule{0pt}{47pt}
\end{array} \right],
\quad
( \ha_{\a} )^{\hi}{}_{\hj }=
\left[ \begin{array}{c|c}
\hz_\a
& 
\rule{30pt}{0pt} 
\hw_{\a \hj'}  
\rule{30pt}{0pt}
\raisebox{-10pt}{\rule{0pt}{0pt}} \\
\hline
(\hw_{\a \hi'})^*
&
\rule{12.5pt}{0pt} 
\left( \hat{a}_{\a } \right)^{ \hat{i}^{\prime} }{}_{\hat{j}^{\prime} } 
\raisebox{-40pt}{\rule{0pt}{80pt}}\rule{12pt}{0pt}
\rule{0pt}{47pt}
\end{array} \right],
\end{equation}
defining the U($1$) gauge fields $z_\a$, $\hz_\a$ and the ``W-bosons''
$w_{\a i'}$, $\hw_{\a \hi'}$.
In the Hamiltonian and the Gauss law constraints, the fields $z$ and $\hz$
appear frequently in the combinations 
\begin{equation}
{z^-}_\a  =z_\a  - \hat{z}_\a ,
\end{equation}
\begin{equation}
{z^+}_\a  =\frac{1}{2}(z_\a  + \hat{z}_\a ).
\end{equation}
In terms of these fields, the Coulomb gauge conditions are
\begin{eqnarray}
&&
{\rm div}\, z^+ =0,
\label{RFGaugeFixRad1}
\\
&&
{\rm div}\, a^{i'}{}_{j'}=0,
\label{RFGaugeFixRad2}
\\
&&
{\rm div}\, \ha^{\hi'}{}_{\hj'}=0.
\label{RFGaugeFixRad3}
\end{eqnarray}

The conditions~(\ref{RFGaugeFixUnitary1})-(\ref{RFGaugeFixUnitary3}) 
and~(\ref{RFGaugeFixRad1})-(\ref{RFGaugeFixRad3}), fix the gauge 
ambiguity (up to residual gauge transformations with constant parameters 
which will be discussed later)~\footnote{Our gauge fixing conditions have 
some similarities to those used in \cite{RBSKim, RBBashkirovKapustin}.}.
The next step is to solve the Gauss law constraints 
using the gauge fixing conditions.
In this way the canonical momenta conjugate to the variables
eliminated by the gauge conditions are rewritten 
in terms of the remaining physical degrees of freedom
of the gauge-fixed theory.
The variables to be eliminated using the Gauss law constraints are 
\begin{equation} 
p_v, 
\ \ \ 
\pi_4{}^{\hat{1}}{}_{i^{\prime}},
\ \ \ 
\pi^{4}{}^{1}{}_{\hat{j}^{\prime}}, 
\ \ \ 
{\rm rot}\, z^-, 
\ \ \ 
{\rm rot}\, a^{i^{\prime}}{}_{j^{\prime}},
\ \ \ 
{\rm rot}\, \hat{a}^{\hat{i}^{\prime}}{}_{\hat{j}^{\prime}}.
\label{RFListVarSolved}
\end{equation} 
The physical variables of the gauge-fixed theory are
\begin{equation}
\phi^{A'i}{}_{\hj},
\ \ \ 
\pi_{A'}{}^{\hi}{}_j, 
\ \ \ 
\psi^{A\hi}{}_j,
\ \ \
u,
\ \ \
p_u,
\ \ \ 
{\rm div}\, z^-,
\ \ \
{\rm rot}\, z^+,
\ \ \
w_{\a i'},
\ \ \
\hw_{\a \hi'},
\ \ \
\phi^{4 i'}{}_{\hj'},
\ \ \
\pi_{4}{}^{\hi'}{}_{j'},
\label{RFListVarPhysical}
\end{equation}
where we have introduced indices $A', B'=1,2,3$.
Once the Gauss laws are solved, one can compute the 
observables in the gauge-fixed theory, such as the Hamiltonian and
various symmetry charges, by substituting the expression for
the variables (\ref{RFListVarSolved}) into the 
original expression of these observables before fixing the gauge.
We note that, for example, ${\rm div}\, z$ and ${\rm rot}\, z$
are canonically conjugate to each other in the following sense.
Expanding the one-form field $z_\a$ using
the one forms constructed from the spherical harmonics, 
$\dr Y_{lm}$ and $*\dr Y_{lm}$ ($l=1,2, \ldots$), one can show that,
up to a numerical factor depending on $l$, 
the coefficients in the expansion are canonically conjugate to each other.
This expansion can be justified by using the standard 
Hodge decomposition theorem~\cite{RBHodgeTheorem},
which states that any differential form can be written as the sum of 
exact (rotationless), co-exact (divergenceless) and harmonic forms.
A harmonic one-form on the sphere is necessarily singular and has the 
form of the gauge field for the Dirac monopole,
which is excluded in our case, as discussed 
below~(\ref{RFBKGGaugeField}).
Hence the one-form fields $z_\a^\pm$ can be specified by giving 
$\rot z^\pm$ and $\div z^\pm$.

The Gauss law constraints are non-linear and 
should be solved in an iterative manner 
in general.
It is convenient to rewrite 
the Gauss law constraints (\ref{RFGaussLaw}), (\ref{RFGaussLawHat}), 
by using (\ref{RFGaugeFixBKG1}), (\ref{RFGaugeFixBKG2}) and 
(\ref{RFBKGGaugeField}), as 
\begin{align}
\frac{k}{2\pi} F_{12}{}^{\rm BKG} + \frac{k}{2\pi} 
\partial^{\prime}_{\underline{1}} a_{\underline{2}} &= i \phi^{4}\pi_4 
- i \pi{}^{4} \phi{}_{4} + \rho_{W}, \label{RFGL1}
\\
\frac{k}{2\pi} \hat{F}_{12}{}^{\rm BKG} + \frac{k}{2\pi}
\partial^{\prime}_{\underline{1}} \hat{a}_{\underline{2}} &
= -i \phi{}_{4}\pi{}^{4} + i \pi_4 \phi^4 -  \hat{\rho}_{W}, \label{RFGL2}
\end{align}
where $F_{12}{}^{\rm BKG}$ and $\hat{F}_{12}{}^{\rm BKG}$ are field 
strengths for the background gauge fields, $B_{\alpha}{}^{i}{}_{j}$ and 
$B_{\alpha}{}^{\hat{i}}{}_{\hat{j}}$, and
$\rho_W$ and $\hrho_W$ are defined as
\begin{align}
\rho_W{}^i{}_j &=
\left( i \phi^{A^{\prime}} \pi_{A^{\prime}} - i \pi^{A^{\prime}} \phi_{A^{\prime}} 
+ \sin \theta \psi^{T}{}_{A} \psi^{A} 
+ i \frac{k}{2\pi} [a_{1}, a_{2}]\right)^i{}_j
, \\
\hat{\rho}_W{}^\hi{}_\hj &=
\left(
i\phi{}_{A^{\prime}} \pi{}^{A^{\prime}} - i \pi_{A^{\prime}} \phi^{A^{\prime}} 
\right)^\hi{}_\hj
- \sin \theta \psi^{T}{}_{A}{}^k{}_\hj \psi^{A\hi}{}_k 
- i \frac{k}{2\pi} [\hat{a}_1, \hat{a}_2]^\hi{}_\hj,
\end{align}
separating the contributions of $\phi^4$ and $\phi^{A'}$ ($A'=1, 2, 3$). 
The $(1,1)$ components of the operators 
$\rho_W$ and $\hrho_W$ may be thought of as
charge densities for the fields, 
$\phi^{A'}$ and $\psi^A$, 
and the ``W-bosons'', $w_{\a i'}$ and $\hw_{\a \hi'}$.
We use the symbol $\partial'$ to 
denote the covariant derivative defined in terms of the background gauge 
fields~(\ref{RFBKGGaugeField}). For example
\begin{align}
& 
\partial'_{\a} \phi^{A'}{}^{1}{}_{\hi'} = \partial_{\alpha} \phi^{A'}{}^{1}{}_{\hi'}
- i B_{\a} \phi^{A'}{}^{1}{}_{\hi'} , \label{RFPartialPrimeEx1}
\\
&
\partial'_{\a} \phi^{A'}{}^{i'}{}_{\hat{1}} = \partial_{\alpha}  
\phi^{A'}{}^{i'}{}_{\hat{1}}
+ i B_{\a}  \phi^{A'}{}^{i'}{}_{\hat{1}},
\\
&
\partial'_{\a} \psi^{A}{}^{\hat{1}}{}_{1} = \partial_{\alpha} 
\psi^{A}{}^{\hat{1}}{}_{1},
\\
&
\partial'_{\a}w_{\b i'} = \partial_{\alpha} w_{\beta i'} 
        -iB_{\a}w_{\b i'}
\\
&
\partial'_{\a} z_{\b} = \partial_{\alpha} z_{\beta} \, . 
\label{RFPartialPrimeEx4}
\end{align}
We use underlined indices to indicate anti-symmetrisation (without any 
normalisation factors). For example 
\begin{equation}
\partial'_{\underline{1}} a_{\underline{2}} = \partial'_{1} a_{2} - 
\partial'_{2} a_{1} \, .
\end{equation}
By adding and subtracting the $(1,1)$ 
components of the two Gauss law constraints
(\ref{RFGL1}) and (\ref{RFGL2}) under the unitary gauge condition
(\ref{RFGaugeFixUnitary1})-(\ref{RFGaugeFixUnitary3}), one obtains, 
respectively,
\begin{align}
p_v
&=
\frac{1}{f+u}
\left(
\frac{k}{2 \pi} q \sin{\th} 
+
\frac{k}{2\pi} \partial_{\underline{1}} z^+_{\underline{2}}
-\frac{1}{2}\rho_W{}^1{}_{1}
+\frac{1}{2}\hrho_W{}^{\hat{1}}{}_{\hat{1}}
\right), \label{RFSolvedpv}
\\
\partial_{\underline{1}} z^-_{\underline{2}}
&=
\frac{2\pi}{k} \left( \rho_W{}^1{}_{1}
+\hrho_W{}^{\hat{1}}{}_{\hat{1}} \right). \label{RFSolvedrotz} 
\end{align}
The right hand sides of these expressions are 
written only in terms of physical variables
and thus $p_v$ and $\partial_{\underline{1}} z^-_{\underline{2}}
\sim{\rm rot}\, z^-$ are solved completely.

From the $(1,i^{\prime})$ and $(i', j')$ 
components of the Gauss law constraints (\ref{RFGL1})-(\ref{RFGL2})
under the unitary gauge condition 
(\ref{RFGaugeFixUnitary1})-(\ref{RFGaugeFixUnitary3}), we obtain   
\begin{align}
\pi_4{}^{\hat{1}}{}_{i^{\prime}} 
&=
-i \frac{\sqrt{2}}{f+u} 
\left(
\frac{k}{2\pi } \partial^{\prime}_{\underline{1}} w_{\underline{2}i^{\prime}} 
- \rho_W{}^{1}{}_{i'}
+ i \pi^{41}{}_{\hj'} \phi_4{}^{\hj'}{}_{i'}  
\right)
\label{RFSolvedpi1}, 
\\
\pi^{{} 4}{}^{1}{}_{\hat{j}^{\prime}}
&=
i \frac{\sqrt{2}}{f+u} 
\left(
\frac{k}{2\pi} \partial'_{\underline{1}} \hat{w}_{\underline{2}\hj'} 
+ \hrho_W{}^{\hat{1}}{}_{\hj'} 
- i \pi_{4}{}^{\hat{1}}{}_{i'} \phi^{4 i'}{}_{\hj'}
\right), 
\label{RFSolvedpi2}
\\
\partial_{\underline{1}}a_{\underline{2}}{}^{i^{\prime}}{}_{j^{\prime}} 
&=
\frac{2\pi}{k} 
\left( 
i\phi^4{}^{i^{\prime}}{}_{\hat{k}^{\prime}} 
\pi_4{}^{\hat{k}^{\prime}}{}_{j^{\prime}} 
-i \pi^{{}}{}^{4}{}^{i^{\prime}}{}_{\hat{k}^{\prime}} 
\phi^{{}}{}_{4}{}^{\hat{k}^{\prime}}{}_{j^{\prime}} + 
\rho_W{}^{i^{\prime}}{}_{j^{\prime}}   \right),
\label{RFSolvedrota} \\
 \partial_{\underline{1}}
 \hat{a}_{\underline{2}}{}^{\hat{i}^{\prime}}{}_{\hat{j}^{\prime}} &=
 \frac{2\pi}{k} \left( -i \phi^{{}}{}_{4}{}^{\hat{i}^{\prime}}{}_{k^{\prime}} 
 \pi^{{}}{}^{4}{}^{k^{\prime}}{}_{\hat{j}^{\prime}}
 + i \pi_4{}^{\hat{i}^{\prime}}{}_{k^{\prime}} 
 \phi^4{}^{k^{\prime}}{}_{\hat{j}^{\prime}} - 
 \hat{\rho}_W{}^{\hat{i}^{\prime}}{}_{\hat{j}^{\prime}}  \right)
 \label{RFSolvedrotahat}.
\end{align}
The right hand sides of these formulae are 
not written solely in terms of the physical variables.
Hence to determine the fields 
$\pi_4{}^{\hat{1}}{}_{i^{\prime}}$,
$\pi^{4}{}^{1}{}_{\hat{j}^{\prime}}$,
${\rm rot}\, a^{i^{\prime}}{}_{j^{\prime}}$ and 
${\rm rot}\, \hat{a}^{\hat{i}^{\prime}}{}_{\hat{j}^{\prime}}$ it is 
necessary to proceed iteratively. The result is an infinite expansion for 
these fields in which the terms produced by each subsequent iteration 
contain a larger number of physical fields.
For example, the right hand side of (\ref{RFSolvedpi1}) contains 
$\pi^{41}{}_{\hj'} \phi_4{}^{\hj'}{}_{i'}$, which should be solved 
again using (\ref{RFSolvedpi2}). This term 
contains quadratic and higher order terms
in the fluctuation 
fields and, in the leading order, can be neglected. Also,
$\rho_W{}^1{}_{i'}$ contains the field $z_\a$,
so part of it should be solved using (\ref{RFSolvedrotz}).

The (iteratively) solved variables should be substituted into the original 
Hamiltonian~(\ref{RFABJMHamiltonianRadialQuantised})
to obtain the Hamiltonian of the gauge-fixed theory.
In general a number of iterations are needed to obtain all the terms in
the Hamiltonian which are necessary to study a given process.

Eliminating the canonical momentum $p_v$ through the Gauss law 
constraint~(\ref{RFSolvedpv}) produces an effective potential for the 
real part of $\phi^{41}{}_{\hat{1}}$. Together with the mass term arising in 
radial quantisation, this results in a version of the Higgs mechanism inducing 
a vacuum expectation value for the real part of $\phi^{41}{}_{\hat{1}}$. This 
effect will play a crucial role in our analysis. The field $p_v$ enters in the 
original Hamiltonian in the term
\begin{equation}
\int \dr\theta \dr\varphi \, \frac{1}{\sin\th}\frac{1}{2}(p_v)^2.
\end{equation}
Using~(\ref{RFSolvedpv}) to express $p_v$ in terms of the physical 
variables produces, among others, the term
\begin{equation}
\int \dr\theta \dr\varphi \, \frac{1}{32 \pi^2} \frac{J^2}{(f+u)^2} \sin \theta \, .
\label{RFCentrifugalPot}
\end{equation}
This term gives an effective potential for the real part of the field
$\phi^{41}{}_{\hat{1}}$, or the $u$ field. 
It is analogous to the centrifugal potential in elementary mechanics.
The potential for the $u$ field contains also a mass term originating from 
the radial quantisation, 
present already in the original Hamiltonian 
(\ref{RFABJMHamiltonianRadialQuantised}).
Hence the total potential for the $u$ field is
\begin{equation}
\int \dr\theta \dr\varphi \, \sin\th \left( \frac{1}{8}(f+u)^2 +
\frac{1}{32 \pi^2} \frac{J^2}{(f+u)^2} \right).
\end{equation}
We fix the value of $f$ by requiring that this potential 
be minimised at $u=0$. It follows that
\begin{equation}
f=\sqrt{\frac{J}{2\pi}}=\sqrt{\frac{kq}{\pi}},
\end{equation}
which gives the vacuum expectation value of
the real part of $\phi^{41}{}_{\hat{1}}$ in the unitary gauge,
or, equivalently, the vacuum expectation value of $|\phi^{41}{}_{\hat{1}}|$.
The value of the potential at the minimum is $J/2$. 
This gives the energy of the perturbative vacuum in the gauge we are using,
\begin{equation}
\Delta = \frac{J}{2},
\end{equation}
reproducing the formula (\ref{RFCFTEnergyGroundState})
for the energy of the BPS ground state.
Substituting $\phi^{4 1}{}_{\hat{1}}=(f+u)/\sqrt{2}$ into the original 
Hamiltonian, one obtains various mass terms and interaction vertices 
containing factors of $f$. This introduces an explicit $J$ dependence in 
the Hamiltonian which will play an important role in the following.

It turns out that in 
the gauge discussed above the states dual to the membrane fluctuations
are created by the $(1, 1)$ diagonal components of the various physical 
fields~\footnote{
We stress that a state with only 
$(1,1)$ excitations can be gauge-invariant,
in the presence of a monopole operator, so long as it satisfies the 
Gauss law constraints. This is different from the situation
in more familiar theories such as $\calN=4$ Super Yang-Mills, 
where gauge-invariant operators are constructed taking 
the trace of products of fields
and hence cannot be built out of single 
components of matrix fields.
}.
Furthermore, the $(i', j')$ components are decoupled from the $(1,1)$ 
components
at least in the first few orders in the approximation that should be valid 
in the large $J$ regime, which will be discussed in 
section~\ref{RSCFTPerturbation}.  
Hence we should focus on the $(1,1)$ diagonal and $(1, i')$ off-diagonal 
components of the physical fields.
The gauge-fixed Hamiltonian quadratic in these components 
is derived substituting the expressions  for the fields in 
(\ref{RFListVarSolved}) obtained solving the Gauss law constraints
into the original Hamiltonian (\ref{RFABJMHamiltonianRadialQuantised}).
The result is
\begin{align}
&~H =   \frac{J}{2}+
\int  \dr\theta \dr\varphi \ \biggl[ 
\frac{1}{\sin \theta} \pi_{A'}{}^{\hat{1}}{}_{1} \pi^{A'}{}^{1}{}_{\hat{1}} 
+\sin\th g^{\alpha \beta}
\partial_{\a} \phi^{A'}{}^{1}{}_{\hat{1}} \partial_{\b} \phi_{A'}{}^{\hat{1}}{}_{1}
+
\frac{1}{4} \sin\th 
    \phi^{A'}{}^{1}{}_{\hat{1}} \phi_{A'}{}^{\hat{1}}{}_{1}
\notag \\
&~
+\frac{1}{\sin \theta}    \left( 
    \pi^{A'}{}^{1}{}_{\hi'} \pi_{A'}{}^{\hi'}{}_{1} 
    +\pi^{A'}{}^{i'}{}_{\hat{1}} \pi_{A'}{}^{\hat{1}}{}_{i'}    \right) 
+\sin \theta g^{\a \b} \left(
    \partial'_{\a} \phi^{A'}{}^{1}{}_{\hi'} \partial'_{\b} \phi_{A'}{}^{\hi'}{}_{1} 
+\partial'_{\a} \phi^{A'}{}^{i'}{}_{\hat{1}} \partial'_{\b} \phi_{A'}{}^{\hat{1}}{}_{i'} 
\right) 
\notag \\
&~
+ \left( q^2 + \frac{1}{4} \right) \sin \theta \left( \phi_{A'}{}^{\hat{1}}{}_{i'} 
\phi^{A'}{}^{i'}{}_{\hat{1}} +
\phi^{A'}{}^{1}{}_{\hi'} \phi_{A'}{}^{\hi'}{}_{1}  \right) 
\notag \\
&~
+\frac{1}{2 \sin \th} p^{2}_{u} + \frac{1}{2} \sin \th  u^2
+\frac{1}{2} \sin \th g^{\alpha \beta} \left(
    \partial_{\alpha} u \partial_{\beta} u 
    +f^2 z^{-}_{\alpha} z^{-}_{\beta} \right)
-\frac{k}{2 \pi f} u \partial_{\underline{1}} z^{+}_{\underline{2}} 
+\frac{k^2}{8 \pi^2 f^2 \sin \theta } (\partial_{\underline{1}} 
z^{+}_{\underline{2}})^2
\notag\\
&~
+\frac{2}{f^2} \left( \frac{k}{2 \pi} \right)^{2} \frac{1}{\sin \theta} \left(
    |\partial'_{\underline{1}} w_{\underline{2} i'} |^{2} 
    +|\partial'_{\underline{1}} \hat{w}_{\underline{2} \hi'}|^{2} \right)
+ \frac{f^2}{2} \sin \theta g^{\alpha \beta} \left( w_{\alpha i'} 
w_{\beta i'}{}^{\ast} 
      + \hat{w}_{\alpha \hi'}  \hat{w}_{\beta \hi'}{}^{\ast} \right) \notag \\
&~
+\sin \theta \psi^{T}{}_{A}{}^{1}{}_{\hat{1}} (\s^r \s^{\a} \partial_{\a} - 1) 
\psi^{A}{}^{\hat{1}}{}_{1} 
\notag\\
&~
+ \sin \theta \biggl\{ \psi^{T}{}_{A}{}^{i'}{}_{\hat{1}} \left( 
\sigma^r \sigma^{\a} \partial'_{\a} -1  \right) \psi^{A}{}^{\hat{1}}{}_{i'} + 
     \psi^{T}{}_{A}{}^{1}{}_{\hi'} \left( \sigma^r \sigma^{\a} \partial'_{\a} -1  
     \right) 
     \psi^{A}{}^{\hi'}{}_{1} \biggl\} \notag \\
&~
- q \sin \theta \left( \psi^{T}{}_{4}{}^{1}{}_{\hi'} \sigma^r \psi^{4}{}^{\hi'}{}_{1} 
-\psi^{T}{}_{4}{}^{i'}{}_{\hat{1}} \sigma^r \psi^{4}{}^{\hat{1}}{}_{i'}  \right) 
+ q \sin \theta \left( \psi^{T}{}_{A'}{}^{1}{}_{\hi'} \sigma^r 
\psi^{A'}{}^{\hi'}{}_{1} -   
     \psi^{T}{}_{A'}{}^{i'}{}_{\hat{1}} \sigma^r \psi^{A'}{}^{\hat{1}}{}_{i'}
     \right)  
\notag\\
&~
- \frac{1}{4} \left( \rho_{W}{}^{1}{}_{1} - \hat{\rho}_{W}{}^{\hat{1}}{}_{\hat{1}}  
\right) 
\biggl] .
\label{RFGaugeFixedHamiltonianCFT}
\end{align}
To obtain this Hamiltonian no iteration 
is actually necessary and
it is sufficient to drop higher order terms in the right hand side of 
(\ref{RFSolvedpi1})-(\ref{RFSolvedrotahat}).
One important step involved in deriving the above expression is
a partial integration, 
\begin{equation}
\int \dr\th \dr\v \, \partial_{\underline{1}} z_{\underline{2}}^+ = 0 \, ,
\end{equation}
which is possible because the fluctuation field $z$ does not 
contain a part proportional to the gauge field
for the Dirac monopole, as discussed below
(\ref{RFBKGGaugeField}). 
Strictly speaking, the term containing $(z^-_\a)^2$ in the fourth line
of (\ref{RFGaugeFixedHamiltonianCFT}) should be understood 
as signifying only the contribution from the 
divergence part of $z^-$. The rotation part of $z^-$  
should be rewritten using (\ref{RFSolvedrotz}) and it
produces only cubic or higher interaction terms.

From this Hamiltonian we have calculated the spectrum.
From the structure of the covariant derivative (\ref{RFCovariantDer}) and
the background field  (\ref{RFBKGGaugeField}), it follows that 
the off-diagonal elements feel the background magnetic flux,
whereas the diagonal elements do not feel the magnetic field,
as exemplified in (\ref{RFPartialPrimeEx1})-(\ref{RFPartialPrimeEx4}).
It follows that the off-diagonal modes have to be expanded in terms of
monopole spherical harmonics, $Y_{qlm}$~\cite{RBWuYangHarmonics} 
(and their generalisations to fermions and vectors),
whereas the diagonal modes should be expanded
in standard spherical harmonics, $Y_{lm}$ (and their generalisations). 
An important point is that,
for the monopole spherical harmonics, the quantum number $l$ starts from
$q=J/(2k)$ (for spin $1/2$ and $1$ fields there are order $1$ shifts),
whereas for the standard spherical harmonics $l$ of course starts from 
$0$ (again with order $1$ shifts for fields with spin).
This effect (the order $q$ shift of the lowest value of $l$ due to 
magnetic flux) combines with the structure of the mass terms 
in~(\ref{RFGaugeFixedHamiltonianCFT}),
which are $J$ dependent due to the Higgs effect. As a result, we find that 
the off-diagonal modes have large frequencies of order $J$,  whereas the 
diagonal modes have small frequencies of order $1$.
We will call them fast (or high-energy) modes and slow (or low-energy)
modes respectively. 

This large separation between the two energy scales naturally leads to
the idea that an approximation of the Born-Oppenheimer type
should be applicable to our system. Namely, the fast modes should be 
integrated out and the effective theory thus obtained will have interactions 
which are suppressed by a power of the ratio between the two energy 
scales, $1/J$. We will discuss in more detail 
how this procedure should be implemented in our formulation in 
section~\ref{RSCFTPerturbation}.
The slow modes ($(1, 1)$ components)
represent physical states and their spectrum should be compared 
to the spectrum of fluctuations on the AdS side studied in 
section~\ref{RSAdSFluctuation}. 

In this paper, we will only sketch the computation 
of the spectrum. We hope to present the details elsewhere.
The results are summarised in 
tables \ref{RTCFTSlowModeSpectrum} and 
\ref{RTCFTFastModeSpectrum}.
These spectra should be considered as  
the leading order result in the Born-Oppenheimer approximation.

For the mass spectrum of the slow modes of
$\phi^{A'}$,
the first and ninth lines in the Hamiltonian 
(\ref{RFGaugeFixedHamiltonianCFT}) contribute.
By expanding $\phi^{A'1}{}_{\hat{1}}$ in spherical harmonics,
one finds that the contribution of the first line to 
the conformal dimension is
\begin{equation}
\sqrt{\left(\frac{1}{2}\right)^2 + l(l+1)} =\frac{1}{2} + l \, .
\label{RFExcitationEnergyWORhoW}
\end{equation}
The first term under the square root 
comes from the mass term, arising from the radial quantisation. 
The second term comes from the kinetic term associated with the  
Laplacian on $S^2$~\footnote{
Expressions such 
as~(\ref{RFExcitationEnergyWORhoW}), which 
can be schematically written as 
$\sqrt{\mbox{const.} + \mbox{(mode no.)}^2}$,
are reminiscent of the spectrum of BMN operators~\cite{RBBMN}.
However, there are important differences.
In our case this formula arises as the leading order term
in the Born-Oppenheimer approximation
and it does not involve the membrane tension.
In the BMN case, the corresponding formula is 
obtained resumming the expansion in the 't Hooft coupling 
and the ratio of the two terms 
under the square root involves the string tension.
}.
The contribution of the ninth line shifts the eigenfrequency by $\pm 1/2$.

The fast mode scalar spectrum is computed using an expansion in 
terms of monopole spherical harmonics. There is a mass term produced 
by the Higgs mechanism originating
from the sextic potential proportional to $f^4$.
The spectrum is integer (or half-integer) valued
which is a consequence of the particular value of the mass term produced
by the Higgs mechanism.

The slow and fast mode fermion spectra can be computed using
the Clebsch-Gordan method.   
For our Hamiltonian obtained in part by the Higgs mechanism
(containing a mass term for the fermions proportional to $f^2$ 
coming from the original Yukawa term),
the Clebsch-Gordan wave functions automatically give eigenmodes
with rational eigenvalues. 
We note that in general for spin $1/2$ fields in a monopole background,
a diagonalisation of $2\times2$ matrices is necessary after the 
Clebsch-Gordan procedure~\cite{RBKazamaYangGoldhaber}.

The slow mode vector fields mix with the scalar $u$.
The computation of the spectrum is done by taking care of this mixing.
For the one-form fields $z^-$ and $z^+$ we used an 
expansion in terms of $\dr Y_{lm}$ and $* \dr Y_{lm}$ respectively,
since the associated physical fields are respectively rotationless and 
divergenceless.

The fast mode vector fields should be solved by expanding  fields
in a basis constructed 
from linear combinations of $\dr'Y_{qlm} $ and 
$*\dr'Y_{qlm}$, where $\dr'$ refers to the 
gauge covariant version of the exterior derivative associated with
the background gauge field.
Special care should be taken for
the low-lying modes with $l=q-1$, for which a
special basis (not written in terms of $\dr'Y $ and $*\dr'Y$) is necessary.
The basis we use is analogous to the monopole vector spherical 
harmonics of \cite{RBWeinberg}.

Since the Hamiltonian on the CFT side corresponds to $\Delta$ whereas 
the Hamiltonian on the AdS side corresponds to $\Delta -J_4/2$, 
it is convenient to compute $\Delta -J_4/2$ to compare the two sides. 
The calculation of the value
of $J_4$ for the various states is non-trivial for reasons associated with
our choice of gauge which we discuss below.
The charge $J_4$ in the Hamiltonian formalism before gauge fixing
is given by
\begin{equation}
J_4 =   \int \dr \theta \dr \varphi\:\Tr \! \left( i\phi^4 \pi_4 - i \pi^4 \phi_4 
- \frac{1}{2} \sin \theta \psi^T{}_{4} \psi^4 
+ \frac{1}{2} \sin \theta \psi^T{}_{A^{\prime}} \psi^{A^{\prime}} 
 \right).  
 \label{RFCFTJ4UngaugeFixed}
\end{equation}
The charge in the gauge-fixed theory, 
obtained by substituting the solved variables, is
\begin{equation}
J_4 = J  +  \int \dr \theta \dr \varphi  
\left( 
           - \frac{1}{2} \rho_W{}^{1}{}_{1} + \frac{1}{2} 
           \hat{\rho}_W{}^{\hat{1}}{}_{\hat{1}}
           \right) 
            + \int \dr \theta \dr \varphi \:
            \Tr\!\left( - \frac{1}{2} \sin \theta \psi^{T}{}_{4} \psi^4 
           + \frac{1}{2} \sin \theta \psi^T{}_{A^{\prime}} \psi^{A^{\prime}}  \right).
           \label{RFCFTJ4GaugeFixed}
\end{equation}
The $J_4$ charges for various slow modes, both before and after  
gauge fixing, can be read off from these expressions 
and are summarised in table \ref{RTCFTJ4}.
\begin{table}[htb] 
\begin{center}
\begin{tabular}[t]{|l||c|c|}
\hline
Field \raisebox{-11pt}{\rule{0pt}{29pt}} &
$J_4$ before gauge fixing &  $J_4$ after gauge fixing
\\
\hline \hline
$\phi^{4}{}^1{}_{\hat{1}}$ \raisebox{-8pt}{\rule{0pt}{22pt}}
& $1$ & $0$
\\
\hline
$\phi^{A'}{}^1{}_{\hat{1}}$ \raisebox{-8pt}{\rule{0pt}{22pt}}
& $0$ & $-1$
\\
\hline
$\psi^{4}{}^{\hat{1}}{}_{1}$ \raisebox{-11pt}{\rule{0pt}{29pt}}
& $\displaystyle \frac{1}{2}$ & $\displaystyle \frac{3}{2}$
\\
\hline
$\psi^{A'}{}^{\hat{1}}{}_{1}$ \raisebox{-11pt}{\rule{0pt}{29pt}}
& $\displaystyle -\frac{1}{2}$ & $\displaystyle \frac{1}{2}$
\\
\hline
\end{tabular}
\caption{$J_4$ charges of matter fields before and after gauge fixing.
The difference is due to a compensating gauge transformation.
The charges of the complex conjugate fields have opposite signs.
}
\label{RTCFTJ4}
\end{center}
\end{table}
The difference between the charges before and after gauge fixing 
may be understood from the following consideration.
The original symmetry transformation associated with the
$J_4$ charge (before gauge fixing) does not preserve the gauge fixing 
condition~(\ref{RFGaugeFixUnitary1}).
This implies the necessity of a compensating gauge transformation,
resulting in a shift of the $J_4$ charges in the gauge-fixed theory.
In the comparison with the AdS side the $J_4$ charge after 
gauge-fixing should be used.

In table \ref{RTCFTSlowModeSpectrum} we have also shown 
$\Delta - J_{4}/2$ for oscillators corresponding to various $(1,1)$ fields.
The results are in complete agreement with the spectrum of fluctuations 
of the spherical membranes on the AdS side, summarised in table
\ref{RTAdSNBPSTreeSpectrum}. We recall that the 
Hamiltonian on the AdS 
side corresponds to $-P_0-P_1=2(\Delta - J_4/2)/R$ because of  
(\ref{RFRelationP1J}) and (\ref{RFRelationP0Delta}).
The agreement verifies the AdS/CFT correspondence in the leading order
in a truly M-theoretic regime for non-BPS observables, which had not 
been studied before. 
The agreement also suggests the existence of an approximation scheme 
on the CFT side corresponding to the perturbative 
expansion on the AdS side discussed in section \ref{RSAdSPerturbation}. 

\begin{table}[htb] 
\begin{center}
\begin{tabular}[t]{|lc||l|c|c|c|}
\hline
Field \raisebox{-11pt}{\rule{0pt}{29pt}} & & Label & 
$\Delta$ & $\displaystyle \Delta-\frac{J_4}{2}$ & Multiplicity 
\\
\hline \hline
\raisebox{-14pt}{Scalars} & 
\hsp{-1.5}$\phi^{A'}{}^1{}_{\hat{1}}$ & $l =0, 1, \ldots$ \rule{0pt}{17pt} &
$l$ & $\displaystyle \frac{1}{2}+l$ & $ 3 \!\times\! (2l+1)$  
\\
\raisebox{-11pt}{\rule{0pt}{0pt}} & \hsp{-1.5}$\phi_{A'}{}^{\hat{1}}{}_{1}$ & 
$l =0, 1, \ldots$ & $1+l$ &
$\displaystyle \frac{1}{2}+l$ & $ 3 \!\times\! (2l+1)$   
\\
\hline 
Vectors  (rot $z^+$, div $z^-$) &  &
$l=0, 1, \ldots$ \rule{0pt}{14pt} & $1+l$ & $1+l$ & $ (2l+1)$  
\\
~~~~~~~~~~~~ and $u$ & & $l=1,2,\ldots 
\raisebox{-7pt}{\rule{0pt}{18pt}}$ & $l$ & $l$ & $(2l+1)$   
\\
\hline
& \hsp{-1.5}$\psi^{A'}{}^{\hat{1}}{}_{1}$ & 
$\displaystyle j=\frac{1}{2}, \frac{3}{2}, \ldots$ \rule{0pt}{20pt} & 
$1+j$ & $\displaystyle \frac{3}{4}+j$ & $3\!\times\! (2j+1)$  
\\  
\raisebox{-14pt}{Fermions} & \hsp{-1.5}$\psi_{A'}{}^1{}_{\hat{1}}$ & 
$\displaystyle j=\frac{1}{2},\frac{3}{2}, \ldots$ \rule{0pt}{20pt} 
& $j$ & $\displaystyle \frac{1}{4}+j$ & $3\!\times\!  (2j+1)$  
\\   
& \hsp{-1.5}$\psi^4{}^{\hat{1}}{}_{1}$ & 
$\displaystyle j=\frac{1}{2}, \frac{3}{2}, \ldots$ & 
$1+j$ & $\displaystyle \frac{1}{4}+j$ & $1\!\times\! (2j+1)$  
\\  
& \hsp{-1.5}$\psi_4{}^1{}_{\hat{1}}$ &
$\displaystyle j=\frac{1}{2},\frac{3}{2}, \ldots$ 
\raisebox{-11pt}{\rule{0pt}{31pt}} &
$j$ & $\displaystyle \frac{3}{4}+j$ & $1\!\times\!  (2j+1)$  \\   
\hline
\end{tabular}
\caption{Mass spectrum for slow modes}
\label{RTCFTSlowModeSpectrum}
\end{center}
\end{table}

\begin{table}[htb] 
\begin{center}
\begin{tabular}[t]{|lc||l|c|c|}
\hline
Field \raisebox{-11pt}{\rule{0pt}{29pt}} & & Label & 
$\Delta$ & Multiplicity   
\\
\hline \hline
\raisebox{-14pt}{Scalars} & 
\raisebox{-14pt}{\!\!\!\!\!\!($\phi^{A'}{}^1{}_{\hi'}$, 
$\phi_{A'}{}^{\hat{1}}{}_{i'}$)} &
$l =q, q+1, \ldots$ \rule{0pt}{17pt} &
$\displaystyle \frac{1}{4}+l$ &
$(N-1)\!\times \!2\! \times\! 3\! \times\! (2l+1)$   
\\
\raisebox{-11pt}{\rule{0pt}{0pt}} &  &
$l =q, q+1, \ldots$&
$\displaystyle \frac{3}{4}+l$ &
$(N-1)\!\times\! 2\! \times\! 3\! \times\! (2l+1)$   
\\
\hline 
\raisebox{-12pt}{\rule{0pt}{29pt}} & & $l=q-1$ & 
$\displaystyle -\frac{1}{4}+q$ & 
$(N-1)\!\times\! 2\!\times\!(2q -1)$  
\\ 
Vectors &\!\!\!\!\!\!\!\!\!\!($w_{\a i}$, ${\hat w}_{\a\hi}$) & 
$l=q,q+1,\ldots$ & $\displaystyle \frac{3}{4}+l$ & 
$(N-1)\!\times\! 2\!\times\!(2l+1)$  
\\
\raisebox{-11pt}{\rule{0pt}{29pt}} & & 
$l=q+1,q+2,\ldots$ & $\displaystyle \frac{1}{4}+l$ & 
$(N-1)\!\times\! 2\!\times\!(2l+1)$  
\\
\hline
\raisebox{-14pt}{Fermions} & 
\raisebox{-14pt}{\!\!\!\!($\psi^{A\,\hat 1}{}_{i'}$, 
$\psi_{A}{}^1{}_{\hi'}$)} & 
$\displaystyle j=q-\frac{1}{2}, q + \frac{1}{2}, \ldots$ \rule{0pt}{17pt} &
$\displaystyle \frac{3}{4}+j$ & 
$(N-1)\!\times\! 2\! \times\!4\!\times\! (2j+1)$  
\\  
\raisebox{-11pt}{\rule{0pt}{0pt}} &  & 
$\displaystyle j=q+\frac{1}{2}, q + \frac{3}{2}, \ldots$ & 
$\displaystyle \frac{1}{4}+j$ &
$(N-1)\!\times\! 2\! \times\!4\!\times\!  (2j+1)$  \\   
\hline
\end{tabular}
\caption{Mass spectrum for fast modes 
}
\label{RTCFTFastModeSpectrum}
\end{center}
\end{table}

The spectrum of the slow modes has a simple interpretation
in terms of the free field theory picture in section \ref{RSCFTBPS}.
As an example we consider a state in which one of the oscillators, with 
mode number $l$, associated with the field $\phi^{A' 1}{}_{\hat{1}}$ is 
excited. Having fixed the background (\ref{RFGaugeFixBKG1}), 
(\ref{RFGaugeFixBKG2}), and (\ref{RFBKGGaugeField}) means 
that we are considering states with fixed $J_M=J$.
Hence in the picture of section \ref{RSCFTBPS} the 
state under consideration corresponds to a state in which the zero-mode of 
$\phi^{4 1 }{}_{\hat{1}}$ is excited $J-1$ times, and the oscillator 
$\phi^{A' 1}{}_{\hat{1}}$ is excited once.
The excitation energy for the latter oscillator in the free field theory picture 
is given by (\ref{RFExcitationEnergyWORhoW}), which corresponds to 
the bare dimension of the $\phi^{A'}$ field with $l$ derivatives acting on it.
We note that in radial quantisation of a free scalar field 
theory, operators such as $\der^l \phi$ 
are mapped to states in which the oscillator 
with angular momentum quantum number $l$
is excited once. 
The energy of this state in the free field theory picture is 
\begin{equation}
\half\times \left(J-1\right)+\left(\half+l\right)=\frac{J}{2}+l.
\end{equation}
By comparing this with the energy of the ground state, $J/2$, we 
see that the excitation energy in this gauge should be $l$,
in agreement with table \ref{RTCFTSlowModeSpectrum}.
This gives a simple interpretation of the 
rationality of the energy spectrum (at tree level) on the AdS side 
in table \ref{RTAdSNBPSTreeSpectrum}, 
which might seem accidental from the point of view of the matrix 
model~\footnote{
In our construction  we use 
operators of the form $\der^l \phi$,
in the sense explained in this paragraph, 
to describe the fluctuations around the 
ground state. This is in marked contrast with 
the BMN sector of the $\calN=4$ supersymmetric Yang-Mills 
theory~\cite{RBBMN}, where the operators involve only insertions 
satisfying $\Delta-J =1$, such as scalar fields without any derivatives.
}.

The gauge fixing conditions we use leave a residual gauge 
freedom corresponding to certain gauge transformations with constant 
parameters on the $S^2$ time slice. 
This translates into the fact that the zero-mode
part of the Gauss law constraint (\ref{RFSolvedrotz}) is not solved.
If one integrates both sides of (\ref{RFSolvedrotz}) over $\th, \v$ the 
left hand side vanishes automatically by partial integration (since $z^-$ 
does not have singularities associated with the Dirac monopole) 
and we obtain the constraint corresponding to the residual gauge symmetry,
\begin{equation}
0= \int \dr\th \dr\v\,  \left( \rho_W{}^1{}_{1}
+\hrho_W{}^{\hat{1}}{}_{\hat{1}} \right). 
\label{RFIntegratedGausslawConstraint11} 
\end{equation}
This condition should be imposed on the states in the gauge-fixed theory.
Similarly, from (\ref{RFSolvedrota}) and (\ref{RFSolvedrotahat}),
we obtain the constraints
\begin{equation}
0 =
\int \dr\th \dr\v \,
\left( 
i\phi^4{}^{i^{\prime}}{}_{\hat{k}^{\prime}} 
\pi_4{}^{\hat{k}^{\prime}}{}_{j^{\prime}} 
-i \pi^{{}}{}^{4}{}^{i^{\prime}}{}_{\hat{k}^{\prime}} 
\phi^{{}}{}_{4}{}^{\hat{k}^{\prime}}{}_{j^{\prime}} + 
\rho_W{}^{i^{\prime}}{}_{j^{\prime}}   \right),
\label{RFIntegratedGausslawConstraintDoublePrime}
\end{equation}
\begin{equation}
0 =
\int \dr\th \dr\v
\left( -i \phi^{{}}{}_{4}{}^{\hat{i}^{\prime}}{}_{k^{\prime}} 
\pi^{{}}{}^{4}{}^{k^{\prime}}{}_{\hat{j}^{\prime}}
+ i \pi_4{}^{\hat{i}^{\prime}}{}_{k^{\prime}} 
\phi^4{}^{k^{\prime}}{}_{\hat{j}^{\prime}} - 
\hat{\rho}_W{}^{\hat{i}^{\prime}}{}_{\hat{j}^{\prime}}  \right) .
\label{RFIntegratedGausslawConstraintDoublePrimeHat}
\end{equation}
These 
constraints do not affect the $(1,1)$ slow modes,
so the comparison to the AdS side is also not affected.
However, the 
constraints~(\ref{RFIntegratedGausslawConstraintDoublePrime}) 
and~(\ref{RFIntegratedGausslawConstraintDoublePrimeHat}) 
impose restrictions on the $(1,i')$ and $(i',j')$ excitations.
We will elaborate further on this point in section \ref{RSCFTPerturbation}.

We will now briefly discuss some aspects of $\calN=6$
supersymmetry in the sector we are considering.
We have fixed the form of the supersymmetry generators in the Hamiltonian
formulation of the radially quantised ABJM theory (before 
gauge fixing)
by requiring that they satisfy the correct superalgebra with the 
Hamiltonian (\ref{RFABJMHamiltonianRadialQuantised})
(at the classical level). The supercharges read
\begin{equation}
Q_{AB}{}_{a} = X_{AB}{}_{a} + \frac{1}{2}\epsilon_{ABCD} 
\left( Y_{CD}{}^{\ast} B  \right)_{a},
\end{equation}
where $B$ is the charge conjugation operator and
\begin{align}
X_{AB}{}_{a} 
& = \int \dr \theta \dr \varphi \; \Tr \biggl[
\psi^{T}{}_{\ul{A}a}{\pi_{\ul{B}}} 
-i \left( \psi^{ T}{}_{\ul{A}}\sigma_r \sigma^{\a} \right)_{a} D_{\a} 
\phi_{\ul{B}} \sin\th 
+\frac{i}{2} \psi^{T}{}_{\ul{A}a} \phi_{\ul{B}} \sin \theta \notag 
\\
&
\ \ \ \ \ \ \ \ \ \ \ \ \ \ \ \ \ \ \ 
+ i \frac{2 \pi}{k} \left( \psi^{T}{}_{\ul{A}} \s_r \right)_{a} \phi_{C} 
\phi^{C} \phi_{\ul{B}} \sin \theta 
-i \frac{2 \pi}{k} \left( \psi^{T}{}_{\ul{A}} \s_r \right)_{a} \phi_{\ul{B}} 
\phi^{C} \phi_{C} \sin \th 
\notag \\
& \ \ \ \ \ \ \ \ \ \ \ \ \ \ \ \ \ \ \ 
-i \frac{4 \pi}{k} \left( \psi^{T}{}_{C} \s_r \right)_{a} \phi_{\ul{A}} \phi^{C} 
\phi_{\ul{B}} \sin \theta 
\biggl],
\end{align}
\begin{align}
Y_{AB}{}_a & = \int \dr \th \dr \varphi \; \Tr \biggl[
i \left(  \psi^{T}{}_{\ul{A}} \s_r  \right)_a \pi_{\ul{B}} 
- \left( \psi^{T}{}_{\ul{A}} \s^{\a} \right)_a D_{\a} \phi_{\ul{B}} \sin \th 
+\frac{1}{2}  \left( \psi^{T}{}_{\ul{A}} \s_r  \right)_a \phi_{\ul{B}} \sin \th 
\notag \\
&
\ \ \ \ \ \ \ \ \ \ \ \ \ \ \ \ \ \ 
-\frac{2 \pi}{k} \psi^{T}{}_{\ul{A}a} \phi_{C} \phi^{C} \phi_{\ul{B}} \sin \th 
+ \frac{2 \pi}{k} \psi^{T}{}_{\ul{A}a}\phi_{\ul{B}} \phi^{C}\phi_{C} \sin \th 
\notag \\
&  
\ \ \ \ \ \ \ \ \ \ \ \ \ \ \ \ \ \ 
+ \frac{4 \pi}{k} \psi^{T}{}_{C a} \phi_{\ul{A}} \phi^{C}\phi_{\ul{B}} \sin \th 
\biggl]. 
\end{align}
The superalgebra is 
\begin{align}
[ Q_{AB}{}_{b}{}^{\ast T}, Q_{CD}{}_{a} ]_{+} 
=& \left( \delta^{A}{}_{C}\delta^{B}{}_{D} - \delta^{A}{}_{D} \delta^{B}{}_{C} 
\right) 
\left( H \delta^{b}{}_{a} - L_i \sigma^{i}{}^{b}{}_{a}  \right) 
\notag \\
& -\left( M^{A}{}_{C} \delta^{B}{}_{D} - M^{A}{}_{D} \delta^{B}{}_{C} - 
M^{B}{}_{C} \delta^{A}{}_{D} + M^{B}{}_{D} \delta^{A}{}_{C} \right) 
\delta^b{}_{a},
\label{RFCFTSuperalgebra}
 \end{align}
where the Hamiltonian, $H$, given in 
(\ref{RFABJMHamiltonianRadialQuantised}) 
can be identified with the dilation operator.
The flavour SU($4$) symmetry generators, 
$M^{A}{}_{B}$, are 
\begin{equation}
M^{A}{}_{B} = \tilde{M}^{A}{}_{B} - \frac{1}{4} \tilde{M}^{C}{}_{C} 
\delta^{A}{}_{B},
\end{equation}
\begin{equation}
\tilde{M}^{A}{}_{B} =
\int \dr\theta \dr\varphi\; \Tr\left(
i \phi^A \pi_B - i \pi^{A} \phi_{B} - \sin \th \psi^{T}{}_{B} \psi^{A}
\right)
\end{equation}
and $L_i$, $i=1,2,3$, are the generators of the SO($3$) rotational symmetry 
acting on the time-slice $S^2$,
\begin{equation}
L_{i} = \int \dr\theta \dr\varphi \; \Tr\biggl[
\pi_A V^{\alpha}{}_{i} D_{\alpha} \phi^A
+ \pi^{ A} V^{\alpha}{}_{i}D_{\alpha} \phi_{A} 
+ \sin \theta \psi^{T}{}_{A} 
\left( i V^{\alpha}{}_{i} D_{\alpha} + \frac{1}{2} \sigma_{i}   \right) \psi^{A}  
\biggl],
\end{equation}
where
\begin{equation}
V^{\alpha}{}_i =
\begin{bmatrix}
V^{\theta}{}_1 & V^{\theta}{}_2 & V^{\theta}{}_3 \\
V^{\varphi}{}_1 & V^{\varphi}{}_2 & V^{\varphi}{}_3
\end{bmatrix}
=
\begin{bmatrix}
\sin \varphi & - \cos \varphi & 0 \\
\cot\th \cos \v & \cot\th \sin\v   & -1
\end{bmatrix}.
\end{equation}

As consistency checks regarding supersymmetry 
and the identification of the vacuum in our gauge
at the quantum level, we have verified the following. 
We have derived the gauge-fixed form of the supercharges
$Q_{AB}$ (by substituting the variables obtained by iteratively solving the
Gauss law constraints) neglecting cubic and higher order terms
in the physical fields.
At this order $Q_{AB}$ involves either terms linear in the fermionic fields or 
quadratic terms containing one fermionic and one bosonic field. 
It turns out that linear terms in the 
fermions are present in $Q_{A'B'}$, but not in $Q_{A'4}$.  
This implies that the vacuum is annihilated by the $Q_{A'4}$
supercharges and thus it is 1/2 BPS. 
We have explicitly verified that the quantum version of the superalgebra 
(\ref{RFCFTSuperalgebra}) is satisfied at the level 
in which one only retains terms quadratic in the fast modes in 
all the generators.
Since the vacuum is annihilated by $Q_{A'4}$, this computation 
ensures that the vacuum energy~(\ref{RFCFTEnergyGroundState}) 
receives no leading order correction, \ie there cannot be a shift from the 
zero-point energy.

\subsection{Perturbation theory}
\label{RSCFTPerturbation}

In this section we discuss the approximation scheme
which we propose to be relevant in the large $J$ sector of the ABJM 
theory. We will present the general features including a diagrammatic 
representation of the approximation for various processes.
We focus on contributions to the energy spectrum and discuss an 
estimate of the dependence on the parameters $N$, $k$ and $J$ for 
some of the leading corrections. We will illustrate a specific 
contribution to the spectrum of scalar modes, which results in the same 
$Nk/J^3$ dependence as  the one-loop correction on the AdS side 
presented in section~\ref{RSAdSPerturbation}, provided that certain 
cancellations, which we expect in view of the large amount of 
supersymmetry in the ABJM theory, take place.
It will be important to explicitly calculate the leading order corrections
following the approach explained below 
and we hope to carry out such calculation in the future.

We focus on the case of a single non-zero GNO charge considered in
the previous subsection. As already explained, 
the large $J$ sector of the ABJM theory involves two types of modes:
the slow modes (diagonal components of the fields), with eigenfrequencies 
of order $1$, and the fast modes (off-diagonal components of the fields), 
with eigenfrequencies of order $J$. 
In general, if there are two types of degrees of freedom 
in a theory with very different energy scales,
one expects that a Born-Oppenheimer type approximation 
-- or low-energy effective description -- should be applicable. 
In the leading order of the Born-Oppenheimer approximation,
one first solves the theory describing the fast modes 
treating the slow modes as fixed parameters. The result is used 
to construct the effective theory for the slow modes.
The coupling of the slow modes in the resulting effective 
theory is suppressed by a power of the ratio of the two energy scales.
The original application of the Born-Oppenheimer approximation was to 
the quantum theory of molecules in which an effective theory for 
the slow motion of the nuclei is obtained after studying the fast motion of 
the electrons in a potential produced by the nuclei with fixed positions.

In the context of the ABJM theory we are interested in, we expect the 
following features to be relevant for the emergence of 
a good approximation scheme for large $J$. 
First, the Abelian part of the action of the ABJM theory is essentially that of 
a free theory, since all couplings among the diagonal fields can be gauged 
away at least classically. Hence direct couplings between the slow modes
associated with the $(1,1)$ components of the fields,
even if they are produced in the iteration process 
described around (\ref{RFSolvedpi1})-(\ref{RFSolvedrotahat}), 
should be unphysical.
Therefore the interaction between slow modes 
should always involve the fast modes. 
Second, since the fast modes by definition have large quadratic terms in 
the action, we expect that their interactions can be treated perturbatively.
Third, supersymmetry should play an important role in controlling the 
behaviour of quantum corrections. Even with the energy gap of order $J$, 
the potentially large zero-point energy could lead to large  
interactions between the slow modes through the fast modes.
However, we expect the leading order contributions to cancel out 
for near-BPS states thanks to supersymmetry.
The remaining terms should be suppressed by a power of $1/J$, which
in the present case is the ratio between low and high energy scales.

These features are analogous to those encountered in the computation of  
scattering amplitudes for D-branes with a small 
relative velocity and a large impact parameter~\cite{RBBachas, 
RBDanielssonFerrettiSundborg, RBKabatPouliot, RBLifschytz, 
RBDKPS}. In the case of this system the potential vanishes for mutually
commuting, diagonal, matrix coordinates of the D-branes, \ie 
there are no direct couplings between the diagonal components.
Interactions between the diagonal components (the positions of the 
D-branes) are only induced by the off-diagonal components (open strings 
stretched between the D-branes). 
Higher order couplings between off-diagonal modes are not the dominant
contribution to the physics in the scattering of D-branes, 
because of the large mass of the open strings.
Supersymmetry implies that 
the leading order terms in the interaction potential
between the diagonal modes mediated by the off-diagonal modes cancel 
out~\footnote{
Actually, this cancellation was discussed before the advent of D-branes
in the matrix model context from the membrane point of view in 
\cite{RBdeWitLuescherNicolai}. The cancellation implies that 
the matrix model has a continuous energy spectrum.
More precisely, it implies the existence of states
with arbitrarily small energy.
This was incorrectly interpreted as signifying an instability  of membranes. 
The interpretation was revised
in recent years~\cite{RBTaylorReview} 
after the D0-brane picture of \cite{RBBFSS}: 
the existence of states with arbitrarily small energy only means that the 
matrix model is a theory which describes multiple membranes, not a 
single membrane. Equivalently, the matrix model is a second-quantised 
rather than a first-quantised theory of membranes
and as such it naturally has a continuous spectrum.}.
The remainder is the 
small interaction between D-branes suppressed 
by a ratio of powers of the small relative velocity and the large impact 
parameter.

A similar approach based on the existence of very different energy
scales is familiar in the context of quantum field theory.
In this case one performs the path integral only over
the high-energy degrees of freedom (the fast modes) to find an effective 
theory governing the dynamics of the low-energy degrees 
of freedom (the slow modes). This ``integrating out'' procedure 
to compute the effective action
has a simple realisation in terms of Feynman diagrams as explained for 
example in~\cite{RBWilsonKogut}. The vertices in the 
effective action are obtained from Feynman diagrams in which all internal 
lines correspond to fast modes and the external lines only involve slow 
modes. We will discuss the low-energy effective description of the ABJM 
theory in the large $J$ regime, which is constructed using this procedure.

A simple way of constructing the path integral of the 
gauge-fixed ABJM theory discussed in section \ref{RSCFTNearBPS} 
is to use a phase space formulation,
in which the functional integration is 
performed over both the canonical variables and their conjugate momenta.
For instance, for a complex scalar field in Euclidean signature,
the Boltzmann factor is 
\begin{equation}
\exp  \int \dr t  \int \dr^2 x  \left( i \pi \frac{\partial}{\partial t} \phi + i \pi^{\ast} 
\frac{\partial}{\partial t} \phi^{\ast} - \mathcal{H} \right),  
\label{RFBoltzmannFactorPhaseSpaceScalarEuclid}
\end{equation}
%This equation number is used as a pointer later so we should be careful
%in case we move this paragraph.
where $\mathcal{H}$ is the Hamiltonian density.

In our gauge
the slow modes are the $(1,1)$ components of the various fields 
and the fast modes are the $(1,i') $ components.
We will comment on the role of $(i', j')$ components later 
in this subsection.
It is in principle reasonable to classify high momentum modes
of the $(1,1)$ components as fast modes as well, since their
eigenfrequencies are of the same order as those of the $(1,i')$ 
components. From this point of view, we obtain a natural UV cut-off for 
the $(1,1)$ slow modes which is reminiscent of the UV cut-off 
arising on the AdS side as a consequence of the fact that 
we consider matrices of large but finite size, as noted 
below~(\ref{RFS7ScalarSpectrum}). 
The difference  between the two prescriptions, 
\ie whether one treats the high momentum $(1,1)$ 
fields as fast or slow,
may  affect the technical details of the 
calculation, but should not produce any difference in the
final low-energy observables. 

In order to obtain the propagators of the fast modes, 
we expand them in a basis 
constructed from the monopole spherical harmonics,
$Y_{qlm}$. For example for the scalar fast modes we use
\begin{eqnarray}
\phi^{A'}{}_{\hat{i}'} &\!\!=\!\!& \sum_{l=q}^{+\infty} \sum_{m=-l}^{+l}
\int \dr \omega 
\left(\phi^{A'}{}_{\hat{i}'}\right)^{ lm \omega} Y_{qlm} e^{i \omega t}, 
\\
\pi_{A'}{}^{\hat{i}'} &\!\!=\!\!& 
\sum_{l=q}^{+\infty} \sum_{m=-l}^{+l}
\int \dr \omega 
\left(\pi_{A'}{}^{\hat{i}'}\right)_{lm\omega} 
\left( Y_{qlm} e^{i\omega t}  \right)^{\ast}\sin \theta.
\end{eqnarray}
In this subsection we omit the colour index $1$ or $\hat{1}$ from the fast 
modes, for brevity.
The propagators can be computed in a standard manner. For instance
one finds
\begin{align}
\biggl \langle \left( \phi_{A'}{}^{\hat{i}'}\right) {}_{lm \omega} \left( 
\phi^{B'}{}_{\hat{j}'}\right){}^{l'm'\omega' }  \biggl \rangle
&= \frac{1}{2\pi} \frac{1}{\left( \omega - \frac{i}{4} \right)^2 + \left( 
l + \frac{1}{2} \right)^2 } \delta (\omega - \omega')
\delta^{B'}{}_{A'} \delta^{\hat{i}'}{}_{\hat{j}'} \delta^{l'}{}_{l} \delta^{m'}{}_{m}.
\label{RFScalarFastModePropagator}
\end{align}
Since we work in the phase space path integral
formalism, there are also other propagators for the scalar fields,
\ie $\langle \phi \pi \rangle$, $\langle \pi \phi \rangle$ and 
$\langle \pi \pi \rangle$.  
The variables $\omega$, $l$, and $m$ can be considered as
the components of 3-momentum
on the space-time $S^2\times \R$. For each loop, 
one has the integration over $\omega$ and summation over $l$ 
and $m$. 
The index $l$ is summed from $q$ (with order $1$ shifts for 
fields with non-zero spin) to infinity and the index $m$ runs from $-l$ to 
$+l$. 

The vertices can be read off
from the gauge-fixed Hamiltonian.
In order to obtain the interaction terms, such as for instance
the cubic and quartic vertices, it is
necessary to iterate equations (\ref{RFSolvedpi1})-(\ref{RFSolvedrotahat})
further than has been done in section \ref{RSCFTNearBPS} 
for the quadratic part of the Hamiltonian. 
Notice that because of the structure of the colour indices, 
all vertices contain an even number of fast modes.

\begin{figure}
\centering
\includegraphics[width=0.65\textwidth]{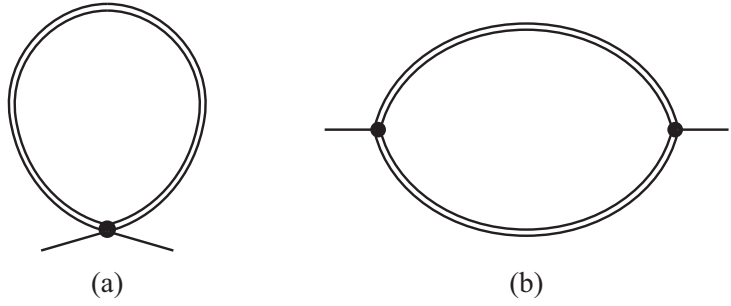}
\caption{One-loop contributions to slow mode quadratic term in the 
effective action. Single lines correspond to $(1,1)$ slow modes and double 
lines to $(1,i')$ fast modes. These processes are expected to 
be subleading because of cancellations due to supersymmetry.}
\label{RPSlow2pt1Loop}
\end{figure}

A possible correction to the
energies of the excited 
states considered in section \ref{RSCFTNearBPS} 
comes from the processes associated with the Feynman diagrams 
depicted in figure \ref{RPSlow2pt1Loop}~\footnote{
We focus on ``one-particle states'',  \ie states in which only one oscillator 
associated with a $(1,1)$ slow mode field
is excited.
For ``multi-particle states''
diagrams with more external lines
should also be considered.
}.
In all the diagrams in this section we
represent the $(1,1)$ slow modes with single lines and
the $(1, i')$ fast modes with double lines. 
These processes produce a
direct radiative correction to 
the slow-slow term in the low-energy effective action.
Even with the requirement that the interaction vertices in 
figure~\ref{RPSlow2pt1Loop} should be only of slow-fast-fast and 
slow-slow-fast-fast kind, there is a very large number of contributions 
to both types of diagrams.
An example of fast-fast-slow-slow vertex is
\begin{equation}
\int \dr^3 x
\left( \frac{ f}{k}  \right)^2 \sin \theta \,
\phi^{A'}{}_{\hat{i}'} 
\phi_{B'}{}^{\hat{i}'} 
\phi^{B' 1}{}_{\hat{1}}
\phi_{A'}{}^{\hat{1}}{}_{1} \,,
\label{RFSSFFScalarVertex}
\end{equation}
where we omitted purely numerical factors, but we kept the 
$k$ dependence. This term, which is produced by the Higgs mechanism 
from the sextic scalar potential, 
is relevant for the diagram in figure~\ref{RPSlow2pt1Loop} (a) 
with two scalar slow modes as the external lines.
The behaviour of this contribution (at leading 
order) can be computed using (\ref{RFSSFFScalarVertex}) and
(\ref{RFScalarFastModePropagator}). We get
\begin{equation}
\left( \frac{f}{k} \right)^2
\times
(N-1)
\times \sum_{l=q}^{\infty} \sum_{m=-l}^{l} \int \dr\omega \,
\frac{1}{\omega^2+l^2} 
\sim
\frac{N f^2}{k^2} \sum_{l=q}^{\infty} \sum_{m=-l}^{l} \frac{1}{l}.
\label{RFEstimateOmega}
\end{equation}
This expression diverges linearly.
This divergence should be cancelled by other contributions 
to the spectrum at the same order. 
Among the additional 
corrections which can contribute to the cancellation are diagrams of the
type in figure \ref{RPSlow2pt1Loop} (a) with different four-point vertices and 
other (vector and fermion) internal lines. Moreover one has to consider the 
``vacuum polarisation'' diagrams of the type in 
figure~\ref{RPSlow2pt1Loop} (b), again with all possible internal lines. All 
these contributions have the same dependence on the parameters, $N$, 
$k$ and $J$, as~(\ref{RFEstimateOmega}). Finally, although the ABJM 
theory is believed to be UV finite, there may be a residual divergence after 
combining all diagrams, which needs to be absorbed into an unphysical -- 
and generally gauge-dependent -- wave function renormalisation. 

It is important that the result of the loop integrals, 
or more precisely of the sums over $l$ and $m$ and the integral over 
$\omega$, is always organised in an expansion in powers 
of $q^{-1}$ and the parameters $J$ or $k$ never appear explicitly.
Assuming there is an $n_0$-fold cancellation
as a result of combining all the above contributions 
in figure \ref{RPSlow2pt1Loop}
and potential unphysical counter terms
(with $n_0=1$ meaning
cancellation of the leading order contribution, 
$n_0=2$ cancellation of the leading and next-to-leading order 
contributions {\it etc.}) we obtain
\begin{equation}
\frac{N f^2}{k^2} q^{1-n_0} \,.
\label{RFLeading2ptSlow1}
\end{equation}
For $n_0=1$ the sum is generically logarithmically 
divergent and we expect $n_0 \ge 2$.
Rewriting~(\ref{RFLeading2ptSlow1}) in terms of $N$, $k$ and $J$, we 
obtain
\begin{equation}
N k^{n_0-3} J^{2-n_0}.
\label{RFLeading2ptSlow2}
\end{equation}
This expression cannot give rise to the same dependence on $N$, $k$ and 
$J$ found on the AdS side, \ie $Nk/J^3$, for any value of $n_0$.
This leads us to conjecture that either $n_0$ is sufficiently 
large,  $n_0\ge 6$, so that this type of correction is 
negligible compared to the 
expected leading order correction 
of order $Nk/J^3$, or the various contributions completely cancel out.
We note that the estimate (\ref{RFLeading2ptSlow1}) is the leading order
term and there are also higher order terms in the expansion in 
inverse powers of $q$.

\begin{figure}[htb]
\centering
\includegraphics[width=0.65\textwidth]{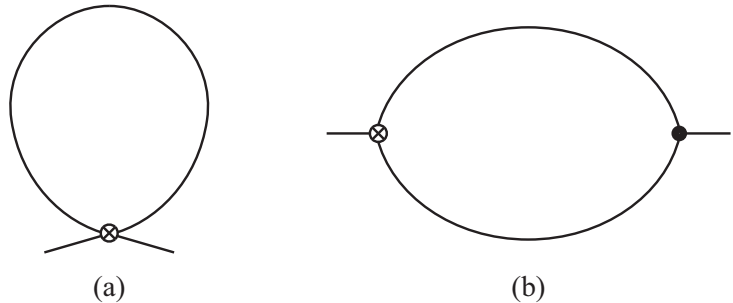}
\caption{Correction to the spectrum at one-loop level in the low-energy 
theory for the $(1,1)$ slow modes. Crossed white dots are effective 
vertices induced by one-loop diagrams in the fast modes.
Black dots represent genuine vertices for the slow modes. These diagrams 
are expected to give the leading correction of order $Nk/J^3$.  
}
\label{RPSlow2ptWithEffVertex}
\end{figure}

We expect the leading order correction to the spectrum to come from 
the processes associated with the Feynman diagrams depicted in  
figure~\ref{RPSlow2ptWithEffVertex}. 
These are one-loop diagrams in the 
low-energy theory for the slow modes involving effective vertices
obtained integrating out fast mode loops. We denote such effective
vertices by crossed white dots. Black dots indicate vertices present in the 
original gauge-fixed Hamiltonian.

\begin{figure}[htb]
\centering
\includegraphics[width=\textwidth]{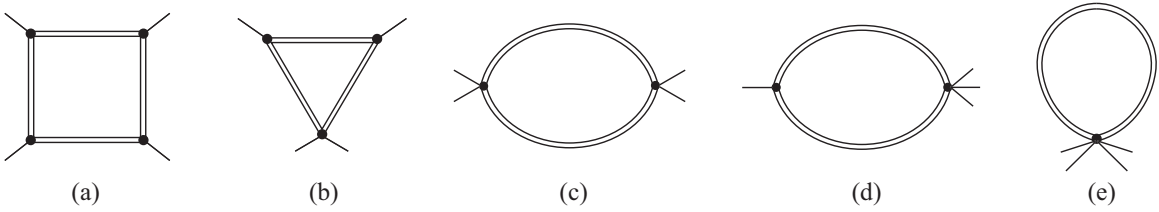}
\caption{One-loop contributions to a quartic effective vertex in the 
low-energy effective action.}
\label{RPSlow4pt1loopFast}
\end{figure}

\begin{figure}[htb]
\centering
\includegraphics[width=0.25\textwidth]{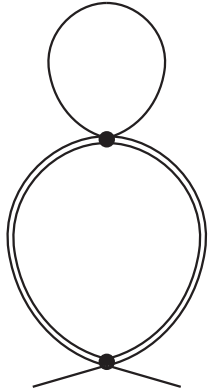}
\caption{A typical leading order correction to the slow mode spectrum} 
\label{RPFishWithBalloon}
\end{figure}

Let us focus for definiteness on corrections to the scalar spectrum. 
In this case the external lines in figure~\ref{RPSlow2ptWithEffVertex} are 
$(1,1)$ components of scalar fields. In diagram (a) the
quartic effective vertex couples the
two scalars to two other slow mode fields which, depending on the type of 
loop, can be two scalars, two vectors or two fermions. The  corresponding 
quartic effective vertices receive contributions from all diagrams in the full 
theory with four external slow-mode lines and internal fast-mode lines. 
Those relevant for the corrections to the spectrum at order $Nk/J^3$ 
involve a single fast-mode loop and are depicted in 
figure~\ref{RPSlow4pt1loopFast}. To determine the vertex relevant for each 
type of slow mode loop in figure~\ref{RPSlow2ptWithEffVertex} (a) one has 
to compute all the contributions to four-point functions from the diagrams in 
figure~\ref{RPSlow4pt1loopFast} where two external lines are slow mode 
scalars and the other two are slow mode scalars, vectors or fermions 
respectively. After performing the loop integrals, one can extract the quartic 
effective vertex for the slow modes. It is straightforward to estimate the 
dependence on $N, k$ and $J$ for 
the diagrams in figure~\ref{RPSlow4pt1loopFast}. 
For instance, for 
a diagram of type (c), in which both vertices are given 
by~(\ref{RFSSFFScalarVertex}) and all internal and external lines 
are scalar fields, the leading order contribution is
\begin{equation}
\left( \frac{f^2}{k^2} \right)^2 
\times (N-1)
\times \sum_{l=q}^{\infty} \sum_{m=-l}^{l} 
\int \dr \omega \left( \frac{1}{ \omega^2 + l^2} \right)^2
\sim \frac{NJ^2}{k^4} \sum_{l=q }^{\infty} (2l+1) \frac{1}{l^{3}} \, .
\label{RFEstimateSlow2ptSlowScalar1Loop1}
\end{equation}
The internal loops in figure~\ref{RPSlow4pt1loopFast} can correspond
to scalars, fermions or vectors. The different contributions can be analysed
in a similar fashion and they all lead to the same dependence on the 
parameters in the scalar quartic effective vertex. 
Assuming again $n_0$-fold cancellations among these diagrams
and possible counter terms, the behaviour
we find is 
\begin{equation}
\frac{NJ^2}{k^4} \sum_{l=q }^{\infty} (2l+1) \frac{1}{l^{3+n_0}} 
\sim \frac{NJ^2}{k^4} \frac{1}{q^{1+n_0}}
\sim \frac{N k^{n_0-3}}{J^{n_0-1}} \, .
\label{RFEstimateSlow2ptSlowScalar1Loop2}
\end{equation}
For $n_0=4$ this expression gives $Nk/J^3$.
This is the same as the weight of the quartic fluctuations about the
fuzzy sphere vacuum relative to the quadratic terms in the matrix 
model Hamiltonian studied in section~\ref{RSAdSPerturbation}. 
One contribution of the type we are describing 
corresponds to the two-loop diagram in the full theory shown in 
figure~\ref{RPFishWithBalloon}.

\begin{figure}[htb]
\centering
\includegraphics[width=0.7\textwidth]{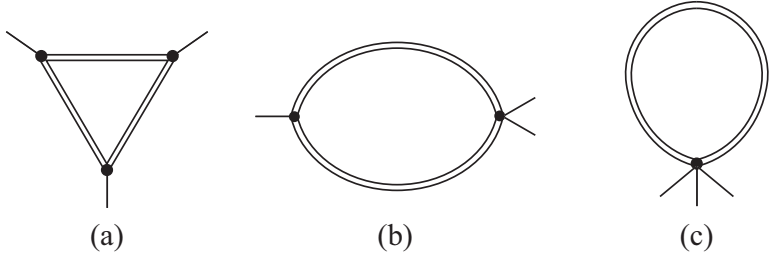}
\caption{One-loop contributions to a cubic effective vertex in the 
low-energy effective action.}
\label{RPSlow3pt1loopFast}
\end{figure}

In the above derivation of the 
estimates~(\ref{RFEstimateSlow2ptSlowScalar1Loop1}) 
and~(\ref{RFEstimateSlow2ptSlowScalar1Loop2}) we considered the case 
of an internal slow mode scalar loop in figure~\ref{RPSlow2ptWithEffVertex} 
(a). Diagrams with an internal fermion or vector loop can also be shown to 
contribute to the two-point function at the same order $Nk/J^3$, again 
assuming appropriate cancellations. 
A method to obtain power counting estimates which can be 
applied to generic diagrams will be outlined 
later in this subsection. 
Another class of leading order corrections to the slow mode spectrum 
is associated with diagrams of the type depicted in 
figure~\ref{RPSlow2ptWithEffVertex} (b). This is a one loop 
diagram in the low-energy theory with one effective vertex and one genuine 
cubic vertex coupling slow modes. The leading contribution to 
the effective cubic vertex is generated by the fast mode one-loop diagrams 
depicted in figure~\ref{RPSlow3pt1loopFast}. 
Again the loop in figure~\ref{RPSlow2ptWithEffVertex} (b) can involve 
scalar, vector or fermion slow mode fields. For each case a suitable cubic 
effective vertex is determined from diagrams of the type in 
figure~\ref{RPSlow3pt1loopFast} with the appropriate external slow mode 
lines.

When combining all the contributions to the two types of diagrams in 
figure~\ref{RPSlow2ptWithEffVertex} to extract the correction to the 
spectrum we expect further cancellations in the slow mode loops, so that
no extra powers of $q$ are produced and the final correction to the 
two-point function is of order $Nk/J^3$. These expected cancellations 
at the level of the slow modes would be analogous to the 
cancellations observed in the pp-wave 
matrix model, which ensure that the sums over intermediate states do not 
produce extra factors of $J/k$.

The corrections to the vector and fermion slow mode spectrum can be 
studied in a similar way. 
We verified by an analogous power-counting that the 
leading non-zero contributions can come from two-point functions of the 
type in figure~\ref{RPSlow2ptWithEffVertex} with vector or fermion external 
lines, assuming again appropriate cancellations.

In the computation of the leading order corrections to the spectrum 
involving the diagrams in figures~\ref{RPSlow2ptWithEffVertex} 
and~\ref{RPSlow4pt1loopFast}, the $(i',j')$ components of the fields are 
indeed unimportant and decouple from the physics of $(1,1)$ modes. 
The integrated Gauss law constraints, 
(\ref{RFIntegratedGausslawConstraint11})-(\ref
{RFIntegratedGausslawConstraintDoublePrimeHat}), imply that
the excitation of $(i', j')$ components 
corresponds to gauge invariant operators constructed from the trace of 
products of matter 
fields and their complex conjugates, \eg $\phi^{A'}$ and $\phi_{A'}$.
The existence of this type of state in the ABJM theory is expected 
from considerations on the gravity side.
A configuration associated with a combination of $(1,1)$ and $(i',j')$ field 
components corresponds to a ``multi-particle'' state in the pp-wave matrix 
model involving a spherical membrane with large $J_4$ and $J_M$ as well 
as gravitons or other particles with vanishing $J_4$ and $J_M$. It is natural 
to expect a suppression in the coupling between the membrane and these 
extra particles because of the large difference in momentum. This supports 
our expectation that the coupling of the $(i',j')$ components to the physical 
$(1,1)$ slow modes should be weak in the large $J$ sector.
It is not straightforward, at this stage, 
to determine the role of the $(i',j')$ fields at higher orders in the 
Born-Oppenheimer approximation. Below we will discuss
a different perspective in which these degrees of freedom can be
understood in a more straightforward fashion.
We also notice that 
as a consequence of the integrated Gauss law constraints single $(1,i')$ 
fast modes cannot be excited and they should always appear in pairs. 
This is related to the fact that all vertices contain an even number of fast 
modes.

The Born-Oppenheimer approach and the description of the physics in 
terms of a low-energy effective action provide a very natural framework in 
which the emergence of a good approximation in the large $J$ sector of 
the ABJM theory is motivated by physical considerations. However, from
a practical point of view it may be technically simpler to compute the 
quantum corrections to the spectrum using the full theory, \ie studying 
contributions to the two-point functions of the $(1,1)$ modes from all 
Feynman diagrams without the restriction that the internal lines be $(1, i')$ 
fields. 
Power counting arguments similar to those presented above can 
be applied in this case as well. The finiteness of the ABJM theory plays 
an essential role in the power counting 
analysis. Since there is no dimensional transmutation in a finite theory, one 
gets a $q$ dependence even for massless propagators because of the 
presence of massive propagators in the diagrams.
Although at first sight there is no reason to expect the
interactions among the $(i',j')$ fields to be suppressed, 
these fields inherit the suppression 
by negative powers of $q$ from the
fast modes.

Computations in the full theory can be 
described in terms of Feynman diagrams using the standard double line 
notation, in which index loops represent sums over the colour index $i$ 
(or $\hi$) from 1 to $N$.
To each index loop in a diagram one has to assign either an index 
taking the value $1$ or a primed index taking the values $2,\ldots, N$. 
This assignment determines which of the internal lines are of $(1,1)$, 
$(1,i')$ or $(i',j')$ type. For each of these internal lines one should use 
the appropriate propagator in the gauge we have fixed.
In a diagram in which $p$  
of the index loops carry a primed index taking  
values $2,\ldots,N$, the colour contractions produce a 
factor $(N-1)^p$. The integer $p$ 
ranges from $0$ to the number of index loops in the diagram.
The latter equals the total number of loops for planar diagrams
and decreases with the degree of non-planarity.
As a result, we are using a large $N$ expansion 
which is different from the 
standard planar expansion.
For example the first subleading term
in our expansion receives contributions from 
planar diagrams for which one index loop carries the index $1$
as well as from the leading non-planar diagrams with 
all index loops carrying primed indices.

Using a method similar to the standard power counting argument 
one can show that the dependence on $N$, $k$ and $q$ of an $L$-loop 
correction to the  $(1,1)$ slow-mode spectrum from arbitrary diagrams is 
given by
\begin{equation}
k^{-L} (N-1)^{p} q^{D-n} \, .
\label{RFPwerCountGaugeConvention}
\end{equation}
Here $D$ is the mass dimension of the coefficient of the quadratic term in 
the action for the field corresponding to the two external lines. 
For example in the case of a scalar $\phi^2$ term one has $D=2$, for a 
fermion $\psi^2$ term $D=1$.
As noted above, the summations and integrals
over loop variables give rise to an expansion in  
inverse powers of $q$. The integer $n$ 
in~(\ref{RFPwerCountGaugeConvention}) 
specifies the order in this expansion. 
Because of the cancellations we expect the integer $n$ 
to be greater than or equal to a certain positive 
integer, $n_0$, which is the order of the cancellation 
used in (\ref{RFLeading2ptSlow1}), (\ref{RFLeading2ptSlow2}) and 
(\ref{RFEstimateSlow2ptSlowScalar1Loop2}).
The integer $p$ denotes the number 
of index loops which are assigned the values $2, \ldots , N$ as explained 
above. By trivial rearranging of terms $(N-1)$ in the above formula can be 
replaced by $N$. We will do this for simplicity below. 
In order to derive this power counting estimate 
it is convenient to rescale the 
variables $\phi$, $\pi$ and $\psi$ in such a way that the action functional in 
the path integral can be written with a common overall factor of $k$. 

In order to obtain a complete understanding of the systematics of the 
perturbative expansion at higher orders, the power counting argument
presented above should be supplemented with precise information about 
the cancellations due to supersymmetry.
For the scattering of D0 branes in the context of the matrix model of 
M-theory -- which, as mentioned earlier, has some close analogies to 
the case we are considering --  the general structure of the expansion in 
terms of powers of the relative velocity and the impact parameter
was discussed in~\cite{RBBecker2PolchinskiTseytlin}.

We conclude this section with a few observations comparing the expansion 
(\ref{RFPwerCountGaugeConvention}) and the results obtained on the 
gravity side from the pp-wave matrix model.
On the AdS side, for the membrane states we have considered in section 
\ref{RSAdS}, there are two coupling constants, $Nk/J^3$ and $J^2/Nk$, 
which are associated with the loop expansion and the 
corrections to the pp-wave approximation.
Hence, for processes in 
which these states are relevant, 
the parameters $N$ and $k$ should always appear with the 
same power. From~(\ref{RFPwerCountGaugeConvention}) we 
find that this is achieved if the parameter $n$ is 
\begin{equation}
n=L+p+D.
\label{RFConditionOnnFromNkPairing}
\end{equation}
Let us focus on these contributions~\footnote{
The fact that $N$ and $k$ always 
appear in the combination $N k$ 
in the corrections discussed in section~\ref{RSAdS} 
has a simple interpretation.
The curvature radius of the \AdStS\ 
background is written only in terms of $Nk$, see (\ref{RFRInTermsOfNk}).
It is natural that local fluctuations of the membranes only 
feel the curvature and do not detect the effect of the 
$\Z_k$ quotient dividing the $S^7$ into $k$ pieces. 
However, in general 
there are other corrections some of which
we expect to depend separately on $N$ or $k$.
Hence the existence of terms in~(\ref{RFPwerCountGaugeConvention}) 
which do not satisfy~(\ref{RFConditionOnnFromNkPairing})
does not necessarily lead to a contradiction.
The appearance of the combination $Nk$ based on considerations on the 
gravity side and its implications for properties of the ABJM theory were 
discussed in~\cite{RBHanadaHoyosShimada}. 
}.
Substituting back into~(\ref{RFPwerCountGaugeConvention}) the order 
estimate becomes
\begin{equation}
(Nk)^{p} J^{-L-p} \, .
\end{equation}
Rewriting this in terms of the two coupling constants on the AdS side 
we obtain
\begin{equation}
\left(\frac{Nk}{J^3}\right)^{\!L-p} \left(\frac{J^2}{Nk}\right)^{\!L-2p}.
\label{RFCFT2CouplingPowers}
\end{equation}
We note that by definition $0\le p \le L$ 
and for smaller $p$ there are more 
$(1,1)$ propagators and the number of possible Feynman diagrams 
decreases.

The leading order term of the pp-wave approximation we have 
considered in section~\ref{RSAdS} corresponds to
\begin{equation}
p=\frac{L}{2}.
\end{equation}
In this case~(\ref{RFCFT2CouplingPowers}) reduces to
\begin{equation}
\left(\frac{Nk}{J^3}\right)^{\!\frac{L}{2}}, 
\label{RFCFTCounterpartOfLoopExpansion}
\end{equation}
\ie the power of $Nk/J^3$ in the expansion 
coincides with half the 
number of loops. This in particular implies 
that, for each given order in the expansion in terms of $Nk/J^3$, 
there is only a finite number of diagrams contributing and hence only a 
finite set of vertices in the gauge-fixed Hamiltonian are necessary.
The processes corresponding to 
 the leading order contribution  of order $Nk/J^3$ 
depicted in figure \ref{RPSlow2ptWithEffVertex}
satisfy $L=2, p=1$, as it should be.

The corrections to the pp-wave approximation come with 
positive powers of $J^2/Nk$ and thus correspond to
diagrams satisfying 
\begin{equation}
p < \frac{L}{2}.
\end{equation}
Therefore, if no terms with $p>L/2$ arise from perturbative calculations, 
(\ref{RFCFT2CouplingPowers}) has a straightforward interpretation as 
dual to the double expansion -- associated with loops and corrections to 
the pp-wave approximation -- discussed below~(\ref{RF1overJ}).
The explanation of terms with $p>L/2$ in~(\ref{RFCFT2CouplingPowers}) 
is less clear, since they are singular for vanishing $J^2/Nk$. However, 
an infinite series in negative powers of $J^2/Nk$ may yield 
a finite non-singular result, which might correspond to a non-perturbative 
correction to the pp-wave approximation in the matrix model.

In general the form of the low-energy effective action or Hamiltonian of 
a theory is strongly constrained by symmetry requirements. This is 
especially the case for supersymmetric theories, 
see~\cite{RBPeskinDuality} for a review. 
For the D0-brane scattering in the matrix model of M-theory this 
has been studied extensively, see for 
example~\cite{RBKazamaMuramatsu} and 
references therein.
At the end of 
section~\ref{RSCFTNearBPS} we have discussed some aspects of the 
supersymmetry algebra of the ABJM theory in the formalism used in this 
paper. It would be very interesting to study the restrictions imposed by 
supersymmetry on the structure of the effective action and on the 
spectrum. 

\section{Multiple membrane case} 
\label{RSMultiMembrane}

As discussed in section \ref{RSAdSBPS}, general zero energy configurations
in the matrix model, obtained solving (\ref{RFFuzzySphereEq}), correspond
to concentric fuzzy spheres with angular momenta $J_{(i)}$ in $S^7$ and
extending in AdS$_4$ with radii $r_{(i)} \approx J_{(i)}/2 \pi T R^2$. In 
order to be able to treat these configurations perturbatively in the pp-wave
approximation, the individual $J_{(i)}$'s should satisfy the condition 
(\ref{REJWindow}). The multi-membrane vacua correspond to states in 
the ABJM theory characterised by GNO charges $q_{(i)} = J_{(i)}/2k$, 
satisfying 
\begin{equation}
\sum_i 2q_{(i)} = \frac{J}{k} \, .
\end{equation}
For simplicity in this section we will focus on the case of two non-zero 
GNO charges, $q_{(1)}\!\!=\!\!J_{(1)}/2k$, 
$q_{(2)}\!\!=\!\!J_{(2)}/2k$, 
$q_{(3)}\!\!=\!\cdots\!=\!\!q_{(N)}\!\!=\!\!0$, 
and we will only briefly comment on generalisations.

From the definition of the covariant derivative (\ref{RFCovariantDer})
it follows that in general the $(i, j)$ component of a bi-fundamental
field has magnetic charge $q_{(i)}-q_{(j)}$. Therefore in the presence of 
two non-zero GNO charges, $q_{(1)}$ and $q_{(2)}$, we have the following 
situation, which, for concreteness, we illustrate in the case of the scalar 
fields, $\phi^{A\,i}{}_{\hat j}$. The other matter fields have a similar structure. 
The (block) diagonal components -- consisting of two $1\times 1$ blocks, 
$\phi^{A\,1}{}_{\hat 1}$ and $\phi^{A\,2}{}_{\hat 2}$, and a 
$(N-2)\times(N-2)$ block, $\phi^{A\,i'}{}_{{\hat j}'}$, ($i',{\hat j}'=3,\ldots,N$) 
-- have zero charge, as in the single membrane case. There are two 
$1\times(N-2)$ blocks, $\phi^{A\,1}{}_{{\hat i}'}$ and 
$\phi^{A\,2}{}_{{\hat i}'}$, and two $(N-2)\times 1$ blocks, 
$\phi^{A\,i'}{}_{\hat 1}$ and $\phi^{A\,i'}{}_{\hat 2}$, whose components 
carry charges $\pm q_{(1)}$ and $\pm q_{(2)}$. Finally the 
$\phi^{A\,1}{}_{\hat 2}$ and $\phi^{A\,2}{}_{\hat 1}$ components have 
charges $\pm(q_{(1)}-q_{(2)})$. 
Thus the 
scalar fields are decomposed as
\begin{equation}
\phi^{A\,i}{}_{\hj} = 
\left[ \begin{array}{c|c|c}
\phi^A{}^1{}_{\hat{1}}
& 
\phi^A{}^1{}_{\hat{2}}
& 
\rule{30pt}{0pt} 
\phi^A{}^1{}_{\hat{j}'}  
\rule{30pt}{0pt}
\raisebox{-10pt}{\rule{0pt}{0pt}} \\
\hline
\phi^A{}^2{}_{\hat{1}}
&
\phi^A{}^2{}_{\hat{2}}
& \rule{30pt}{0pt}
\phi^A{}^2{}_{\hat{j}'}  
\rule{30pt}{0pt}
\raisebox{-10pt}{\rule{0pt}{25pt}} \\
\hline 
\phi^A{}^{i'}{}_{\hat{1}}
\raisebox{-40pt}{\rule{0pt}{80pt}} & 
\phi^A{}^{i'}{}_{\hat{2}}
\raisebox{-40pt}{\rule{0pt}{80pt}} & 
\rule{12.5pt}{0pt} 
\phi^A{}^{i'}{}_{\hj'}
\raisebox{-40pt}{\rule{0pt}{80pt}}\rule{12pt}{0pt}
\rule{0pt}{47pt}
\end{array} \right] .
\end{equation}
From the discussion in the previous section, one would expect that all the 
off-diagonal components should be identified as fast modes and integrated 
out, while the $(1,1)$ and $(2,2)$ diagonal components should 
correspond to slow modes associated with membrane excitations. 
However, the condition (\ref{REJWindow}) for $J_{(1)}$ and $J_{(2)}$  
implies $q_{(1)}\gg 1$ and $q_{(2)}\gg 1$, but in general it is possible to 
have $q_{(1)}-q_{(2)}\sim O(1)$ (and even 
$q_{(1)}-q_{(2)}=0$)~\footnote{We use the symbol $O(1)$ to signify 
that the quantity in question is much smaller than $J/2k$.}. In this case 
the Born-Oppenheimer approximation requires that the $(1,2)$ and
$(2,1)$ components of the ABJM fields  be treated as slow modes,
since they feel a magnetic charge $q_{(1)}-q_{(2)}$ 
and therefore their expansion in monopole spherical harmonics starts 
with quantum number $l=|q_{(1)}-q_{(2)}|\sim O(1)$. 

The simplest states in this class of $(1,2)$ slow modes correspond to 
excitations of the complex scalar fields $(\phi^{A'})^i{}_{\hat{j}}$, $A'=1,2,3$ 
with $i=1,\hat{j}=2$ or $i=2,\hat{j}=1$. 
The spectrum for these states can be computed 
in a gauge similar to that used in section \ref{RSCFTNearBPS} 
in which we set 
$\phi^{4\,1}{}_{\hat{2}}=0$ and $\phi^{4\,2}{}_{\hat{1}}=0$. 
The calculation is very similar to that for the $(1, i')$ fast modes,
for which the spectrum is given in table \ref{RTCFTFastModeSpectrum}.
For these $(1,2)$ scalars the resulting spectrum is 
\begin{equation}
\D-\frac{J_4}{2} = \frac{1}{2}+l \, ,
\label{RF12ScalarSlowSpectrum}
\end{equation}
where $l=|q_{(1)}-q_{(2)}|, |q_{(1)}-q_{(2)}|+1,\ldots$ and the multiplicity is 
$2\times 3\times (2l+1)$, with the factor of $2$ due to the fact that the 
fields are complex. 

Vector and fermion excitations contain extra slow modes as well. These 
can be studied in a similar fashion, however, their analysis requires a
lengthier computation which we have not completed 
and thus we will not present the details here.

The generalisation to the case of three or more non-zero GNO charges 
is straightforward. For example in the case of three GNO charges, 
$q_{(1)}\sim q_{(2)}\sim q_{(3)}$, there are extra slow modes associated 
with the $(1,2)$, $(1,3)$, $(2,3)$ and $(2,1)$, $(3,1)$, $(3,2)$ components 
of the fields. 

Having found this new set of low-energy excitations in the 
ABJM theory, we should be able to identify dual configurations on the 
matrix model side, corresponding to excitations of the multi-membrane 
vacua. Focussing again on the two-membrane case, we recall that the 
vacuum in the matrix model is described by block-diagonal matrices with 
blocks given in (\ref{RFiBlockMMSol})-(\ref{RFMultiM2Radii}), 
corresponding to SU(2) irreducible representations of dimension 
$J_{(1)}/k$ and $J_{(2)}/k$, with $J_{(1)}+J_{(2)}=J$. When considering 
fluctuations around these configurations one turns on entries in the entire 
matrices, including the off-diagonal blocks, which correspond to 
$(J_{(1)}/k)\times (J_{(2)}/k)$ rectangular matrices. These rectangular 
matrices are the natural candidates to describe excitations dual to the 
slow modes associated with the $(1,2)$ and $(2,1)$ components 
of the ABJM fields. The corresponding spectrum was computed 
in~\cite{RBDSVR1} using a Clebsch-Gordan method. 
The basis used in the computation of the spectrum of fluctuations in the
rectangular off-diagonal blocks was further systematically studied 
in~\cite{RBIshikiShimasakitakayamaTsuchiya},
where a direct correspondence between this basis 
and the monopole spherical harmonics was pointed out.
More specifically in~\cite{RBIshikiShimasakitakayamaTsuchiya} it was 
shown that rectangular $(J_{(1)}/k)\times (J_{(2)}/k)$ matrices can be
expanded in a basis consisting of a discretised version of the monopole
spherical harmonics with charge 
$q_{(1)}-q_{(2)}=(J_{(1)}-J_{(2)})/2k$.
The scalar fluctuations coming from $S^7$ 
directions are  $(X^n)^u{}_v$, where $n=4,\ldots,9$
and the matrix indices, $u$ and $v$, 
 span the off-diagonal (rectangular) blocks.
Their energies are given by~\cite{RBDSVR1} 
\begin{equation}
\omega = \frac{2}{R} \left(\half+l\right) \, ,
\label{RFOffDiagonalS7Spectrum}
\end{equation}
where the quantum number $l$ takes values
\begin{equation}
\fr{2k} |J_{(1)}-J_{(2)}| \le l \le \fr{2k} (J_{(1)} + J_{(2)}) - 1 \, .
\label{RFOffDiagonalRange}
\end{equation}
There are six polarisations, corresponding to $n=4,\ldots,9$, hence
for each $l$ in the range (\ref{RFOffDiagonalRange}) the multiplicity is 
$6\times(2l+1)$. 

Using $\omega=(2\Delta-J)/R$ and $q_{(i)}=J_{(i)}/2k$, $i=1,2$, the matrix 
model spectrum 
(\ref{RFOffDiagonalS7Spectrum})-(\ref{RFOffDiagonalRange}) agrees 
with the result 
(\ref{RF12ScalarSlowSpectrum}) for the slow modes associated with the
$(1,2)$ and $(2,1)$ components of the scalars $\phi^{A'}$ in 
the ABJM theory, verifying the AdS/CFT duality for this particular set 
of states. 

Notice that in the matrix model there is a built-in upper 
bound in the range (\ref{RFOffDiagonalRange}) for the quantum number 
$l$. As remarked at the end of section 
\ref{RSAdSPerturbation} and in section \ref{RSCFTPerturbation} 
after (\ref{RFBoltzmannFactorPhaseSpaceScalarEuclid}), 
in view of the 
approximation schemes that we are using on the two sides of the 
correspondence, we can only expect good quantitative 
agreement for low-lying states in the spectra, with quantum number 
$l\ll J/2k$. Hence the absence of a corresponding upper bound on the 
CFT side would not necessarily lead to a contradiction.
However, the Born-Oppenheimer scheme does indeed suggest 
the existence of a similar upper bound, 
as it is natural not to consider modes with large $l$
-- and in particular $l\gtrsim J/2k=(J_{(1)}+J_{(2)})/2k$ -- as slow 
modes~\footnote{
The numerical coefficient in the expression for the cut-off 
should not be taken too seriously.
As is always the case with low-energy effective 
descriptions, the significance of such a bound is only in setting a 
separation between states with quantum numbers much 
below and much above a certain value.}.

The agreement between the spectra of these low-energy off-diagonal 
modes has interesting implications. The off-diagonal blocks in the 
regularised multi-membrane sectors have no obvious interpretation 
in the conventional 
continuum membrane theory. Thus the matrix model contains
additional degrees of freedom with no counterpart in the 
membrane theory. The fact that, at least when $|J_{(i)}-J_{(j)}|\ll J$, 
these fluctuations have corresponding low-energy states in the 
ABJM theory -- and the spectra on the two sides match -- indicates 
that these are genuine M-theory degrees of freedom and not an 
artefact of the matrix regularisation. Therefore our results provide 
an explicit and concrete example showing that the matrix model can 
capture aspects of the dynamics of M-theory beyond the 
conventional supermembrane theory~\cite{RBBergshoeffSezginTownsend}. 
The existence of the extra degrees of freedom, appearing when
the two membranes are close to each other (as the 
condition $|J_{(i)}-J_{(j)}|\ll J$ implies), can be thought of
as a manifestation of the non-Abelian nature of 
membranes, analogous to that of D-branes~\footnote{
This non-Abelian character is manifest in the Bagger-Lambert and 
ABJM theories~\cite{RBBaggerLambert1,RBBaggerLambert2,RBABJM}, 
which were proposed as low-energy descriptions of multiple membranes. 
The possibility of interpreting the block off-diagonal components in 
the pp-wave matrix model as the non-Abelian degrees of freedom
of membranes was suggested in~\cite{RBDSVR1}.
In~\cite{RBShimadaBetaMM} it was pointed out 
that the non-Abelian nature 
of membranes may explain certain interesting properties 
of stable solutions (corresponding to membranes with torus topology) in 
a deformed version of the matrix model, where 
configurations of membranes characterised by different winding numbers
in the continuum theory become indistinguishable in the matrix model
description.}. It would be interesting to verify that the 
agreement discussed above between the energies of these particular 
states on the two sides of the AdS/CFT duality persists after the inclusion 
of quantum corrections. We hope to investigate this issue in the future.

\section{Conclusions and Discussion} 
\label{RSConclusion}

In this paper we have studied the AdS$_4$/CFT$_3$ duality proposed 
in~\cite{RBABJM} in an M-theoretic regime in which neither the 
ten-dimensional type IIA string limit nor the low-energy 
eleven-dimensional supergravity 
approximation are applicable. In order to make it possible to quantitatively 
study the correspondence in this regime, we have focussed on a special 
sector associated with a large quantum number, $J$. On the gravity side 
$J$ is an orbital angular momentum and for large $J$ the membrane 
configurations we consider can be described using the pp-wave matrix 
model.
On the CFT side the dual sector involves monopole operators which 
are conveniently studied using a Hamiltonian formulation within the 
framework of radial quantisation. In this approach we consider states 
satisfying a Gauss law constraint associated with the presence of a large 
flux, controlled by the parameter $J$, through 
the $S^2$ corresponding to fixed-time slices in radial 
quantisation. In the large $J$ regime we identified approximation schemes 
which are simultaneously valid on both sides of the duality. On the one 
hand the pp-wave matrix model is weakly coupled and therefore a standard 
quantum mechanical perturbative expansion is applicable. On the other 
hand in the ABJM theory the presence of a large parameter makes it 
possible to give a (weakly coupled) effective description of the physical 
degrees of freedom dual to M-theory states using a Born-Oppenheimer 
approach. The choice of a suitable gauge is a crucial element of our 
analysis on the CFT side. Another essential ingredient is a version of the 
Higgs mechanism, which, together with the presence of a large magnetic 
flux, leads to a separation between low and high energy states thus 
allowing us to identify the physical degrees of freedom. 

When using radial quantisation and the state-operator map, the AdS/CFT 
dictionary directly relates energy spectra on the two sides of the duality. We 
have verified the agreement between these  spectra in the large $J$ sector 
at leading order for both BPS and near-BPS states. 
This provides a very non-trivial test of the AdS$_4$/CFT$_3$ duality 
of~\cite{RBABJM} in an 
M-theoretic regime which had not been accessible so far. At the same time, 
by independently reproducing the membrane spectrum from the dual CFT, 
our results provide strong support for the validity of the matrix model 
approach to M-theory.  

The AdS/CFT dictionary for the large $J$ sector we discussed
is summarised in table \ref{dictionary}.

\begin{table}[htb]
\begin{center}
\begin{tabular}[t]{|c||c||c|}
\hline
& AdS side \raisebox{-11pt}{\rule{0pt}{29pt}} & CFT side \\
\hline \hline
Framework\raisebox{-11pt}{\rule{0pt}{29pt}} 
& pp-wave matrix model %(\ref{RFMMHamiltonian1}) 
& Radial quantisation with large flux 
%(\ref{RFABJMHamiltonianRadialQuantised}) 
\\
\hline
Approximation \raisebox{-18pt}{\rule{0pt}{40pt}} &
\parbox{4cm}{\centering pp-wave approximation and loop expansion} & 
Born-Oppenheimer \\
\hline
\parbox{3cm}{\centering BPS \\ ground states} 
\raisebox{-18pt}{\rule{0pt}{40pt}}
& Collection of fuzzy spheres %(\ref{RFVacuumClassificationMM})  
& Flux characterised by GNO charges 
%(\ref{RFVacuumClassificationABJM}) 
\\
\hline 
& \raisebox{-11pt}{\rule{0pt}{29pt}} $6$ real scalars from $S^7$   
& 3 complex scalars $\phi^{A'}$ ($A'=1,2,3$)   
\\
\parbox{3cm}{\centering Near BPS \\ fluctuations}
& $3$ real scalars from AdS$_4$  
& $\phi^4$ and gauge fields  \\
& \raisebox{-11pt}{\rule{0pt}{27pt}} $16$ real fermions 
& 4 complex spinors $\psi^{A}$  
\\
\hline
\end{tabular}
\caption{Dictionary for M-theoretic AdS$_4$/CFT$_3$ in the large $J$ 
sector. The pp-wave matrix model Hamiltonian is (\ref{RFMMHamiltonian1}).
The Hamiltonian of ABJM theory in our gauge,
which incorporates the effect of the magnetic flux, 
is (\ref{RFGaugeFixedHamiltonianCFT}).
The BPS ground states are classified on both sides of the duality  
by a partition of integers satisfying  
(\ref{RFVacuumClassificationMM}) on the AdS side
and (\ref{RFVacuumClassificationABJM}) on the CFT side.
The near-BPS fluctuation spectrum on the AdS side is summarised in 
table \ref{RTAdSNBPSTreeSpectrum}. On the CFT side, the corresponding 
degrees of freedom are $(1,1)$ component of various fields 
and their spectrum is given in 
table~\ref{RTCFTSlowModeSpectrum}. 
}
\label{dictionary}
\end{center}
\label{RTDictionary}
\end{table}

The starting point of our analysis -- \ie the observation that focussing 
on a sector characterised by a large quantum number leads to a 
simplification in the study of the AdS/CFT duality -- is similar to the premise 
of the work of BMN in the context of the AdS$_5$/CFT$_4$ 
correspondence~\cite{RBBMN}. More generally there are 
analogies between our construction and that of~\cite{RBBMN}. 
However, the final picture that emerges from our investigation is 
fundamentally different from the one proposed by BMN. This is the 
manifestation of the fact that we have applied similar ideas to the 
description of a very different physical system -- membranes rather than 
strings.

The relationship between the AdS and CFT sides of the duality we have 
studied in this paper seems to be remarkable for its directness. In 
particular, in comparing the two sides of the correspondence the 
sphere introduced on the CFT side as a tool in the radial quantisation 
can almost be identified with the sphere representing the minimal energy 
configuration for membranes on the matrix model side.
The implication of this observation is that the states on the two sides 
are naturally described in terms of the same (monopole) spherical 
harmonics, making the definition of the map between bulk and boundary 
observables more straightforward. This may not be so surprising since the 
important degrees of freedom in the bulk of \AdStS\ are membranes and 
the boundary ABJM theory describes the low-energy dynamics of 
membranes, so that on both sides one focusses on the same 
kind of objects. This is in strong contrast with more familiar examples of 
AdS/CFT duality and in particular the canonical AdS$_5$/CFT$_4$ case, 
where the bulk degrees of freedom are closed fundamental strings and the 
boundary theory describes the low-energy degrees of freedom of 
D3-branes.~\footnote{
A relation between configurations of D3-branes (the so-called giant 
gravitons) extended in AdS$_5$ and states in the radially quantised 
$\mathcal{N}=4$ Super Yang-Mills theory similar to our construction 
was considered in~\cite{RBHashimotoHiranoItzhaki}.
}

It is essential that we use 
large but finite $J/k \times J/k$ matrices 
on the AdS side. 
This is in particular crucial in establishing the map relating BPS states 
on the two sides of the duality, which are classified by a set of integers -- 
associated respectively with the angular momenta of individual membranes 
in the matrix model and with GNO charges of monopole operators in the 
CFT. 
The fact that we can formulate a duality with the ABJM theory using
finite dimensional matrices is interesting.
In this respect our construction is 
different from the standard approach to the 
matrix model description of membranes~\cite{RBGoldstoneHoppe, 
RBdWHN}, in which the size of the matrices plays the role of a regularisation 
parameter and should be taken to infinity.
The matrix model seems to describe a 
theory in which the membranes are discretised.
This is reminiscent of the description of gauge invariant operators 
dual to closed strings 
in terms of a discrete spin chain in versions of the AdS/CFT correspondence
in which the gravity dual is a string theory.
A consequence of working with matrices of finite size
is the presence of an upper bound on the mode numbers in the 
expansion of the fluctuations in spherical harmonics. 
We have seen that a corresponding cut-off naturally arises on the CFT side
in the context of the Born-Oppenheimer approximation. 
Here it follows from the fact that it is not completely justified 
to treat as low-energy modes the high momentum components 
of the slow modes with energies higher than the mass of the fast modes.

The implications of the direct nature of the 
duality we have presented are particularly intriguing in the 
case studied in section \ref{RSMultiMembrane}, where there are multiple
concentric spherical membranes 
of approximately equal radius. In this situation we have 
seen that the block off-diagonal degrees of freedom of the matrix model
have as counterpart in the dual ABJM theory certain 
off-diagonal components of the fields. The block off-diagonal degrees of 
freedom do not exists in the conventional 
continuum membrane theory, which does not take into account the
possibility that membranes possess non-Abelian degrees of freedom. 
The fact that their spectrum appears to be reproducible in the CFT 
suggests that these degrees of freedom should not be considered as 
spurious, or a kind of ``lattice artefact''.  Instead they
seem to be the manifestation of a genuinely non-Abelian nature of 
membranes in M-theory. This is a new and non-trivial insight into the 
dynamics of M-theory that can be deduced from the study of the AdS/CFT 
correspondence. It would be also interesting to understand this non-Abelian 
nature of membranes directly from the matrix model without relying on 
the AdS/CFT correspondence. This presumably will help to shed light on a 
possible non-Abelian Born-Infeld type description of membranes.

Another interesting feature of the ABJM theory which emerges from our 
analysis is the following. Let us consider the case in which only one 
membrane is present on the AdS side and correspondingly only 
the first GNO charge is non-zero on the CFT side.
In this case, as we have seen in section \ref{RSCFTNearBPS}, 
the excitation of the $(1,1)$ components of the ABJM fields 
is identified with the excitation of phonons on the stable spherical 
membrane. On the other hand, the excitations of (diagonal) components in 
the lower right $(N-1)\times (N-1)$ block 
would in general give rise to other non-zero GNO charges.
We should interpret this as the creation of additional 
membranes~\footnote{Strictly speaking, this argument is partially based on 
an extrapolation of the results of our analysis valid for $J\gg 1$, 
as the momenta/GNO charges of the created membranes may not be 
large.}. Thus the ABJM theory combines 
the features of a first-quantised and a second-quantised
description of membranes in this manner. 

At the end of section~\ref{RSCFTPerturbation} we have presented the 
estimate~(\ref{RFCFT2CouplingPowers}) for the behaviour of loop 
corrections in the 
ABJM theory in terms of powers of $N$. However, an intriguing possibility is 
to retain the $(N-1)$ combination, which has an interesting explanation in 
terms of the gravity dual. In our interpretation of the ABJM theory in the 
large $J$ sector the slow modes correspond to the membrane fluctuations 
studied in section~\ref{RSAdS}. The \AdStS\ background is obtained as the 
near horizon geometry of a stack of $N$ membranes. One can think of the 
fluctuations as coming from excitations of the original $N$ background 
membranes. Then for a state containing a single excited membrane the 
background comprises only the remaining $(N-1)$ membranes. It may be 
possible to interpret the $(N-1)$ fast modes as corresponding to the 
background associated with these $(N-1)$ membranes. This picture is also 
consistently generalised to the case of multi-membrane configurations on 
the gravity side, which is related to a sector of the CFT with multiple 
non-zero GNO charges. For instance in the case of two membranes/GNO 
charges briefly discussed in section~\ref{RSMultiMembrane} -- at least if 
the two membranes have comparable angular momenta -- the fast modes 
can be combined into groups of $(N-2)$ fields. This corresponds to the fact 
that on the AdS side we have two excited states leaving a background of 
$(N-2)$ membranes.

The general analysis of perturbative corrections in 
section~\ref{RSCFTPerturbation} suggests the possibility of the emergence 
of a novel type of large $N$ expansion in the ABJM theory for $J\gg1$. We 
have provided a prescription for determining the dependence on powers of 
$(N-1)$ in the single membrane sector. This involves drawing Feynman 
diagrams in the standard double line notation and then specifying for all 
index loops whether they carry a colour index $1$ or a primed index taking 
values $2,\ldots,N$. The different perturbative contributions can be 
classified according to the power of $(N-1)$ they produce. This power is 
given by the number of index loops carrying primed colour indices. 
The resulting large $N$ expansion is different from 
the standard 't Hooft expansion.
Moreover, the general considerations on the structure of the 
diagrammatic corrections to the spectrum discussed in 
section~\ref{RSCFTPerturbation} -- and specifically the expected 
cancellations due to supersymmetry -- suggest a relation between the 
order in the loop expansion and the powers of $(N-1)$, which is inherently 
new (see for example 
(\ref{RFCFT2CouplingPowers})-(\ref{RFCFTCounterpartOfLoopExpansion})).
It is well-known that if one focusses on the 
contribution of planar diagrams in the 't Hooft expansion,
a theory often simplifies and shows various special properties.
It would be interesting to study whether the leading order contributions
in this new type of large $N$ expansion have similar 
special properties.
It is intriguing to speculate that the emergence of this new 
type of large $N$ expansion may be related to the fact that we are 
considering a genuinely M-theoretic regime. In the sector under 
consideration the elementary degrees of freedom on the gravity side are 
not strings, whereas the standard 't Hooft expansion suggests strongly a 
stringy interpretation for the fundamental degrees of freedom.

The most important next step in our program will be to 
compute the higher order corrections in the spirit of the
Born-Oppenheimer approximation discussed in 
section~\ref{RSCFTPerturbation}. The calculation
is quite involved 
as the gauge-fixed Hamiltonian, obtained iteratively solving the Gauss law 
constraints, contains a very large number of interaction vertices. Therefore 
it will be crucial to develop techniques to 
simplify the computations. 

Although the gauge adopted in this paper seems to be well-suited to
clarify the structure of the physical degrees of freedom, there may be 
more convenient choices for explicit loop 
computations. The situation may be analogous to the well-known case of
Yang-Mills theories when the gauge symmetry is spontaneously broken via 
the Higgs mechanism. In that case, the unitary gauge is well-suited 
for studying the spectrum of the theory, but there are other gauge choices
which are more convenient to perform loop 
computations. 

The spectrum of the slow modes and the effective theory governing their 
dynamics discussed in section~\ref{RSCFT} 
should be highly constrained by supersymmetry. It is important 
to concretely study the restrictions imposed by supersymmetry.
This should also facilitate the explicit computation of quantum corrections. 
A formulation using superfields, adapted to the large $J$ sector, might be 
useful in this respect.
As discussed in section~\ref{RSCFTPerturbation}, the results on the AdS 
side of the correspondence indicate the presence of cancellations leading to 
certain patterns in the structure of higher order perturbative corrections in 
the CFT. It will be important to explore this structure directly in the ABJM 
theory by carrying out calculations of loop corrections to the spectrum.

An efficient computational scheme may arise from the adaptation of the 
methods based on localisation to the study of the sector we focussed 
on. The localisation approach relies on the existence of a nilpotent 
supercharge which annihilates the observables under consideration. As 
such the method is only applicable to BPS quantities which are invariant 
under at least one supersymmetry. To implement the method one deforms 
the ABJM theory by the addition of terms invariant under the relevant 
supercharge. 
The deformation is controlled by a parameter in such a way 
that when the parameter is sent to infinity a saddle point approximation
becomes exact, allowing an explicit evaluation of the observables. 
The main focus of our 
investigation are non-BPS states and for this reason in order to study their 
spectrum we have relied on a different approximation scheme that arises 
for large $J$. However, the observables we have considered -- similarly to 
the BMN operators in the $\calN=4$ supersymmetric Yang-Mills theory -- 
are near BPS, with $1/J$ acting as the parameter measuring their deviation 
from exactly BPS observables. In view of this it might be possible to 
generalise the localisation approach to make it applicable to near-BPS 
observables. In this case the deformation parameter cannot be sent strictly 
to infinity, because the observables are not invariant under the deformation. 
However, for a carefully chosen deformation, we expect the variation 
of near BPS observables such as those studied in this paper to be 
suppressed by a power of $1/J$. It should then be possible to take the 
deformation parameter to be as large as a positive power of $J$ and 
evaluate near BPS observables using a saddle point approximation, 
including higher order terms in an expansion in $1/J$.
This approach may potentially have many applications beyond the large 
$J$ sector of the ABJM theory and it would be interesting to test this idea in 
simple models.

The sector we have discussed provides a setting for the study of 
the interactions responsible for processes involving the splitting or joining of 
membranes. This is a central and still little understood aspect of the 
dynamics of M-theory and the pp-wave matrix model, together with its dual 
description in terms of the ABJM theory, appears to be particularly suited 
to investigate it.
The perturbative vacua on the gravity side,
which correspond to  configurations 
with varying numbers of membranes, 
actually belong to the same Hilbert space of the matrix model. 
In other words, as is well-known, the matrix model
should be interpreted as a second-quantised theory of 
membranes~\cite{RBBFSS, RBTaylorReview}. 
It should be possible to compute 
transition amplitudes between two states (either ``vacua'' or excited states)  
characterised by different sets of integers as 
already mentioned in section \ref{RSAdSBPS}.
These transition amplitudes are analogous to the string field theory 
vertices in the ten-dimensional
pp-wave background and thus should provide 
the building blocks for the computation, from the dual 
gravitational perspective, of $n$-point correlation functions of the 
operators we defined via the state-operator map. 
Analogous computations for the three-point functions of BMN operators 
have been done in string theory. For a recent reference, 
see~\cite{RBGrignaniZayakin}. For an analysis of the relation between the 
transition amplitudes (or the string vertex) and the CFT OPE coefficients, 
see~\cite{RBShimada3}.
It is important to determine the coupling 
constant governing these processes, which should 
correspond to tunnelling amplitudes.
In this paper we have  assumed this coupling to be 
small. It is tempting to conjecture that it may be given by a 
certain combination of powers of $N, k$ and $J$.
It is also an interesting problem to compute three-point functions
of operators with non-zero monopole charges, such as those considered 
in this paper, directly on the CFT side. 
This is presumably related to the tunnelling process discussed at the end of 
section~\ref{RSCFTBPS}.

In this paper we studied the M-theory regime 
in which the parameter $k$ is of order $1$. 
It is of some interest to consider whether there is a 
type IIA regime ($N \gg 1, k\gg 1$ with $N/k$ fixed) in which a
description similar to the one given in this paper based on 
the pp-wave approximation is possible.
An essential difference   
in the type IIA regime is that, since the M-theory circle 
is small, M2-branes wrapped on the M-theory circle should 
also be considered. To incorporate these degrees of freedom it seems 
appropriate to use the matrix string formulation~\cite{RBMotl, 
RBDijkgraafVerlindeVerlinde}. 
Several works have studied aspects which are relevant for this
line of investigation. 
A direct map identifying the 
degrees of freedom associated with wrapped 
membranes in matrix string theory was discussed 
in~\cite{RBSekinoYoneya}. The matrix string theory on a type IIA 
supersymmetric pp-wave background was constructed 
in~\cite{RBBonelli, RBSugiyamaYoshidaMatrixString}.
An M2-brane solution in the type IIA regime, which is wrapped around the 
M-theory circle and has torus topology, was found 
in~\cite{RBNishiokaTakayanagiTorus}.
Wrapped M5-brane solutions related to 
the wrapped M2-brane solutions
were discussed in~\cite{RBHerreroLozanoPicos}.
Considerations on the CFT counterparts of these solutions 
were presented in \cite{RBBerensteinPark, 
RBEzhuthachanShimasakiYokoyama}.
There may be connections to BPS solutions of the membrane theory 
on the pp-wave background with arbitrary genus found 
in~\cite{RBBakKimLee}.
We also note that the pp-wave approximation for
string states in the type IIA limit with zero monopole charge
was studied in \cite{RBNishiokaTakayanagiPP, RBGaiottoGiombiYin}.

In recent years methods derived from the study of integrable systems 
have played an important role in the computation of corrections to 
the spectrum on both sides of the AdS/CFT 
correspondence~\cite{RBIntegrabilityReview}. A question that arises is 
whether integrability can be relevant in the sector of the AdS$_4$/CFT$_3$ 
duality that we considered in this paper. The consensus is that integrability 
in the AdS/CFT correspondence is a feature arising only in the planar 
approximation. This seems to indicate that it should not be expected in the 
M-theoretic regime. However, since we are dealing with a theory of 
membranes rather than strings, the significance of the planar 
approximation is unclear. 
An interesting possibility is that 
integrability might arise in the large $J$ sector of the ABJM theory if one 
focusses on the leading contributions in the novel large $N$ expansion that 
we described above.
Moreover, extending the ideas developed in 
the context of stringy examples of AdS/CFT duality, it is natural to expect 
that the relevant integrable systems in a case involving membrane degrees 
of freedom might be $2+1$ dimensional. These considerations lead us to 
suspect that, if integrability can play a role in the present 
context, it should present interesting new features.

As already noted above constraints from supersymmetry will presumably 
play a crucial role in better understanding the structure of the ABJM theory 
in the large $J$ regime. For $k=1, 2$ supersymmetry is expected to be
enhanced to $\calN=8$. The extra supersymmetries are related to the 
presence of monopole operators and some of the associated R-currents 
are already known~\cite{RBBashkirovKapustin,RBBennaKlebanovKlose}. 
More concretely, the part of the $\calN=8$ supersymmetry algebra broken 
for $k\neq 1, 2$ corresponds to generators transforming under $J_M$ with 
charge $\pm 2$. Since the monopole charge, $J_M$, is a multiple of $k$, 
these charges cannot exists for $k \ge 3$ and this explains why the 
$\calN=8$ supersymmetry is broken down to $\calN=6$ for $k \ge 3$. In 
section~\ref{RSCFTNearBPS} we have focused on states with given $J_M$ 
in the gauge-fixed theory. It seems to be straightforward to relax this 
restriction. Since $J_M$ is a conserved charge, the Hilbert space of the 
ABJM theory can be viewed as the direct sum of the vector spaces of 
states with fixed $\Jt$. The action of the full superalgebra, including the 
supercharges that change the value of $J_M$ by $\pm2$ units, spans the 
entire Hilbert space of the theory relating states with different quantum 
number $J_M$. Since our gauge has the advantage of being very explicit, 
it should be possible to use it to write down the supercharges for the full 
$\calN =8$ supersymmetry, at least at the classical level. It would be 
interesting to do this and study their commutators to explicitly verify the 
closure of the $\mathcal{N}=8$ superalgebra.

The mechanism of the breaking of  $\calN=8$ supersymmetry explained 
above helps to clarify why the pp-wave matrix model has 
32 real supersymmetries~\cite{RBBMN, RBDSVR2}.
At leading order in the pp-wave approximation order $1$ differences in 
the value of $J_M \sim J$ cannot be detected. Hence it is natural to 
expect that the $\calN=8$ supersymmetry of the matrix model for arbitrary 
$k$ should be interpreted as an approximate symmetry, which would be 
broken by the inclusion of corrections to the pp-wave approximation.
In a similar fashion the low-energy sector that we identified in the ABJM 
theory for large $J$ should possess an approximate $\calN=8$ 
supersymmetry even for $k\ge 3$. It would be interesting to study 
concretely these aspects, analysing the corrections to the pp-wave 
approximation on the gravity side and the symmetries of the low-energy 
effective theory for the slow modes on the CFT side.

Following the ABJM proposal there have been 
many generalisations leading to other examples of AdS$_4$/CFT$_3$ 
dualities with less supersymmetry. 
It should be possible to extend our 
analysis to these cases as well. The approach and the techniques 
we have developed in this paper
will also be useful more generally in the study of various  properties of 
three-dimensional
conformal Chern-Simons-matter theories, irrespective of
whether or not they have gravity duals. More specifically, 
the Born-Oppenheimer type approximation we have discussed in 
section~\ref{RSCFT} provides a new approach to the computation of 
conformal dimensions of 
various types of operators with large monopole charge, which may be
applicable in situations where a conventional perturbative expansion is 
not justified. 

The work presented in this paper has interesting connections to the 
little understood $(2,0)$ superconformal field theory in six dimensions. This 
theory, which is believed to describe the low-energy dynamics of a stack of 
M5-branes, is expected to have as gravity dual M-theory in an 
AdS$_7\times S^4$ background. It is interesting to notice that the 
application of the pp-wave approximation to this background leads to the 
same geometry as the one obtained from 
AdS$_4\times S^7$~\cite{RBBMN,RBppBlauHulletal2}. This suggests that 
the six-dimensional $(2,0)$ theory should contain a 
sector dual to the matrix model we have considered for which a weak 
coupling description might be possible.
Since M2- and M5-branes are electromagnetic duals of each other in 
the eleven dimensional target space of M-theory, one may expect the 
M2-brane excitations discussed in section~\ref{RSAdSFluctuation} to be 
captured by solitonic degrees of freedom in the $(2,0)$ 
theory.  

According to the proposal 
of~\cite{RBMaldacenaSheikhJabbariVanRaamsdonk} there are
states in the pp-wave matrix model which have dual descriptions 
as M5- or M2-branes. For instance a configuration 
characterised by a partition  $J\!=\!1+1+\cdots+1$ can be identified with a 
single M5-brane. The pp-wave approximation should be applicable to such 
M5-branes as discussed in section \ref{RSAdS}, since they carry large 
angular momentum and their size is small. Even if they cannot be treated 
perturbatively in the matrix model, it is possible that they can be studied by 
a Monte-Carlo simulation or by devising an appropriate approximation 
scheme such as a variational method.
It would be interesting if one could gain any insights into the dynamics 
of M5-branes using the pp-wave matrix model and its dual description 
in terms of the ABJM theory.

We have seen that the large $J$ limit seems to provide a good 
framework in which concepts from M-theory -- and its matrix formulation -- 
and the AdS/CFT duality work together. We hope that the interplay of these 
ideas may lead to a better understanding of both M-theory and the 
AdS/CFT correspondence.

\newpage
%\vsp{0.7}
\ndt
{\bf Acknowledgments}

\vsp{0.3}
\ndt
We would like to thank 
S.~Ananth, J.~Armas, 
J.~Bhattacharya, 
S.~Frolov,
M.~Hanada, K.~Hashimoto, S.~Hirano, M.~Honda, J.~Hoppe, K.~Hosomichi, 
H.~Kanno, H.~Kawai, Y.~Kazama, Y.~Kimura, Y.~Kikukawa, S.~Komatsu, 
T.~Kuroki, 
T.~McLoughlin, S.~Moriyama, Y.~Mitsuka,
W.~Nahm,
N.~Obers, K.~Oda, Y.~Okawa, T.~Okazaki, T.~Onogi,   
K.~Sakai, T.~Sakai, S.~Shatashvili, R.~Suzuki, F.~Sugino, S.~Sugimoto, 
T.~Takayanagi, A.~Tanaka, S.~Terashima, S.~Theisen,
S.~Yamaguchi and T.~Yoneya for encouragement, discussions and 
comments.
Y.S. is very grateful to the Yukawa Institute for Theoretical Physics, Nagoya 
University and the Niels Bohr Institute, where part of this work was done, 
for their warm hospitality.
Y.S. would like to acknowledge financial support from the Hakubi Centre for 
Advanced Research.
Part of this work was done while H.S. was a member of the Niels Bohr 
International Academy and he would like to thank the institute and his 
colleagues there.

%%%%%%%%%%%%%%%%%%%%%%%%%%%%%%%%%%%%%%%
%%%%%%%%%%%%%%%%%%%%%%%%%%%%%%%%%%%%%%%
%%%%%%%%%%%%%%%%%%%%%%%%%%%%%%%%%%%%%%%


\begin{thebibliography}{3}


%\cite{Hull:1994ys}
%\bibitem{Hull:1994ys}
\bibitem{RBHullTownsend}
  C.~M.~Hull and P.~K.~Townsend,
  ``Unity of superstring dualities'',
  {\it Nucl. Phys.} {\bf B438} (1995) 109
  [hep-th/9410167].
  %%CITATION = HEP-TH/9410167;%%
  
%\cite{Witten:1995ex}
%\bibitem{Witten:1995ex}
\bibitem{RBWitten}
  E.~Witten,
  ``String theory dynamics in various dimensions'',
  {\it Nucl. Phys.} {\bf B443} (1995) 85
  [hep-th/9503124].
  %%CITATION = HEP-TH/9503124;%%

\bibitem{RBGoldstoneHoppe}{J.~Goldstone, unpublished (1982). \\
J.~Hoppe, PhD Thesis (M.I.T. 1982).}

%\cite{THU-88-15}
%\bibitem{THU-88-15}
\bibitem{RBdWHN}
  B.~de Wit, J.~Hoppe and H.~Nicolai,
  ``On the Quantum Mechanics of Supermembranes'',
  {\it Nucl. Phys.} {\bf B305} (1988) 545.
        %%CITATION = NUPHA,B305,545;%%

%\cite{Townsend:1995kk}
%\bibitem{Townsend:1995kk}
\bibitem{RBTownsendD0}
  P.~K.~Townsend,
  ``The eleven-dimensional supermembrane revisited'',
  {\it Phys. Lett.} {\bf B350} (1995) 184
  [hep-th/9501068].
  %%CITATION = HEP-TH/9501068;%%

\bibitem{RBBFSS}
%\cite{Banks:1996vh}
%\bibitem{Banks:1996vh}
  T.~Banks, W.~Fischler, S.~H.~Shenker and L.~Susskind,
   ``M theory as a matrix model: A Conjecture'',
   {\it Phys. Rev.} {\bf D55} (1997) 5112
        [hep-th/9610043].
          %%CITATION = HEP-TH/9610043;%%

%\cite{deWit:1989vb}
%\bibitem{deWit:1989vb}
\bibitem{RBdeWitMarquardNicolai}
  B.~de Wit, U.~Marquard and H.~Nicolai,
   ``Area Preserving Diffeomorphisms And Supermembrane 
   Lorentz Invariance'',
    {\it Commun. Math. Phys.}  {\bf 128} (1990) 39.
        %%CITATION = CMPHA,128,39;%%

%\cite{Ezawa:1997da}
%\bibitem{Ezawa:1997da}
\bibitem{RBEzawaMatsuoMurakami}
  K.~Ezawa, Y.~Matsuo and K.~Murakami,
  ``Lorentz symmetry of supermembrane in light cone gauge formulation'',
    {\it Prog. Theor. Phys.}  {\bf 98} (1997) 485
        [hep-th/9705005].
          %%CITATION = HEP-TH/9705005;%%

%\cite{Fujikawa:1997jt}
\bibitem{RBFujikawaOkuyama}
  K.~Fujikawa and K.~Okuyama,
   ``On a Lorentz covariant matrix regularization of membrane theories'',
     {\it Phys. Lett.} {\bf B411} (1997) 261
        [hep-th/9706027].
          %%CITATION = HEP-TH/9706027;%%

%\cite{Hoppe:2010ed}
%\bibitem{Hoppe:2010ed}
\bibitem{RBHoppeLorentz}
  J.~Hoppe,
   ``Matrix Models and Lorentz Invariance'',
    {\it J. Phys.} {\bf A44} (2011) 055402
        [arXiv:1007.5505 [hep-th]].
          %%CITATION = ARXIV:1007.5505;%%

%\cite{Hoppe:2011ef}
%\bibitem{Hoppe:2011ef}
\bibitem{RBHoppeTrzetrzelewski}
  J.~Hoppe and M.~Trzetrzelewski,
   ``Lorentz-invariant membranes and finite matrix approximations'',
    {\it Nucl. Phys.} {\bf B849} (2011) 628
        [arXiv:1101.4403 [hep-th]].
          %%CITATION = ARXIV:1101.4403;%%

\bibitem{RBTaylorReview}
%\cite{Taylor:2001vb}
%\bibitem{Taylor:2001vb}
  W.~Taylor,
    ``M(atrix) theory: Matrix quantum mechanics as a fundamental theory'',
      {\it Rev. Mod. Phys.}  {\bf 73} (2001) 419
        [hep-th/0101126].
          %%CITATION = HEP-TH/0101126;%%

%\cite{Maldacena:1997re}
%\bibitem{Maldacena:1997re}
\bibitem{RBMaldacena}
  J.~M.~Maldacena,
   ``The Large N limit of superconformal field theories and supergravity'',
   {\it Adv. Theor. Math. Phys.}  {\bf 2} (1998) 231
        [hep-th/9711200].
          %%CITATION = HEP-TH/9711200;%%

%\cite{arXiv:0806.1218}
%\bibitem{arXiv:0806.1218}
\bibitem{RBABJM}
  O.~Aharony, O.~Bergman, D.~L.~Jafferis and J.~Maldacena,
  ``N=6 superconformal Chern-Simons-matter theories, M2-branes 
  and their gravity duals'',
    {\it JHEP} {\bf 0810} (2008) 091
        [arXiv:0806.1218 [hep-th]].
          %%CITATION = JHEPA,0810,091;%%

%\cite{Schwarz:2004yj}
%\bibitem{Schwarz:2004yj}
\bibitem{RBSchwarzCS}
  J.~H.~Schwarz,
  ``Superconformal Chern-Simons theories'',
  {\it JHEP} {\bf 0411} (2004) 078
  [hep-th/0411077].
  %%CITATION = HEP-TH/0411077;%%

%\cite{Bagger:2006sk}
%\bibitem{Bagger:2006sk}
\bibitem{RBBaggerLambert1}
  J.~Bagger and N.~Lambert,
   ``Modeling Multiple M2's'',
   {\it Phys. Rev.} {\bf D75} (2007) 045020
        [hep-th/0611108].
          %%CITATION = HEP-TH/0611108;%%

%\cite{Bagger:2007jr}
%\bibitem{Bagger:2007jr}
\bibitem{RBBaggerLambert2}
  J.~Bagger and N.~Lambert,
  ``Gauge symmetry and supersymmetry of multiple M2-branes'',
  {\it Phys. Rev.} {\bf D77} (2008) 065008
        [arXiv:0711.0955 [hep-th]].
          %%CITATION = ARXIV:0711.0955;%%

%\cite{Bagger:2007vi}
%\bibitem{Bagger:2007vi}
\bibitem{RBBaggerLambert3} 
  J.~Bagger and N.~Lambert,
  ``Comments on multiple M2-branes'',
  {\it JHEP} {\bf 0802} (2008) 105
  [arXiv:0712.3738 [hep-th]].
  %%CITATION = ARXIV:0712.3738;%%
  
%\cite{Gustavsson:2007vu}
%\bibitem{Gustavsson:2007vu}
\bibitem{RBGustavsson1}
  A.~Gustavsson,
  ``Algebraic structures on parallel M2-branes'',
    {\it Nucl. Phys.} {\bf B811} (2009) 66
        [arXiv:0709.1260 [hep-th]].
          %%CITATION = ARXIV:0709.1260;%%

%\cite{Gustavsson:2008dy}
%\bibitem{Gustavsson:2008dy}
\bibitem{RBGustavsson2}
  A.~Gustavsson,
  ``Selfdual strings and loop space Nahm equations'',
  {\it JHEP} {\bf 0804} (2008) 083
        [arXiv:0802.3456 [hep-th]].
          %%CITATION = ARXIV:0802.3456;%%

%\cite{VanRaamsdonk:2008ft}
\bibitem{RBVanRaamsdonkBiFundamental}
  M.~Van Raamsdonk,
  ``Comments on the Bagger-Lambert theory and multiple M2-branes'',
  {\it JHEP} {\bf 0805} (2008) 105
  [arXiv:0803.3803 [hep-th]].
  %%CITATION = ARXIV:0803.3803;%%
  
%\cite{Hosomichi:2008jd}
%\bibitem{Hosomichi:2008jd}
\bibitem{RBHosomichiLeeLeeLeePark}
  K.~Hosomichi, K.~-M.~Lee, S.~Lee, S.~Lee and J.~Park,
  ``N=4 Superconformal Chern-Simons Theories with Hyper and 
  Twisted Hyper Multiplets'',
   {\it JHEP} {\bf 0807} (2008) 091
        [arXiv:0805.3662 [hep-th]].
          %%CITATION = ARXIV:0805.3662;%%

%\cite{Gaiotto:2007qi}
%\bibitem{Gaiotto:2007qi}
\bibitem{RBGaiottoYin}
  D.~Gaiotto and X.~Yin,
  ``Notes on superconformal Chern-Simons-Matter theories'',
  {\it JHEP} {\bf 0708} (2007) 056
  [arXiv:0704.3740 [hep-th]].
  %%CITATION = ARXIV:0704.3740;%%

%\cite{Nilsson:1984bj}
%\bibitem{Nilsson:1984bj}
\bibitem{RBNilssonPope}
  B.~E.~W.~Nilsson and C.~N.~Pope,
 ``Hopf Fibration Of Eleven-dimensional Supergravity'',
 {\it Class. Quant. Grav.}  {\bf 1} (1984) 499.
  %%CITATION = CQGRD,1,499;%%
  
%\cite{hep-th/0202021}
%\bibitem{hep-th/0202021}
\bibitem{RBBMN}
  D.~E.~Berenstein, J.~M.~Maldacena and H.~S.~Nastase,
   ``Strings in flat space and pp waves from N=4 superYang-Mills'',
    {\it JHEP} {\bf 0204} (2002) 013
        [hep-th/0202021].
          %%CITATION = JHEPA,0204,013;%%

%\cite{Gubser:2002tv}
%\bibitem{Gubser:2002tv}
\bibitem{RBGKP2}
  S.~S.~Gubser, I.~R.~Klebanov and A.~M.~Polyakov,
  ``A Semiclassical limit of the gauge / string correspondence'',
  {\it Nucl. Phys.} {\bf B636} (2002) 99
  [hep-th/0204051].
  %%CITATION = HEP-TH/0204051;%%

%\cite{Frolov:2002av}
%\bibitem{Frolov:2002av}
\bibitem{RBFrolovTseytlin}
  S.~Frolov and A.~A.~Tseytlin,
   ``Semiclassical quantization of rotating superstring in AdS(5) x S**5'',
    {\it JHEP} {\bf 0206} (2002) 007
        [hep-th/0204226].
          %%CITATION = HEP-TH/0204226;%%

%\cite{Dasgupta:2002hx}
%\bibitem{Dasgupta:2002hx}
\bibitem{RBDSVR1}
  K.~Dasgupta, M.~M.~Sheikh-Jabbari and M.~Van Raamsdonk,
  ``Matrix perturbation theory for M theory on a PP wave'',
  {\it JHEP} {\bf 0205} (2002) 056
  [hep-th/0205185].
  %%CITATION = HEP-TH/0205185;%%

%\cite{Dasgupta:2002ru}
%\bibitem{Dasgupta:2002ru}
\bibitem{RBDSVR2}
  K.~Dasgupta, M.~M.~Sheikh-Jabbari and M.~Van Raamsdonk,
   ``Protected multiplets of M theory on a plane wave'',
    {\it JHEP} {\bf 0209} (2002) 021
        [hep-th/0207050].
          %%CITATION = JHEPA,0209,021;%%

%\cite{Kim:2002if}
%\bibitem{Kim:2002if}
\bibitem{RBKimPlefka}
  N.~Kim and J.~Plefka,
    ``On the spectrum of PP wave matrix theory'',
    {\it Nucl. Phys.} {\bf B643} (2002) 31
        [hep-th/0207034].
          %%CITATION = HEP-TH/0207034;%%


%\cite{'tHooft:1977hy}
%\bibitem{'tHooft:1977hy}
\bibitem{RBtHooftMonopoleOperators}
  G.~'t Hooft,
    ``On the Phase Transition Towards Permanent Quark Confinement'',
      {\it Nucl. Phys.} {\bf B138} (1978) 1.
        %%CITATION = NUPHA,B138,1;%%

%\cite{Borokhov:2002ib}
%\bibitem{Borokhov:2002ib}
\bibitem{RBBorokhovKapustinWu1}
  V.~Borokhov, A.~Kapustin and X.~-k.~Wu,
  ``Topological disorder operators in three-dimensional conformal 
  field theory'',
  {\it JHEP} {\bf 0211} (2002) 049
  [hep-th/0206054].
  %%CITATION = HEP-TH/0206054;%%

%\cite{Borokhov:2002cg}
\bibitem{RBBorokhovKapustinWu2}
  V.~Borokhov, A.~Kapustin and X.~-k.~Wu,
   ``Monopole operators and mirror symmetry in three-dimensions'',
   {\it JHEP} {\bf 0212} (2002) 044
        [hep-th/0207074].
          %%CITATION = HEP-TH/0207074;%%

%\cite{Goddard:1976qe}
\bibitem{RBGNO}
  P.~Goddard, J.~Nuyts and D.~I.~Olive,
    ``Gauge Theories and Magnetic Charge'',
    {\it Nucl. Phys.} {\bf B125} (1977) 1.
        %%CITATION = NUPHA,B125,1;%%

%\cite{Dirac:1931kp}
%\bibitem{Dirac:1931kp}
\bibitem{RBDiracMonopole}
  P.~A.~M.~Dirac,
    ``Quantized Singularities in the Electromagnetic Field'',
      {\it Proc. Roy. Soc. Lond.}  {\bf A 133} (1931) 60.
        %%CITATION = PRSLA,A133,60;%%

%\cite{Wu:1975es}
% \bibitem{Wu:1975es}
\bibitem{RBWuYangQuantisationCondition}
  T.~T.~Wu and C.~N.~Yang,
    ``Concept of Nonintegrable Phase Factors and Global Formulation 
    of Gauge Fields'',
     {\it Phys. Rev.} {\bf D12} (1975) 3845.
        %%CITATION = PHRVA,D12,3845;%%


%\cite{Kim:2009wb}
%\bibitem{Kim:2009wb}
\bibitem{RBSKim}
  S.~Kim,
   ``The Complete superconformal index for N=6 Chern-Simons theory'',
   {\it Nucl. Phys.} {\bf B821} (2009) 241
        [arXiv:0903.4172 [hep-th]].
          %%CITATION = ARXIV:0903.4172;%%

%\cite{SheikhJabbari:2009kr}
%\bibitem{SheikhJabbari:2009kr}
\bibitem{RBSheikhJabbariSimon}
  M.~M.~Sheikh-Jabbari and J.~Simon,
   ``On Half-BPS States of the ABJM Theory'',
    {\it JHEP} {\bf 0908} (2009) 073
        [arXiv:0904.4605 [hep-th]].
          %%CITATION = ARXIV:0904.4605;%%

%\cite{Benna:2009xd}
%\bibitem{Benna:2009xd}
\bibitem{RBBennaKlebanovKlose}
  M.~K.~Benna, I.~R.~Klebanov and T.~Klose,
    ``Charges of Monopole Operators in Chern-Simons Yang-Mills 
    Theory'',
    {\it JHEP} {\bf 1001} (2010) 110
        [arXiv:0906.3008 [hep-th]].
          %%CITATION = ARXIV:0906.3008;%%


%\cite{Berenstein:2009sa}
%\bibitem{Berenstein:2009sa}
\bibitem{RBBerensteinPark}
  D.~Berenstein and J.~Park,
    ``The BPS spectrum of monopole operators in ABJM:
    Towards a field theory description of the giant torus'',
      {\it JHEP} {\bf 1006} (2010) 073
        [arXiv:0906.3817 [hep-th]].
          %%CITATION = ARXIV:0906.3817;%%

%\cite{Bashkirov:2010kz}
%\bibitem{Bashkirov:2010kz}
\bibitem{RBBashkirovKapustin}
  D.~Bashkirov and A.~Kapustin,
   ``Supersymmetry enhancement by monopole operators'',
   {\it JHEP} {\bf 1105} (2011) 015
        [arXiv:1007.4861 [hep-th]].
          %%CITATION = ARXIV:1007.4861;%%


%\cite{Minahan:2008hf}
\bibitem{RBMinahanZarembo}
  J.~A.~Minahan and K.~Zarembo,
    ``The Bethe ansatz for superconformal Chern-Simons'',
      {\it JHEP} {\bf 0809} (2008) 040
        [arXiv:0806.3951 [hep-th]].
          %%CITATION = ARXIV:0806.3951;%%

%\cite{Gaiotto:2008cg}
\bibitem{RBGaiottoGiombiYin}
  D.~Gaiotto, S.~Giombi and X.~Yin,
    ``Spin Chains in N=6 Superconformal Chern-Simons-Matter Theory'',
      {\it JHEP} {\bf 0904} (2009) 066
        [arXiv:0806.4589 [hep-th]].
          %%CITATION = ARXIV:0806.4589;%%

%\cite{Grignani:2008is}
\bibitem{RBGrignaniHarmarkOrselli}
  G.~Grignani, T.~Harmark and M.~Orselli,
    %``The SU(2) x SU(2) sector in the string dual of N=6 superconformal Chern-Simons theory,''
      {\it Nucl. Phys.} {\bf B810} (2009) 115
        [arXiv:0806.4959 [hep-th]].
          %%CITATION = ARXIV:0806.4959;%%

%\cite{Klose:2010ki}
\bibitem{RBKlose}
  T.~Klose,
    ``Review of AdS/CFT Integrability, Chapter IV.3: N=6 Chern-Simons and Strings on AdS4xCP3,''
      {\it Lett. Math. Phys.} {\bf 99} (2012) 401
        [arXiv:1012.3999 [hep-th]].
          %%CITATION = ARXIV:1012.3999;%%

%\cite{Beisert:2010jr}
%\bibitem{Beisert:2010jr}
\bibitem{RBIntegrabilityReview}
  N.~Beisert, C.~Ahn, L.~F.~Alday, Z.~Bajnok, J.~M.~Drummond, 
  L.~Freyhult, N.~Gromov, R.~A.~Janik {\it et al.},
  ``Review of AdS/CFT Integrability: An Overview'',
  {\it Lett. Math. Phys.}  {\bf 99} (2012) 3
  [arXiv:1012.3982 [hep-th]].
  %%CITATION = ARXIV:1012.3982;%%

%\cite{Nishioka:2008gz}
\bibitem{RBNishiokaTakayanagiPP}
  T.~Nishioka and T.~Takayanagi,
    ``On Type IIA Penrose Limit and N=6 Chern-Simons Theories'',
      {\it JHEP} {\bf 0808} (2008) 001
        [arXiv:0806.3391 [hep-th]].
          %%CITATION = ARXIV:0806.3391;%%



%\cite{Kapustin:2009kz}
\bibitem{RBKapustinWilletYaakov}
  A.~Kapustin, B.~Willett and I.~Yaakov,
    ``Exact Results for Wilson Loops in Superconformal Chern-Simons Theories with Matter'',
      {\it JHEP} {\bf 1003} (2010) 089
        [arXiv:0909.4559 [hep-th]].
          %%CITATION = ARXIV:0909.4559;%%

%\cite{Drukker:2010nc}
\bibitem{RBDrukkerMarinoPutrov}
  N.~Drukker, M.~Marino and P.~Putrov,
    ``From weak to strong coupling in ABJM theory'',
      {\it Commun. Math. Phys.}  {\bf 306} (2011) 511
        [arXiv:1007.3837 [hep-th]].
          %%CITATION = ARXIV:1007.3837;%%

%\cite{Drukker:2011zy}
\bibitem{RBDrukkerMarinoPutrov2}
  N.~Drukker, M.~Marino and P.~Putrov,
    ``Nonperturbative aspects of ABJM theory'',
      {\it JHEP} {\bf 1111} (2011) 141
        [arXiv:1103.4844 [hep-th]].
          %%CITATION = ARXIV:1103.4844;%%
            
%\cite{Fuji:2011km}
\bibitem{RBFujiHiranoMoriyama}
  H.~Fuji, S.~Hirano and S.~Moriyama,
    ``Summing Up All Genus Free Energy of ABJM Matrix Model'',
      {\it JHEP} {\bf 1108} (2011) 001
        [arXiv:1106.4631 [hep-th]].
          %%CITATION = ARXIV:1106.4631;%%

%\cite{Klebanov:1996un}
%\bibitem{Klebanov:1996un}
\bibitem{RBKlebanovTseytlin}
  I.~R.~Klebanov and A.~A.~Tseytlin,
    ``Entropy of near extremal black p-branes,''
    {\it Nucl. Phys.} {\bf B475} (1996) 164
        [hep-th/9604089].
          %%CITATION = HEP-TH/9604089;%%

\bibitem{RBPenrose}
R. Penrose, “Any spacetime has a plane wave as a limit”, in 
{\it Differential geometry and relativity}, pp. 271-275, M. Cahen and 
M. Flato editors (1976).

%\cite{Gueven:2000ru}
%\bibitem{Gueven:2000ru}
\bibitem{RBGueven}
  R.~Gueven,
  ``Plane wave limits and T duality'',
  {\it Phys. Lett.}  {\bf B482} (2000) 255
  [hep-th/0005061].
  %%CITATION = HEP-TH/0005061;%%

%\cite{Blau:2002dy}
\bibitem{RBppBlauHulletal2}
  M.~Blau, J.~M.~Figueroa-O'Farrill, C.~Hull and G.~Papadopoulos,
  ``Penrose limits and maximal supersymmetry'',
  {\it Class. Quant. Grav.}  {\bf 19} (2002) L87
  [hep-th/0201081].
  %%CITATION = HEP-TH/0201081;%%
   
%\bibitem{Dobashi:2002ar}
\bibitem{RBDobashiShimadaYoneya}
  S.~Dobashi, H.~Shimada and T.~Yoneya,
   ``Holographic reformulation of string theory on AdS(5) x S**5 
   background in the PP wave limit'',
    {\it Nucl. Phys.} {\bf B665} (2003) 94
        [hep-th/0209251].
          %%CITATION = HEP-TH/0209251;%%

%\cite{Shimada:2004sw}
%\bibitem{Shimada:2004sw}
\bibitem{RBShimada3}
H.~Shimada,
  ``Holography at string field theory level: Conformal three point 
  functions of BMN operators'',
   {\it Phys. Lett.} {\bf B647} (2007) 211
      [hep-th/0410049].
        %%CITATION = HEP-TH/0410049;%%
        
%\cite{Gubser:1998bc}
%\bibitem{Gubser:1998bc}
\bibitem{RBGKP}
  S.~S.~Gubser, I.~R.~Klebanov and A.~M.~Polyakov,
   ``Gauge theory correlators from noncritical string theory'',
    {\it Phys. Lett.} {\bf 428} (1998) 105
        [hep-th/9802109].
          %%CITATION = HEP-TH/9802109;%%

%\cite{Witten:1998qj}
%\bibitem{Witten:1998qj}
\bibitem{RBWittenHolography}
  E.~Witten,
   ``Anti-de Sitter space and holography'',
    {\it Adv. Theor. Math. Phys.}  {\bf 2} (1998) 253
        [hep-th/9802150].
          %%CITATION = HEP-TH/9802150;%%

%\cite{Asano:2003xp}
%\bibitem{Asano:2003xp}
\bibitem{RBAsanoSekinoYoneya}
  M.~Asano, Y.~Sekino and T.~Yoneya,
   ``PP wave holography for Dp-brane backgrounds'',
    {\it Nucl. Phys.} {\bf 678} (2004) 197
        [hep-th/0308024].
          %%CITATION = HEP-TH/0308024;%%

%\cite{Tsuji:2006zn}
\bibitem{RBTsuji}
  A.~Tsuji,
    ``Holography of Wilson loop correlator and spinning strings'',
      {\it Prog. Theor. Phys.}  {\bf 117} (2007) 557
        [hep-th/0606030].
          %%CITATION = HEP-TH/0606030;%%

%\cite{Janik:2010gc}
\bibitem{RBJanikSurowkaWereszczynski}
  R.~A.~Janik, P.~Surowka and A.~Wereszczynski,
    ``On correlation functions of operators dual to classical spinning 
    string states'',
      {\it JHEP} {\bf 1005} (2010) 030
        [arXiv:1002.4613 [hep-th]].
          %%CITATION = ARXIV:1002.4613;%%

%\cite{KowalskiGlikman:1984wv}
%\bibitem{KowalskiGlikman:1984wv}
\bibitem{RBKowalskiGlikman}
  J.~Kowalski-Glikman,
   ``Vacuum States in Supersymmetric Kaluza-Klein Theory'',
    {\it Phys. Lett.} {\bf B134} (1984) 194.
        %%CITATION = PHLTA,B134,194;%%

%\cite{Sugiyama:2002rs}
\bibitem{RBSugiyamaYoshida1}
  K.~Sugiyama and K.~Yoshida,
   ``Supermembrane on the PP wave background'',
    {\it Nucl. Phys.} {\bf B644} (2002) 113
        [hep-th/0206070].
          %%CITATION = HEP-TH/0206070;%%

%\cite{deWit:1998yu}
\bibitem{RBdeWitPeetersPlefkaSevrin}
  B.~de Wit, K.~Peeters, J.~Plefka and A.~Sevrin,
   ``The M theory two-brane in AdS(4) x S**7 and AdS(7) x S**4'',
    {\it Phys. Lett.} {\bf B443} (1998) 153
        [hep-th/9808052].
          %%CITATION = HEP-TH/9808052;%%

%\cite{Susskind:1997cw}
\bibitem{RBSusskindDLCQ}
  L.~Susskind,
   ``Another conjecture about M(atrix) theory'',
    hep-th/9704080.
        %%CITATION = HEP-TH/9704080;%%

%\cite{Shimada:2008xy}
\bibitem{RBShimadaBetaMM}
  H.~Shimada,
    ``Beta-deformation for matrix model of M-theory'',
     {\it Nucl. Phys.}  {\bf B813} (2009) 283
        [arXiv:0804.3236 [hep-th]].
          %%CITATION = ARXIV:0804.3236;%%

%\cite{Yee:2003ge}
%\bibitem{Yee:2003ge}
\bibitem{RBYeeYi}
J.~-T.~Yee and P.~Yi,
``Instantons of M(atrix) theory in PP wave background'',
{\it JHEP} {\bf 0302} (2003) 040
[hep-th/0301120].
%%CITATION = HEP-TH/0301120;%%

%\cite{Bachas:2000dx}
\bibitem{RBBachasHoppePioline}
C.~Bachas, J.~Hoppe and B.~Pioline,
 ``Nahm equations, N=1* domain walls, and D strings in AdS(5) x S(5)'',
  {\it JHEP} {\bf 0107} (2001) 041
   [hep-th/0007067].
          %%CITATION = HEP-TH/0007067;%%

\bibitem{RBMaldacenaSheikhJabbariVanRaamsdonk}
%\cite{Maldacena:2002rb}
%\bibitem{Maldacena:2002rb}
  J.~M.~Maldacena, M.~M.~Sheikh-Jabbari and M.~Van Raamsdonk,
  ``Transverse five-branes in matrix theory'',
  {\it JHEP} {\bf 0301} (2003) 038
  [hep-th/0211139].
  %%CITATION = HEP-TH/0211139;%%

%\cite{Lozano:2013sra}
%\bibitem{Lozano:2013sra}
\bibitem{RBLozanoPrinsloo}
  Y.~Lozano and A.~Prinsloo,
  ``$S^2 \times S^3$ geometries in ABJM and giant gravitons'',
  {\it JHEP} {\bf 04} (2013) 148
  [arXiv:1303.3748 [hep-th]].
  %%CITATION = ARXIV:1303.3748;%%

%\cite{Balasubramanian:2002sa}
%\bibitem{Balasubramanian:2002sa}
\bibitem{RBBalasburamanianEtal}
  V.~Balasubramanian, M.~-x.~Huang, T.~S.~Levi and A.~Naqvi,
  ``Open strings from N=4 superYang-Mills'',
  {\it JHEP} {\bf 0208} (2002) 037
  [hep-th/0204196].
  %%CITATION = HEP-TH/0204196;%%


%\cite{Fairlie:1988qd}
%\bibitem{Fairlie:1988qd}
\bibitem{RBFairlieFletcherZachos1}
  D.~B.~Fairlie, P.~Fletcher and C.~K.~Zachos,
    ``Trigonometric Structure Constants for New Infinite Algebras'',
      {\it Phys. Lett.}  {\bf B218} (1989) 203.
        %%CITATION = PHLTA,B218,203;%%

%\cite{Fairlie:1989vv}
%\bibitem{Fairlie:1989vv}
\bibitem{RBFairlieFletcherZachos2}
  D.~B.~Fairlie and C.~K.~Zachos,
    ``Infinite Dimensional Algebras, Sine Brackets and SU(Infinity)'',
     {\it Phys. Lett.}  {\bf B224} (1989) 101.
        %%CITATION = PHLTA,B224,101;%%
 
%\cite{Floratos:1989au}
%\bibitem{Floratos:1989au}
\bibitem{RBFloratos}
  E.~G.~Floratos,
    ``The Heisenberg-Weyl Group On The Z(n) X Z(n) Discretized 
    Torus Membrane,''
      {\it Phys. Lett.} {\bf B228} (1989) 335.
        %%CITATION = PHLTA,B228,335;%%

%\cite{Bandres:2008ry}
%\bibitem{Bandres:2008ry}
\bibitem{RBBandresLipsteinSchwarz}
  M.~A.~Bandres, A.~E.~Lipstein and J.~H.~Schwarz,
   ``Studies of the ABJM Theory in a Formulation with Manifest 
   SU(4) R-Symmetry'',
    {\it JHEP} {\bf 0809} (2008) 027
        [arXiv:0807.0880 [hep-th]].
          %%CITATION = ARXIV:0807.0880;%%


%\cite{Gustavsson:2009pm}
\bibitem{RBGustavssonRey}
  A.~Gustavsson and S.~-J.~Rey,
    ``Enhanced N=8 Supersymmetry of ABJM Theory on R**8 and R**8/Z(2)'',
      arXiv:0906.3568 [hep-th].
        %%CITATION = ARXIV:0906.3568;%%


%\cite{Berenstein:2008dc}
%\bibitem{Berenstein:2008dc}
\bibitem{RBBerensteinTrancanelli}
  D.~Berenstein and D.~Trancanelli,
    ``Three-dimensional N=6 SCFT's and their membrane dynamics'',
     {\it Phys. Rev.} {\bf D78} (2008) 106009
        [arXiv:0808.2503 [hep-th]].
          %%CITATION = ARXIV:0808.2503;%%

%\cite{Kim:2009ia}
%\bibitem{Kim:2009ia}
\bibitem{RBKimMadhu}
  S.~Kim and K.~Madhu,
    ``Aspects of monopole operators in N=6 Chern-Simons theory'',
     {\it JHEP} {\bf 0912} (2009) 018
        [arXiv:0906.4751 [hep-th]].
          %%CITATION = ARXIV:0906.4751;%%


\bibitem{RBGolubVanLoan}
See for example:\\
G.~H.~Golub and C.~F.~Van Loan,
``Matrix computations'', Johns Hopkins University Press (1996).

%\cite{ITP-SB-76-5}
%\bibitem{ITP-SB-76-5}
\bibitem{RBWuYangHarmonics}
  T.~T.~Wu and C.~N.~Yang,
   ``Dirac Monopole Without Strings: Monopole Harmonics'',
    {\it Nucl. Phys.} {\bf B107} (1976) 365.
        %%CITATION = NUPHA,B107,365;%%

%\cite{SheikhJabbari:2004ik}
\bibitem{RBSheikhJabbari}
  M.~M.~Sheikh-Jabbari,
    ``Tiny graviton matrix theory: DLCQ of IIB plane-wave string theory, 
    a conjecture'',
      {\it JHEP} {\bf 0409} (2004) 017
        [hep-th/0406214].
          %%CITATION = HEP-TH/0406214;%%

\bibitem{RBDiracConstraint}
    P.~A.~M.~Dirac, ``Lectures on quantum mechanics'',
    Yeshiva Univ., N. Y. (1964); Dover (2001).

%\cite{Goddard:1973qh}
%\bibitem{Goddard:1973qh}
\bibitem{RBGGRT}
  P.~Goddard, J.~Goldstone, C.~Rebbi and C.~B.~Thorn,
    ``Quantum dynamics of a massless relativistic string'',
     {\it Nucl. Phys.} {\bf B56} (1973) 109.
        %%CITATION = NUPHA,B56,109;%%

%\cite{Eguchi:1980jx}
\bibitem{RBHodgeTheorem}
See for example:\\
  T.~Eguchi, P.~B.~Gilkey and A.~J.~Hanson,
    ``Gravitation, Gauge Theories and Differential Geometry'',
      {\it Phys. Rept.}  {\bf 66} (1980) 213. \\
        %%CITATION = PRPLC,66,213;%%
%\cite{Nakahara:2003nw}
%\bibitem{Nakahara:2003nw}
%\bibitem{RBNakahara}
M.~Nakahara, ``Geometry, Topology and Physics'',
IOP Publishing (1990).

%\cite{ITP-SB-76/63}
%\bibitem{ITP-SB-76/63}
\bibitem{RBKazamaYangGoldhaber}
        Y.~Kazama, C.~N.~Yang and A.~S.~Goldhaber,
        ``Scattering of a Dirac Particle with Charge Ze by a Fixed 
        Magnetic Monopole'',
        {\it Phys. Rev.} {\bf D15} (1977) 2287.
        %%CITATION = PHRVA,D15,2287;%%

%\cite{Weinberg:1993sg}
\bibitem{RBWeinberg}
  E.~J.~Weinberg,
   ``Monopole vector spherical harmonics'',
   {\it Phys. Rev.} {\bf D49} (1994) 1086
        [hep-th/9308054].
          %%CITATION = HEP-TH/9308054;%%

%\cite{Bachas:1995kx}
%\bibitem{Bachas:1995kx}
\bibitem{RBBachas}
  C.~Bachas,
   ``D-brane dynamics'',
   {\it Phys. Lett.} {\bf B374} (1996) 37
        [hep-th/9511043].
          %%CITATION = HEP-TH/9511043;%%

%\cite{Danielsson:1996uw}
%\bibitem{Danielsson:1996uw}
\bibitem{RBDanielssonFerrettiSundborg}
  U.~H.~Danielsson, G.~Ferretti and B.~Sundborg,
   ``D particle dynamics and bound states'',
    {\it Int. J. Mod. Phys.} {\bf A11} (1996) 5463
        [hep-th/9603081].
          %%CITATION = HEP-TH/9603081;%%

%\cite{Kabat:1996cu}
%\bibitem{Kabat:1996cu}
\bibitem{RBKabatPouliot}
  D.~N.~Kabat and P.~Pouliot,
   ``A Comment on zero-brane quantum mechanics'',
   {\it Phys. Rev. Lett.}  {\bf 77} (1996) 1004
        [hep-th/9603127].
          %%CITATION = HEP-TH/9603127;%%

%\cite{Lifschytz:1996iq}
%\bibitem{Lifschytz:1996iq}
\bibitem{RBLifschytz}
  G.~Lifschytz,
   ``Comparing d-branes to black-branes'',
   {\it Phys. Lett.} {\bf B388} (1996) 720
        [hep-th/9604156].
          %%CITATION = HEP-TH/9604156;%%

%\cite{Douglas:1996yp}
%\bibitem{Douglas:1996yp}
\bibitem{RBDKPS}
  M.~R.~Douglas, D.~N.~Kabat, P.~Pouliot and S.~H.~Shenker,
   ``D-branes and short distances in string theory'',
   {\it Nucl. Phys.} {\bf B485} (1997) 85
        [hep-th/9608024].
          %%CITATION = HEP-TH/9608024;%%

%\cite{de Wit:1988ct}
%\bibitem{de Wit:1988ct}
\bibitem{RBdeWitLuescherNicolai}
  B.~de Wit, M.~L\"{u}scher and H.~Nicolai,
    ``The Supermembrane Is Unstable'',
      {\it Nucl. Phys.} {\bf B320} (1989) 135.
        %%CITATION = NUPHA,B320,135;%%

%\cite{Wilson:1973jj}
\bibitem{RBWilsonKogut}
  K.~G.~Wilson and J.~B.~Kogut,
    ``The Renormalization group and the epsilon expansion'',
     {\it Phys. Rept.}  {\bf 12} (1974) 75.
        %%CITATION = PRPLC,12,75;%%

%\cite{Becker:1997xw}
%\bibitem{Becker:1997xw}
\bibitem{RBBecker2PolchinskiTseytlin}
  K.~Becker, M.~Becker, J.~Polchinski and A.~A.~Tseytlin,
  ``Higher order graviton scattering in M(atrix) theory'',
  {\it Phys. Rev.} {\bf D56} (1997) 3174
  [hep-th/9706072].
  %%CITATION = HEP-TH/9706072;%%

%\cite{Hanada:2011yz}
%\bibitem{Hanada:2011yz}
\bibitem{RBHanadaHoyosShimada}
  M.~Hanada, C.~Hoyos and H.~Shimada,
   ``On a new type of orbifold equivalence and M-theoretic 
   $AdS_4/CFT_3$ duality'',
     {\it Phys. Lett.} {\bf B707} (2012) 394
        [arXiv:1109.6127 [hep-th]].
          %%CITATION = ARXIV:1109.6127;%%

%\cite{Peskin:1997qi}
%\bibitem{Peskin:1997qi}
\bibitem{RBPeskinDuality}
  M.~E.~Peskin,
  ``Duality in supersymmetric Yang-Mills theory'',
  hep-th/9702094.
  %%CITATION = HEP-TH/9702094;%%

%\cite{Kazama:2002jm}
%\bibitem{Kazama:2002jm}
\bibitem{RBKazamaMuramatsu}
  Y.~Kazama and T.~Muramatsu,
    ``Power of supersymmetry in D particle dynamics'',
      {\it Nucl. Phys.} {\bf B656} (2003) 93
        [hep-th/0210133].
          %%CITATION = HEP-TH/0210133;%%

\bibitem{RBIshikiShimasakitakayamaTsuchiya}
%\cite{Ishiki:2006yr}
%\bibitem{Ishiki:2006yr}
  G.~Ishiki, S.~Shimasaki, Y.~Takayama and A.~Tsuchiya,
    ``Embedding of theories with SU(2$|$4) symmetry into the plane 
    wave matrix model'',
      {\it JHEP} {\bf 0611} (2006) 089
        [hep-th/0610038].
          %%CITATION = HEP-TH/0610038;%%

%\cite{Bergshoeff:1987cm}
%\bibitem{Bergshoeff:1987cm}
\bibitem{RBBergshoeffSezginTownsend}
  E.~Bergshoeff, E.~Sezgin and P.~K.~Townsend,
    ``Supermembranes and Eleven-Dimensional Supergravity'',
     {\it Phys. Lett.} {\bf B189} (1987) 75.
        %%CITATION = PHLTA,B189,75;%%
\\        
%\cite{Bergshoeff:1987qx}
%\bibitem{Bergshoeff:1987qx}
  E.~Bergshoeff, E.~Sezgin and P.~K.~Townsend,
    ``Properties of the Eleven-Dimensional Super Membrane Theory'',
     {\it Annals Phys.}  {\bf 185} (1988) 330.
        %%CITATION = APNYA,185,330;%%

%\cite{Hashimoto:2000zp}
%\bibitem{Hashimoto:2000zp}
\bibitem{RBHashimotoHiranoItzhaki}
A.~Hashimoto, S.~Hirano and N.~Itzhaki,
``Large branes in AdS and their field theory dual'',
{\it JHEP} {\bf 0008} (2000) 051
[hep-th/0008016].
%%CITATION = HEP-TH/0008016;%%

%\cite{Grignani:2012ur}
\bibitem{RBGrignaniZayakin}
  G.~Grignani and A.~V.~Zayakin,
   ``Three-point functions of BMN operators at weak and strong 
   coupling II. One loop matching'',
      {\it JHEP} {\bf 1209} (2012) 087
        [arXiv:1205.5279 [hep-th]].
          %%CITATION = ARXIV:1205.5279;%%

%\cite{Motl:1997th}
\bibitem{RBMotl}
L.~Motl,
    ``Proposals on nonperturbative superstring interactions'',
hep-th/9701025.

%\cite{Dijkgraaf:1997vv}
\bibitem{RBDijkgraafVerlindeVerlinde}
R.~Dijkgraaf, E.~P.~Verlinde and H.~L.~Verlinde,
    ``Matrix string theory'',
{\it Nucl. Phys.} {\bf B500} (1997) 43
[hep-th/9703030].

%\cite{Sekino:2001ai}
\bibitem{RBSekinoYoneya}
  Y.~Sekino and T.~Yoneya,
    ``From supermembrane to matrix string'',
      {\it Nucl. Phys.} {\bf B619} (2001) 22
        [hep-th/0108176].
          %%CITATION = HEP-TH/0108176;%%

%\cite{Bonelli:2002mb}
\bibitem{RBBonelli}
  G.~Bonelli,
    ``Matrix strings in pp wave backgrounds from deformed 
    superYang-Mills theory'',
      {\it JHEP} {\bf 0208} (2002) 022
        [hep-th/0205213].

%\cite{Sugiyama:2002tf}
\bibitem{RBSugiyamaYoshidaMatrixString}
  K.~Sugiyama and K.~Yoshida,
    ``Type IIA string and matrix string on PP wave'',
      {\it Nucl. Phys.} {\bf B644} (2002) 128
        [hep-th/0208029].
          %%CITATION = HEP-TH/0208029;%%

%\cite{Nishioka:2008ib}
\bibitem{RBNishiokaTakayanagiTorus}
  T.~Nishioka and T.~Takayanagi,
    ``Fuzzy Ring from M2-brane Giant Torus'',
      {\it JHEP} {\bf 0810} (2008) 082
        [arXiv:0808.2691 [hep-th]].
          %%CITATION = ARXIV:0808.2691;%%

%\cite{Herrero:2011bk}
%\bibitem{Herrero:2011bk}
\bibitem{RBHerreroLozanoPicos}
  M.~Herrero, Y.~Lozano and M.~Picos,
    ``Dielectric 5-Branes and Giant Gravitons in ABJM'',
      {\it JHEP} {\bf 1108} (2011) 132
        [arXiv:1107.5475 [hep-th]].
          %%CITATION = ARXIV:1107.5475;%%

%\cite{Ezhuthachan:2011kf}
\bibitem{RBEzhuthachanShimasakiYokoyama}
  B.~Ezhuthachan, S.~Shimasaki and S.~Yokoyama,
    ``BPS solutions in ABJM theory and Maximal Super Yang-Mills 
    on $R\times S^2$'',
      {\it JHEP} {\bf 1112} (2011) 048
        [arXiv:1107.3545 [hep-th]].
          %%CITATION = ARXIV:1107.3545;%%

%\cite{Bak:2005jh}
\bibitem{RBBakKimLee}
  D.~Bak, S.~Kim and K.~-M.~Lee,
    ``All higher genus BPS membranes in the plane wave background'',
      {\it JHEP} {\bf 0506} (2005) 035
        [hep-th/0501202].
          %%CITATION = HEP-TH/0501202;%%

\end{thebibliography}
\end{document}